%% file: main.tex
\documentclass[10pt,lettersize,journal]{IEEEtran}
\usepackage{amssymb,amsmath,amsthm,amsfonts}
\usepackage{graphicx}

\usepackage{float}
\usepackage{color}
\usepackage{xcolor}
\definecolor{skyblue}{RGB}{135, 206, 235}
\definecolor{lightcoral}{RGB}{240, 128, 128}
\usepackage{stmaryrd}
\usepackage{url,hyperref}
\usepackage{hyperref}
\usepackage{multirow}
\usepackage{threeparttable}
\usepackage{braket}
\usepackage{stfloats}

\usepackage{subcaption}
\usepackage{array}
\usepackage{makecell} 
\usepackage{parskip}
\usepackage{proof}
\usepackage{algorithm,algorithmic}
\usepackage{adjustbox}

\usepackage{qcircuit}
\usepackage[compat=0.4]{yquant}
\usepackage{tikz}

\usetikzlibrary{calc,patterns}
\usetikzlibrary{shapes.geometric, arrows, positioning}

\tikzstyle{startstop} = [rectangle, rounded corners, minimum width=1cm, minimum height=1cm, text centered, draw=black]
\tikzstyle{process} = [rectangle, minimum width=1cm, minimum height=1cm, text centered, draw=black]
\tikzstyle{decision} = [diamond, minimum width=1cm, minimum height=1cm, text centered, draw=black]
\tikzstyle{arrow} = [thick,->,>=stealth]

\usepackage{pgfplots}
\pgfplotsset{compat=1.16}
\usepackage{xcolor}
\definecolor{darkorange}{rgb}{1.0, 0.55, 0.0} 

\tikzset{
    DasAtomStyle/.style={color={rgb,255:red,0;green,0;blue,0}, ultra thick},
    TetrisStyle/.style={color={rgb,255:red,92;green,167;blue,199}, ultra thick, dotted}, 
    EnolaStyle/.style={color={rgb,255:red,251;green,53;blue,45}, ultra thick, loosely dashed}, 
    AtomiqueStyle/.style={color={rgb,255:red,192;green,96;blue,0}, ultra thick, loosely dashdotdotted}
}

\newcommand{\define}{\ensuremath{\triangleq}}

\newcommand{\mathceil}[1]{\left\lceil #1 \right\rceil}

\newtheorem{example}{Example}
\newtheorem{remark}{Remark}

\theoremstyle{definition}
\newtheorem{definition}{Definition}

\usepackage{comment}

\newcommand{\blue}[1]{{\color{black}{#1}}}

\usepackage[normalem]{ulem}

\begin{document}

\title{DasAtom: A Divide-and-Shuttle Atom Approach to Quantum Circuit Transformation\thanks{Work partially supported by the Australian Research Council (DP220102059), the National Science Foundation of China (12071271, 12471437), the Beijing Nova Program (20220484128), the Innovation Program for Quantum Science and Technology (2021ZD0302901) and the Young Scientists Fund of the Natural Science Foundation of Jiangsu Province (BK20240536)}}
\author{Yunqi Huang, Dingchao Gao, Shenggang Ying, and Sanjiang Li$^*$
\thanks{Yunqi Huang and Sanjiang Li are with Centre for Quantum Software and Information (QSI), Faculty of Engineering and Information Technology, University of Technology Sydney, NSW 2007, Australia. Dingchao Gao and Shenggang Ying are with Key Laboratory of System Software (Chinese Academy of Sciences) and State Key Laboratory of Computer Science, Institute of Software, Chinese Academy of Sciences.}
\thanks{Corresponding author (E-mail: Sanjiang.Li@uts.edu.au)}
}
\maketitle

\begin{abstract}
Neutral atom (NA) quantum systems are emerging as a leading platform for quantum computation, offering superior or competitive qubit count and gate fidelity compared to superconducting circuits and ion traps. However, the unique features of NA devices, such as long-range interactions, long qubit coherence time, and the ability to physically move qubits, present distinct challenges for quantum circuit compilation. In this paper, we introduce DasAtom, a novel divide-and-shuttle atom approach designed to optimise quantum circuit transformation for NA devices by leveraging these capabilities. DasAtom partitions circuits into subcircuits, each associated with a qubit mapping that allows all gates within the subcircuit to be directly executed. The algorithm then shuttles atoms to transition seamlessly from one mapping to the next, enhancing both execution efficiency and overall fidelity. For a 30-qubit Quantum Fourier Transform (QFT), DasAtom achieves a 415.8x improvement in fidelity over the move-based algorithm Enola and a 10.6x improvement over the SWAP-based algorithm Tetris. Notably, this improvement is expected to increase exponentially with the number of qubits, positioning DasAtom as a highly promising solution for scaling quantum computation on NA platforms.
\end{abstract}

\textbf{Keywords}: neutral atom quantum computing, quantum circuit transformation, divide-and-conquer (DAC), subgraph isomorphism

\section{Introduction}

\IEEEPARstart{Q}{uantum} computing has the potential to revolutionise various fields, including cryptography \cite{Shor94}, chemistry \cite{cao2019quantum}, and machine learning \cite{schuld2015introduction}. Among the diverse quantum hardware platforms under development, neutral atom (NA) quantum systems have garnered significant attention due to their inherent advantages in scalability, qubit connectivity, and gate fidelity \cite{henriet2020quantum,bluvstein2022quantum,radnaev2024universal}. Unlike other quantum platforms such as superconducting circuits, NA systems can leverage long-range interactions and native multi-qubit gates, which allow for the execution of complex quantum operations with fewer resources. Additionally, the ability to physically move qubits within NA systems introduces a new dimension in quantum circuit optimisation that is not present in more rigid architectures \cite{Lengwenus10coherent,bluvstein2022quantum}.

Despite these advantages, the unique characteristics of NA devices present novel challenges in quantum circuit compilation. Specifically, the need to efficiently map qubits and execute operations while improving overall fidelity is a non-trivial problem. Existing quantum circuit compilation methods, such as SWAP-based \cite{Baker21isca-neutral_atom,Patel22isca-geyser,Li23tetris} and move-based \cite{Tan22reconfigurable,Tan2024compilingquantum,wang2024atomique,Wang23Q-Pilot,tan2024compilation} algorithms, are not fully optimised for the capabilities of NA systems, often leading to suboptimal performance in terms of overall fidelity. 

The SWAP-based method, exemplified by Tetris  \cite{Li23tetris}, employs a fixed-atom array and transforms quantum circuits in a manner similar to compilers for superconducting devices \cite{Zulehner+18_Astar,Li+19-sabre,Sivarajah20qst_tket,LiZF21_fidls,ZhouFL20_MCTS_iccad}. This approach leverages the long-range interactions of NA systems to achieve denser qubit connectivity, which contrasts with state-of-the-art IBM superconducting quantum computers that typically have sparse qubit connections, with an average degree between 2 and 3. Tetris operates as a heuristic greedy algorithm, addressing both qubit connection constraints and parallel execution constraints. 

On the other hand, Enola \cite{Tan2024compilingquantum}, the most recent and advanced move-based compiler, assumes sufficient separation between atoms to eliminate parallel execution constraints. As \cite{Tan22reconfigurable,Tan2024compilingquantum,Wang23Q-Pilot,wang2024atomique}, Enola exploits gate commutability in QAOA \cite{farhi2014quantum} circuits, aiming to scheduling gates from a commutation group in near-optimal number of Rydberg stages. For generic quantum circuits like Quantum Fourier Transform (QFT), CZ gates are often blocked by single-qubit gates (cf. Fig.~\ref{fig:qft_cz}). Enola executes each layer of CZ gates \emph{collectively}. 
Empirical results on QFT circuits indicate that the slow atom movement (partially due to the large atom distance) is the primary factor contributing to overall fidelity loss. 

However, even these advanced methods do not fully capitalise on the strengths of the NA platform, particularly the synergistic use of long-range interactions and atom shuttling, which can significantly enhance the efficiency and fidelity of quantum circuit execution. This gap highlights the need for a more comprehensive approach, leading to the development of DasAtom---a divide-and-shuttle atom algorithm specifically designed for NA systems. DasAtom combines the advantages of both Tetris and Enola while avoiding their shortcomings. Tetris effectively uses long-range interactions to enable dense qubit connectivity, but it does not utilise the ability to move atoms, limiting its flexibility. On the other hand, Enola leverages atom shuttling to adapt qubit mappings dynamically, yet it cannot take advantage of long-range interactions as atoms are far apart. DasAtom integrates both of these key capabilities: it partitions circuits into subcircuits, assigns an optimal qubit mapping for each subcircuit, and then shuttles atoms to smoothly transition between mappings. By doing so, it ensures that every gate in a subcircuit is directly executable, thereby enhancing overall fidelity and efficiency.

We conducted empirical comparisons of DasAtom's performance against Tetris, Enola, \blue{and Atomique} using the benchmark circuits used in \cite{Li23tetris}. This set of 33 benchmark circuits includes RevLib circuits as well as the quantum circuits Bernstein–Vazirani (BV), Quantum Volume (QV), and QFT, which range from 5 to 16 qubits with up to 3,089 CZ gates. In addition, we evaluated Deutsch-Jozsa (DJ), 3-regular MaxCut QAOA, Greenberger–Horne–Zeilinger (GHZ), QFT,  QV, {two}-local ansatz, and W-state circuits with qubit counts ranging from 5 to 50. Our experiments demonstrate that DasAtom consistently delivers significant performance gains in both overall fidelity and compiler runtime. 
For instance, DasAtom achieves a \blue{415.8x, 10.6x, 16.1x} improvement in fidelity over, \blue{respectively, Enola, Tetris, and Atomique} on a 30-qubit QFT, while the runtimes of \blue{Enola, Tetris, and Atomique are, {respectively, 9,851x, 384x, 2.6x} that} of DasAtom. Moreover, these improvements are expected to scale exponentially with the number of qubits, making DasAtom a highly promising solution for quantum computation on NA platforms.

The remainder of this paper is organised as follows: Section II provides a brief background and discusses related work. Sections III and IV offer a detailed description of the DasAtom algorithm, including its implementation and a comprehensive performance analysis across various quantum circuits. In Section V, we explain why DasAtom outperforms other algorithms and explore potential avenues for optimisation. The last section concludes the paper.

\section{Background and Related Work}

\subsection{Neutral atom quantum hardware}

In neutral atom quantum hardware, neutral atoms are trapped in arrays of tweezers and the computational states $\ket{0}$ and $\ket{1}$ are encoded in the hyperfine ground states of an alkali or alkaline-earth-like atom ~\cite{grimm2000optical,henriet2020quantum,Schmid24qst}. 
These atoms can be arranged in one, two, or even three-dimensional configurations~\cite{endres2016atom,barredo2016atom,barredo2018synthetic,Schlosser23scalable}. In this work, we focus on a $b\times b$ regular grid $G(b,b)$ with constant distance $d>0$. Fig.~\ref{fig:AG_diff_rini} shows a $3\times 3$ grid $G(3,3)$. In the following, we write $q,q',q_i$ for program qubits in a circuit, and write $p,p',p_i$ for nodes or their coordinates in $G(b,b)$.

Single-qubit gates are implemented through individual or global optical addressing of the atoms, and two-qubit gates are realised by exciting the atoms into a Rydberg state using laser beams.
The excitation to Rydberg states induces a strong dipole-dipole interaction between the atoms~\cite{jaksch2000fast}. 
This interaction is governed by an interaction radius $R_{\text{int}}$, within which a CZ gate on two atoms $p_i, p_j$ can be performed if the distance $D(p_i, p_j) \le R_{\text{int}}$, where $D$ represents the Euclidean distance. 

One significant advantage of NA platform is its ability for long-range interaction. Two qubits can interact even if they are not neighbours in the grid. For instance, the interaction radius $R_{\text{int}}$ can be $r_{\text{int}}\times d$ for 
\blue{some $r_{\text{int}}\geq 1$}.
Qubit connectivity of a quantum architecture is often captured by its architecture graph $AG = (V, E)$, where the nodes in $V$ correspond to the physical qubits (i.e., trapped atoms here), while the edges indicate the qubits capable of interacting with each other. For NA platform, each physical qubit $p$ in $V$ is assigned to the trap coordinates $(x,y)$, and the edges $E$ can be defined as:
\begin{equation}\label{eq:qubit_connectivity}
    E = \{(p_i, p_j) | p_i, p_j \in V, D(p_i, p_j) \le R_{\text{int}}\}. 
\end{equation}
In the $3\times 3$ grid shown in Fig.~\ref{fig:AG_diff_rini}, the black, red dotted, and blue edges are, respectively, those with distances $d$,  $\sqrt{2}d$, $2d$.

\begin{figure}
    \centering
    \scalebox{0.5}{\input{tex/arch_ex}}
    \caption{The architecture graph of a neutral atom  quantum hardware, where $(x,y)$ ($0\leq x,y\leq 2$) denotes the location of an atom, and two atoms are connected if their distance is smaller than $R_\text{int}=r_\text{int}\times d$, where $r_\text{int}\in \set{1,\sqrt{2},2}$.}
    \label{fig:AG_diff_rini}
\end{figure}

To minimise crosstalk between gates, \emph{parallel gate execution} is feasible only if qubits corresponding to different CZ gates maintain a distance of at least the restriction radius $R_{\text{restr}} \ge R_{\text{int}}$ from  all qubits involved in other simultaneously executed two-qubit gates. Specifically, for two CZ gates $g$ on $p_i, p_j$ and $g'$ on $p_a, p_b$ to be executed in parallel, the conditions 
\begin{equation}
    D(p_u, p_v) > R_{\text{restr}}
\end{equation} 
must be satisfied for any  $u\in\set{i,j}$ and any $v\in\set{a,b}$. {For example, suppose $R_\text{int}=R_\text{restr}= d$ in Fig.~\ref{fig:AG_diff_rini}. Let $p_{i,j}$ be a physical qubit located at $(i,j)$. Then  $CZ(p_{0,2},p_{1,2})$ can be executed in parallel with $CZ(p_{0,0},p_{1,0})$, but not in parallel with $CZ(p_{1,1},p_{1,0})$.}

\begin{figure*}[thb]
    \centering
    \input{tex/mov_fig}
        
    \caption{\blue{Illustration of atom movement in NAQC: (a) program qubits are mapped to a $4\times 4$ SLM array, (b) a $2\times 2$ AOD array is activated, and two SLM atoms, carrying $q_1,q_2$, are loaded (i.e., transferred) to the AOD array, (c) AOD rows (columns) are moved to the right (upward), carrying $q_1$  and $q_2$ to positions (2,3) and (3,2). To apply a CZ gate between $q_0$ and $q_1$, we offload $q_1$ to the SLM array and then apply an individual Rydberg laser to interact $q_0$ and $q_1$. If $R_\text{restr}=R_\text{int}=d$, then $CZ(q_0,q_1)$ and $CZ(q_2,q_3)$ can be applied in parallel. However, this is not possible when $R_\text{restr}=\sqrt{2}d$ and $R_\text{int}=d$.}
    } 
    \label{fig:atom_move}
    \end{figure*}
    
The ability to move qubits is the most distinctive advantage of NA platforms~\cite{Lengwenus10coherent,bluvstein2022quantum,Schmid24qst}. In NA systems, qubits are captured in two types of traps (cf. Fig.~\ref{fig:atom_move}). A spatial light modulator (SLM) generates an array of static traps, while a 2D acousto-optic deflector (AOD) creates mobile traps that can move within the plane. The AOD traps are formed at the intersection of a set of rows and columns.
Atom movement is a high-fidelity operation and an atom can traverse a region for 2,000 qubits with only 0.1\% coherence time \cite{bluvstein2022quantum}. \blue{Fig.~\ref{fig:atom_move} illustrates the procedure of atom movement in neutral atom quantum computing (NAQC): At each movement stage, a $k\times k$ AOD array is activated for some integer $k>0$, atoms are loaded from the SLM array to the AOD array, and the AOD array is moved,  subject to the constraint that different columns (rows) must not cross each other. After the movement, the AOD atoms are offloaded to the SLM array and the AOD array is deactivated. During each movement, the relative positions of AOD atoms are preserved, but the relative positions of SLM atoms are changed after the movement. }

\subsection{Quantum circuit compilation}

\begin{figure*}
    \centering
\scalebox{0.7}{
\begin{tikzpicture}
    \begin{yquant}
    qubit {\Large $\reg_{\idx}$} q[5];
    h q[4];
    box {$P(\frac{\pi}{2})$} q[4] | q[3];
    box {$P(\frac{\pi}{4})$} q[4] | q[2]; 
    box {$P(\frac{\pi}{8})$} q[4] | q[1]; 
    box {$P(\frac{\pi}{16})$} q[4] | q[0]; 
    h q[3];
    box {$P(\frac{\pi}{2})$} q[3] | q[2]; 
    box {$P(\frac{\pi}{4})$} q[3] | q[1]; 
    box {$P(\frac{\pi}{8})$} q[3] | q[0]; 
    h q[2];
    box {$P(\frac{\pi}{2})$} q[2] | q[1]; 
    box {$P(\frac{\pi}{4})$} q[2] | q[0]; 
    h q[1];
    box {$P(\frac{\pi}{2})$} q[1] | q[0]; 
    h q[0];
    
    \end{yquant}
\end{tikzpicture}}
   \caption{The QFT-5 circuit, where $H$ is the Hadamard gate, $P(\theta)$ is a phase gate with phase $\exp(i\theta)$.}
\label{fig:qft_5_ori}
\end{figure*}

In quantum computing, a quantum circuit serves as a model that represents the flow of information and operations within a quantum algorithm. A quantum circuit is composed of qubits and quantum gates that manipulate the states of these qubits. In this work, we denote a quantum circuit by $C$, which consists of a sequence of quantum gates $\{g_1, \ldots,g_m\}$ acting on qubits in $Q=\{q_1,\ldots,q_n\}$. Fig.~\ref{fig:qft_5_ori} shows the well-known QFT circuit on five qubits. Note that, for clarity, we have omitted the SWAP gates at the end of the circuit.

\begin{figure*}
    \centering
    \scalebox{0.7}{
    \begin{tikzpicture}
    \begin{yquant}
    qubit {\Large $\reg_{\idx}$} q[5];
    box {$R$} q[0,1,2,3,4];
    box {$R$} q[0,1,2,3,4];
    zz (q[3], q[4]);
    box {$R$} q[3]; 
    zz (q[3], q[4]);
    box {$R$} q[3]; box {$R$} q[4];
    box {$R$} q[3]; zz (q[2], q[4]);
    box {$R$} q[2]; zz (q[2], q[4]);
    box {$R$} q[2]; box {$R$} q[4];
    zz (q[2],  q[3]); zz (q[1], q[4]);
    box {$R$} q[1]; box {$R$} q[2];
    zz (q[2], q[3]); zz (q[1], q[4]);
    box {$R$} q[1]; box {$R$} q[2]; box {$R$} q[3]; box {$R$} q[4];
    box {$R$} q[2];
    zz (q[1], q[3]); zz (q[0], q[4]);
    box {$R$} q[0]; box {$R$} q[1];
    zz (q[1], q[3]); zz (q[0], q[4]);
    box {$R$} q[0]; box {$R$} q[1]; box {$R$} q[3];
    zz (q[1], q[2]); zz (q[0], q[3]);
    box {$R$} q[0]; box {$R$} q[1];
    zz (q[1], q[2]); zz (q[0], q[3]);
    box {$R$} q[0];box {$R$} q[1];box {$R$} q[2]; box {$R$} q[1];
    zz (q[0], q[2]); box {$R$} q[0];
    zz (q[0], q[2]); box {$R$} q[0];
    zz (q[0], q[1]); box {$R$} q[0];
    zz (q[0], q[1]);
    box {$R$} q[0];
    \end{yquant}
    \end{tikzpicture}
    }
\caption{The QFT-5 circuit decomposed in gate set $\set{CZ, R_x,R_y,R_z}$, where each $R$ denotes an $R_x$, $R_y$, or $R_z$ gate.}
\label{fig:qft_cz}
\end{figure*}

Circuits like QFT-5 in Fig.~\ref{fig:qft_5_ori} often contain gates that are not native in a target quantum device. This means we need to decompose non-native gates in $C$ into native gates. For example, let $\mathcal{G}=\set{R_x,R_y,R_z, CZ}$ be the set of native gates of the target quantum device. For QFT-5 in Fig.~\ref{fig:qft_5_ori}, we need decompose $H$ and control phase gate $CP(\theta)$ in gates in $\mathcal{G}$. The result is shown in Fig.~\ref{fig:qft_cz}. 

Due to limited qubit connectivity, it is often not possible to execute the synthesised circuit directly, as some CZ gates may act on two far away device qubits. For example, \blue{Fig.~\ref{fig:atom_move}a illustrates a $4\times 4$ SLM array and an initial mapping $f$ that maps program qubits $q_i$ ($0\leq i\leq 4)$ within $G(4,4)$. Assuming $R_\text{int}=2d$, a CZ gate $CZ(q_i,q_j)$ can be executed if the distance $D(f(q_i),f(q_j))\leq 2d$. Clearly, $CZ(q_1,q_2)$ cannot be executed directly in this case, as $D(f(q_1),f(q_2))>2d$.}

\begin{figure*}
    \centering
\scalebox{0.7}{
\begin{tikzpicture}
    \begin{yquant}
    qubit {\Large $\reg_{\idx}$} q[5];
    zz (q[3], q[4]); barrier (q);
    zz (q[3], q[4]); barrier (q);
    zz (q[2], q[4]); barrier (q);
    zz (q[2], q[4]); barrier (q);
    zz (q[2], q[3]); zz (q[1], q[4]);
    barrier (q);
    zz (q[2], q[3]); zz (q[1], q[4]);
    barrier (q);
    zz (q[1], q[3]); zz (q[0], q[4]);
    barrier (q);
    zz (q[1], q[3]); zz (q[0], q[4]);
    barrier (q);
    zz (q[1], q[2]); zz (q[0], q[3]);
    barrier (q);
    zz (q[1], q[2]); zz (q[0], q[3]);
    barrier (q);    barrier (q);
    zz (q[0], q[2]); barrier (q);
    zz (q[0], q[2]); barrier (q);
    zz (q[0], q[1]); barrier (q);
    zz (q[0], q[1]);    
    \end{yquant}
    \end{tikzpicture}} 
    \caption{The decomposed QFT-5 circuit with all single-qubit gates removed. The CZ circuit is partitioned in 12 CZ layers and divided into two subcircuits with the second consisting of the last four CZ gates.}
    \label{fig:qft_cz_only}
\end{figure*}

We next outline the general procedures for quantum circuit transformation. To execute a decomposed circuit $C$ on an NA device, each program qubit $q$ in $C$ is first mapped to a physical atom located at $p_{i,j}$ in the device's SLM array. With this initial mapping $f$, not all CZ gates are directly executable. If a CZ gate $g$ cannot be executed directly, the two qubits involved must be moved close together. In NA platforms, this can be achieved in two ways: by inserting SWAP gates or by atom shuttling (cf. \cite{Schmid24qst}). For instance, in the mapping shown in Fig.~\ref{fig:atom_move}a, executing $CZ(q_1,q_2)$ requires either swapping $q_1$ with $q_4$ or $q_2$ with $q_4$ using SWAP gates. Alternatively, we can move the atom carrying $q_1$ from $p_{0,2}=f(q_1)$ to one of the three neighbouring positions around $p_{1,0}=f(q_2)$. If a neighbouring position is occupied, the occupying atom must be moved first.

\subsection{Related work} \label{sec:related_works}
Quantum circuit transformation (QCT) is the process of converting a program circuit into a form that is executable on a target quantum device, whether it be 
a superconducting or NA quantum device. QCT is a crucial component of quantum circuit compilation. Since IBM launched its cloud-based quantum computing services, numerous   algorithms have been proposed in the literature, including \cite{Zulehner+18_Astar,Li+19-sabre,Sivarajah20qst_tket,Zhou+20_SAHS,ZhouFL20_MCTS_iccad,LiZF21_fidls,lao22tcad-timing,zhu22iterated_local_search}. 

The unique features of NA devices introduce distinct challenges for quantum circuit compilation. Early efforts sought to leverage these features within fixed atom arrays. Baker et al. proposed the first compiler for NA devices, which accounted for long-range interaction  \cite{Baker21isca-neutral_atom}. Multi-qubit gate support was subsequently incorporated in \cite{Patel22isca-geyser}. Inspired by the popular block puzzle game, Li et al. \cite{Li23tetris} proposed the heuristic algorithm \emph{Tetris}, which effectively reduces qubit idle time while exploiting the rich qubit connections in NA device and adhering to parallel execution constraints. 

Unlike the above SWAP-gate based shuttling approach, Tan et al. \cite{Tan22reconfigurable,Tan2024compilingquantum} developed OLSQ-DPQA, an SMT solver-based compiler for the dynamically field-programmable qubit array (DPQA) architecture {\cite{bluvstein2022quantum}}, which employs atom shuttling.  
To address the scalability challenge of OLSQ-DPQA, several follow-up compilers have been introduced, including Atomique \cite{wang2024atomique}, Q-Pilot \cite{Wang23Q-Pilot}, and Enola \cite{tan2024compilation}, all designed for DPQA. Enola's compilation process is divided into scheduling, placement, and routing. It schedules a commutation group of CZ gates in a near-optimal number of Rydberg stages and, for generic quantum circuits, schedules each layer of parallel CZ gates as a Rydberg stage. Enola offers two placement strategies: dynamic placement, which generates a new qubit mapping for each Rydberg stage, and static placement, which retains the same mapping throughout. Compared to OLSQ-DPQA, Atomique, and Q-Pilot, Enola demonstrates superior or comparable fidelity improvements \cite{tan2024compilation}.   

Combining the two methods described above is natural. Brandhofer et al. \cite{Brandhofer_iccad21} proposed a compiler that integrates SWAP gates with a special atom movement technique called one-dimensional displacements, aimed at reducing circuit depth and improve circuit fidelity. Another hybrid compiler was proposed in \cite{schmid2023hybrid}, where the NA hardware is grid-based, and atom can be moved from one grid point to another only if the target grid point is unoccupied. Experiments in \cite{schmid2023hybrid} show that, for shuttling-favoured NA devices, the hybrid approach is essentially the same as the atom shuttling method in performance.

Interested readers may consult the recent review paper  \cite{Schmid24qst} for more information.

Divide-and-conquer (DAC) is a widely used approach in quantum circuit transformation that several algorithms employ to enhance efficiency. Siraichi et al. \cite{siraichi+19_bmt} introduced the BMT algorithm, which partitions a circuit’s gate list into sublists, aligns them through subgraph isomorphism and token swapping, and determines an optimal transformation path with dynamic programming. Similarly, Wu et al. \cite{iccad22/Wu_robust} developed an SMT-based DAC method that partitions circuits and links sublist embeddings with SWAP gates to minimise transformation costs. Both methods, however, rely on exhaustive search, limiting scalability to circuits with 20 or more qubits. Recently, Huang et al. \cite{huang2024adac} proposed the adaptive DAC algorithm, which dynamically partitions circuits and optimises routing heuristics, achieving substantial performance gains, including nearly 50\% improvement on the IBM Tokyo architecture.

\section{DasAtom Transformation Framework}
\blue{Before describing our algorithm, we recall that qubit interaction in a quantum circuit can be represented by a graph.}
\begin{definition}[Interaction Graph]
For a quantum circuit $C$ with qubit set $Q$, its interaction graph, denoted as $IG(C)$, is an undirected graph $(Q,E_\text{int})$, where:
\begin{itemize}
    \item each node represents a qubit of $C$. 
    \item Two nodes $q,q'\in Q$ are connected if there is a two-qubit gate in $C$ acting on $q,q'$. 
\end{itemize}
\end{definition}
If the interaction graph of $C$ matches well with the architecture graph $AG=(V,E)$, there is no need to route the qubits: what we need is an embedding $f$ from $IG(C)$ to $AG$. Here a 1-1 mapping $f:Q\to V$ is an embedding if $(f(q),f(q'))$ is an edge in $AG$ for any edge $(q,q')$ in $IG(C)$. In this case, $f$ is also called a subgraph isomorphism.

\subsection{Overview}
Let $\mathcal{G}$ be an NA-native gate set that includes the two-qubit CZ gate and local addressed rotation gates $R_z$, as well as (locally or globally addressed) $R_x$ and $R_y$. \blue{Given an input circuit $C$ on $n$ qubits, we first synthesise $C$ using gates from $\mathcal{G}$ (cf. \cite{nottingham24circuit}).}  We assume that single-qubit gates and CZ gates are executed in separate stages, meaning that in each stage, either single-qubit gates or CZ gates are executed, but not both. Whenever there is a single-qubit gate in the front layer, we execute (and remove) it to form a new front layer. \blue{This ensures that, at some point, we obtain a front layer consisting only of CZ gates, which can be executed collectively, provided the parallel gate execution constraints are satisfied.}

\blue{In the following, we assume single-qubit gates are removed before calling DasAtom, a convention also adopted in OLSQ-DPQA, Atomique, Q-Pilot, and Enola. For simplicity, we continue to refer to the resulting CZ-only circuit as $C$.} 

\blue{Let $d$ denote the atom distance, $R_\text{int} \geq d$ the interaction radius, and $R_\text{restr}\geq R_\text{int}$ the  restriction radius. We assume that CZ gates are executed individually using Rydberg lasers. Our algorithm, DasAtom, proceeds as follows \blue{(cf. Fig.~\ref{fig:DAS_flow} for the flowchart)}:}
\begin{enumerate}
    \item Partition the CZ circuit $C$ into layers $L_1,\ldots,L_m$, where each $L_i$ contains only CZ gates.
    \item Divide the CZ circuit $C$ into subcircuits such that (i) each layer $L_i$ is fully contained within a subcircuit; (ii) the interaction graph  of each subcircuit is embeddable in Grid$(b,b)$, where $b = \mathceil{\sqrt{n}}$. 
    \item For each subcircuit $C_i$, construct a mapping $f_i$ that embeds the interaction graph $IG(C_i)$ to Grid$(b,b)$. For each CZ gate $g=CZ(q,q')$ in $C_i$, we execute $g$ by interacting $f_i(q)$ and $f_i(q')$. Since $D(f_i(q),f_i(q')) \leq R_{\text{int}}$, the gate can be executed directly. However, due to parallel execution constraints, not all gates in the same layer can be executed simultaneously. 
    \item Perform routing based on atom shuttling (instead of inserting SWAP gates). This can be achieved by modifying the routing algorithm provided in \cite{Tan2024compilingquantum,tan2024compilation} or 
    \cite{schmid2023hybrid}. This process transforms $f_i$ into $f_{i+1}$.
\end{enumerate}

\begin{figure}
\centering
\scalebox{0.78}{
\begin{tikzpicture}[node distance=1.5cm]

\node (start) [startstop] {Input: CZ circuit $C$, $d$, $R_\text{int}$, $R_\text{restr}$};
\node (partition1) [process, below of=start] {Divide $C$ into CZ layers $L_1,...L_m$};
\node (partition2) [process, below of=partition1] {Divide $C$ into $C_1, ..., C_k$ and find an embedding $f_i$ for each $C_i$};
\node (execute) [process, below of=partition2] {Start with $i=1$ and execute $C_i$ with $f_i$};
\node (decision) [decision, below of=execute, yshift=-0.5cm] {$i = k?$};
\node (route) [process, below of=decision, yshift=-0.5cm] {Route $f_i$ to $f_{i+1}$};
\node (increment) [process, left of=route, xshift=-2.5cm] {$i = i + 1$};
\node (end) [startstop, right of=decision, xshift=1.5cm] {End};

\draw [arrow] (start) -- (partition1);
\draw [arrow] (partition1) -- (partition2);
\draw [arrow] (partition2) -- (execute);
\draw [arrow] (execute) -- (decision);
\draw [arrow] (decision) -- node[anchor=east] {no} (route);
\draw [arrow] (route) -- (increment);
\draw [arrow] (increment) |- (execute.west);
\draw [arrow] (decision) -- node[anchor=south] {yes} (end);

\end{tikzpicture}}
\caption{The flowchart of DasAtom}
\label{fig:DAS_flow}
\end{figure}

In the following, we discuss the implementation of the above procedures in detail.

\begin{figure*}
\centering
\subfloat[]
{
\scalebox{0.5}{\input{tex/qft5_embedding1}}
\label{subfig:qft5_emb1}
}
\hspace{1.5em}
\subfloat[]
{
\scalebox{0.5}{\input{tex/qft5_interaction1}}
\label{subfig:ig_p1}
}
\hspace{1.5em}
\subfloat[]
{
\scalebox{0.5}{\input{tex/qft5_embedding2}}
\label{subfig:qft5_emb2}
}
\hspace{1.5em}
\subfloat[]
{
\scalebox{0.5}{\input{tex/qft5_interaction2}}
\label{subfig:ig_p2}
}
\caption{The embeddings in $G(3,3)$ and interaction graphs of the two subcircuits of QFT-5 (shown in Fig.~\ref{fig:qft_cz_only}).}
\label{fig:qft5_p2}
\end{figure*}

\subsection{Circuit division}

After decomposing gates in circuit $C$ into NA-native gates and removing single-qubit gates, we partition $C$ into CZ layers $L_1,\cdots,L_m$. We further divide $C$ into subcircuits $C_1,\ldots,C_k$, where each $C_i$ consists of consecutive CZ layers from $C$, and the interaction graph of $C_i$ can be embedded in $AG$ with a mapping $f_i$. We denote $C_i = C[\ell_i : m_i]$ if $C_i$ is composed of layers $L_{\ell_i},\ldots,L_{m_i}$. Thus, we have $1=\ell_1\leq m_1<\ell_2\leq m_2 < \ldots < \ell_k \leq m_k=m$.

\blue{Unlike the gate-by-gate partitioning methods used in \cite{siraichi+19_bmt} and \cite{iccad22/Wu_robust}, our division scheme is coarse-grained. This design has two implications: first, it significantly reduces the number of costly subgraph isomorphism checks required by our algorithm; second, it presents opportunities for further optimsation.} For circuit division, we employ the Rustworkx implementation of VF2 \cite{Cordella+04-vf2}. When partitioning the circuit into layers, the number of VF2 calls is at most $O(depth)$, where $depth$ is the number of CZ layers in $C$. \blue{In contrast, when using a gate-by-gate partition, the number of VF2 calls grows exponentially in $depth$.}

\begin{algorithm}
\caption{Circuit division}
\label{Alg:cir_div}
\begin{algorithmic}[1]
\REQUIRE A CZ circuit $C$ and an architecture graph $AG$
\ENSURE Subcircuits $SC$  and embeddings $EM$
\STATE Initialize two empty sequences $SC$ and $EM$
\STATE $i_\text{last} \leftarrow 0$ 
\FOR{$0 < i \le $ depth of $C$}
\IF{$IG(C[i_\text{last}:i])$ is embeddable in $AG$} 
\STATE continue
\ELSE
\STATE Append $C[i_\text{last}:i-1]$ to $SC$
\STATE Find an embedding $f_i$ for $IG(C[i_\text{last}:i-1])$ and append it to $EM$
\STATE $i_\text{last} \leftarrow i$
\ENDIF
\ENDFOR
\STATE return $SC$ and $EM$
\end{algorithmic}
\end{algorithm}

For our running example, if we set $R_\text{int}=\sqrt{2}d$, then the QFT-5 circuit can be partitioned into two subcircuits, see Fig.~\ref{fig:qft_cz_only}. The interaction graphs and their embeddings to $AG=G(3,3)$ of these two subcircuits are shown in Fig.~\ref{fig:qft5_p2}. To minimise qubit idle time, it is desirable to execute as many CZ gates in parallel as possible, ensuring they satisfy the previously mentioned restriction constraint. In our example, $R_{\text{restr}} = 2\times R_{\text{int}} = 2\sqrt{2}d$. The first sub-circuit includes, for instance, a CZ layer consisting of gates $CZ(q_1, q_3)$ and $CZ(q_0, q_4)$, with the qubit mapping $f$ shown in Fig.~\ref{fig:qft5_p2}(a). Since $D(f(q_3), f(q_4)) = d < R_{\text{restr}}$, $CZ(q_1, q_3)$ and $CZ(q_0, q_4)$ cannot be executed in parallel due to the restriction constraint.

\subsection{\blue{Atom shuttling}}

Suppose we have divided the CZ circuit $C$ into subcircuits $C_1,\ldots,C_m$ with corresponding embeddings $f_1,\ldots,f_m$. How do we transition from one mapping to the next? In superconducting quantum devices, this is typically done by inserting SWAP gates. \blue{In this section, we show how this can be completed in NAQC by atom shuttling. The routing method described below is adapted from \cite{tan2024compilation} for our purpose.}

Let $Q=\set{q_0,\ldots,q_{n-1}}$ be the set of program qubits in the input circuit. Suppose $f,f'$ are two 1-1 mappings from $Q$ to grid points in $G(b,b)$ (where $b = \mathceil{\sqrt{n}}$). Our goal is to transition from $f$ to $f'$. To do this, we need to move each $q_i\in Q$ from its current position $f(q_i)=(x_i,y_i)$ in $G(b,b)$ to a new position $f'(q_i)=(x'_i,y'_i)$, also within $G(b,b)$. 

\blue{Formally, we represent each move from $f(q_i)$ to $f'(q_i)$ as a 4-tuple $m_i\define (x_i,y_i,x'_i,y'_i)$. We consider a move $m_i$ to be trivial if $x'_i=x_i$ and $y'_i=y_i$, meaning the qubit remains stationary. Define $M=\{m_i\mid m_i \ \text{is non-trivial}, 0\leq i\leq n-1\}$. Moves in $M$ can be implemented \emph{serially} (cf. Fig.~\ref{fig:atom_move}): for each move $m_i$, we activate a $1\times 1$ AOD at the starting point $(x_i,y_i)$, load $q_i$, move the AOD to $(x'_i,y'_i)$, and then offload $q_i$ at its new position. If the target position is occupied by another qubit $q_j$ (for some $j>i$), we temporarily move $q_j$ aside, place $q_i$ at the target, and delay movement $m_j$ until later.}

\blue{While this approach accomplishes the transition, it is not ideal since atom movement is slow and can reduce overall fidelity. In many cases, moves can be executed in parallel. Fig.~\ref{fig:atom_move} actually illustrates how two moves, $m_1=(0,2,2,3)$ and $m_2=(1,0,3,2)$, can be implemented simultaneously. However, not all moves are eligible for parallel execution; legal parallel moves must satisfy the constraint that different AOD columns (or rows) do not cross or overlap.} 

\blue{
To formalise the concept of legal parallel moves, we introduce the notion of \emph{conflicting moves}. Two moves $m_i\define (x_i,y_i,x'_i,y'_i)$ and $m_j\define(x_j,y_j,x'_j,y'_j)$ in $M$ are considered \emph{in conflict} if they fail to meet the following condition for some $*$ in $\set{<,=,>}$:
\begin{itemize}
    \item $(x_i * x_j) \Longleftrightarrow (x'_i * x'_j)$ and $(y_i * y_j) \Longleftrightarrow (y'_i * y'_j)$.
\end{itemize}
If $m_i$ and $m_j$ do not conflict, they are deemed \emph{compatible}. A set of moves $M'=\set{m_{i_1},\ldots,m_{i_\ell}}$ is compatible if every pair within the set is compatible. In this case, the moves in $M'$ can be implemented in parallel by  activating a $k\times k$ AOD array for some $k>0$, positioning all qubits corresponding to the moves in $M'$ at the grid points, loading them from the SLM to the AOD, moving them in parallel to their destinations, and offloading them. Our objective is to find a sequence of parallel, compatible moves that transports each $(x_i,y_i)$ to $(x'_i,y'_i)$ efficiently, minimising the total movement time.
}

\vspace*{2mm}
\blue{
\begin{example}\label{ex:M}
\blue{Fig.~\ref{fig:qft5_p2} shows two  mappings $f$ and $f'$ in $G(3,3)$ from $Q=\set{q_1,\ldots,q_5}$. In this instance, we have $M=\{m_0\define (0,2,0,1), m_1\define (1,0,0,2), m_4 \define (0,1,1,0)\}$. The three moves $m_0,m_1,m_4$ in $M$ are pairwise conflicting. For example, $m_0=(0,2,0,1)$ and $m_1=(1,0,0,2)$ are conflicting as $x_0=0<1=x_1$ but $x'_0=0=0=x'_1$, violating the constraint $(x_i * x_j) \Longleftrightarrow (x'_i * x'_j)$ for $*$ being $<$ or $=$.  This means that we should implement them sequentially. Fig.~\ref{fig:mov&conflictgraph}a shows such a legal implementation, where $m_1$ is implemented first, followed by $m_0$, and lastly, $m_4$.}
\end{example}
To address the general problem, we construct a conflict graph for $M$ following the approach in \cite{tan2024compilation}. In this graph, each node represents a move in $M$, and an edge connects two nodes if their moves are incompatible. Fig.~\ref{fig:mov&conflictgraph}b illustrates the conflict graph for the movement set in Example~\ref{ex:M}. Any set of compatible moves forms an independent set of vertices, reducing the problem to finding a maximum independent set (MIS) in the conflict graph. Given that the MIS problem is NP-hard, an approximate solution can be obtained by searching for a maximal independent set \cite{tan2024compilation}. This approach allows us to identify  a sequence of compatible moves that transition from $f$ to $f'$ in a reasonably short total movement time, as the longest move in each compatible set determines the AOD movement time for the set \cite{tan2024compilation}.
}

\blue{
\begin{remark}
Several existing routing algorithms, such as those in \cite{tan2024compilation,schmid2023hybrid}, are suitable for our purposes. To facilitate comparison with Enola, we adapt the routing algorithm from \cite{tan2024compilation} to meet our requirements, with only a minor adjustment. In Enola, the routing task is slightly different: at each Rydberg stage, a set $S$ of CZ gates is selected, where no two gates share a common qubit. The objective is to shuttle atoms so that the two qubits $p_i,p_j$ involved in each CZ gate in $S$ are aligned at the same grid point.\footnote{In DPQA, each grid point in the SLM array can hold two traps.} Given a current mapping $f$, this alignment only requires moving $f(p_i)\define (x_i,y_i)$ to $f(p_j)\define (x_j,y_j)$, or vice versa. Thus, any CZ gate $CZ(p_i,p_j)$ in $S$ involves two dual moves: $m=(x_i,y_i,x_j,y_j)$ and $m'=(x_j,y_j,x_i,y_i)$. By contrast, DasAtom does not include dual movements within its conflict graph.
\end{remark}
}

\begin{figure}
    \centering
    \subfloat[]{
    \scalebox{0.55}{\input{tex/qft5_move}}
    }
    %
    \hspace{1.5em}
    \subfloat[]{
    {\raisebox{3em}{\scalebox{0.8}{\input{tex/mov_conflict_graph}}}}
    }
    \caption{\blue{(a) Atom movement between the two mappings in Fig.~\ref{fig:qft5_p2}, (b) the conflict graph for $M=\set{m_0,m_1,m_4}$, where $m_0=(0,2,0,1)$, $m_1=(1,0,0,2)$, and $m_4=(0,1,1,0)$. }}
    \label{fig:mov&conflictgraph}
\end{figure}

\section{Evaluation and Analysis}

We implemented our proposed algorithm in Python. 
All experiments for DasAtom and Enola were conducted on {a MacBook Pro featuring a 2.3 GHz Intel Core i5 processor and 16 GB memory}, while all experiments for Tetris \blue{and Atomique}
were conducted on an Ubuntu 20.04 server with 40 cores of Intel Xeon Gold 5215 @ 2.50GHz, 512 GB of RAM. 
DasAtom is open source at https://github.com/Huangyunqi/DasAtom.

In our evaluation, we assume that the input circuit is decomposed in the same manner, with all single-qubit gates  removed before applying the algorithms. 

\blue{For readers' convenience, we summarise all key parameters, symbols, and their default values (if applicable) in Table~\ref{tab:parameter}.
}
\begin{table}[tbh]
    \caption{Key parameters, symbols, and default values}\label{tab:parameter}
    \centering
    \scalebox{0.98}{
    \blue{
    \begin{tabular}{m{0.6cm}<{\centering}|m{2cm}<{\centering}|m{1.2cm}<{\centering}||m{0.6cm}<{\centering}|m{2cm}<{\centering}}
        Symb.\ & Meaning & default value & Symb.\ & Meaning \\ \hline
        $T_2$ & coherence time & 1.5s&  $n$ &  qubit count \\ \hline
        
        $f_{\text{cz}}$ &  CZ gate fidelity & 0.995 
        & $m$ & CZ gate count
        \\ \hline
        $f_{\text{trans}}$ & atom transfer fidelity & 1  
        & $s$ & atom transfer count \\ \hline        

        $t_{\text{cz}}$ &  CZ gate duration & 0.2 $\mu$s  
        & $h$ & compiled circuit depth \\ \hline
        
        $t_{\text{trans}}$ & atom transfer duration & 20 $\mu$s 
        & $D$ & total maximal movement distance \\ \hline
        $v$ & atom moving speed & 0.55 $\mu\text{m}/\mu\text{s}$ & $T$ & compiled circuit execution time \\ \hline
        $d$ & atom distance &  3 $\mu$m  & $T_\text{idle}$ & qubit idle time\\ \hline
        $R_\text{int}$ & interaction radius & 6 $\mu$m  & $r_\text{int}$ & interaction factor \\ \hline
        $R_\text{restr}$ & restriction radius & 12 $\mu$m  & $r_\text{restr}$ & restriction factor \\ 
    \end{tabular}
    }
    }
\end{table}

\subsection{Fidelity}\label{sec:fidelity}

For easy of comparison across algorithms  on different NAQC platforms, and following \cite{Schmid24qst}, we use the \emph{approximate success probability} as a proxy for circuit fidelity. \blue{We assume single-qubit gates to be error-free in this study, disregarding their contribution to the overall error analysis. This is justified by their typically much higher fidelity compared to CZ gates.}

Let $T_2$ be the atom dephasing time, $n,m$ the numbers of qubits and CZ gates in the (compiled) circuit, $T$ the total circuit execution time, $s$ the number of atom transfers, $v$ the speed of atom movement, and $t_\text{cz}$ and $t_\text{trans}$ the durations of a $CZ$ gate execution and and atom transfer, $f_\text{cz}$ and $f_\text{trans}$ the fidelities of gate $CZ$ and atom transfer. The approximate success probability of the compiled circuit is 
    \begin{equation}\label{eq:appr.success_probability}
                P(C) = \exp\bigg(\!-\frac{T_{\text{idle}}}{T_2}\bigg) \times f_\text{cz}^m \times f_\text{trans}^{s},
    \end{equation}
where 
\begin{equation}\label{eq:T_idle}
    T_\text{idle}=n\times T - m\times t_\text{cz}.
\end{equation}

\blue{
We note that $T$ is calculated as 
\begin{align}\label{eq:T}
    T=h\times t_\text{cz}+s\times t_\text{trans}+D/v,
\end{align} 
where $h$ is the depth of the compiled circuit and $D$ the sum of the maximal movement distance of each (parallel) movement over all atom movements. As Tetris uses no atom movement, the execution time of Tetris' compiled circuit is simply  $h\times t_\text{cz}$. 
}

\subsection{\blue{Architecture assumptions}}

\blue{
Various architectures have been proposed for NAQC \cite{bluvstein2022quantum,Evered23high,Graham22multi,radnaev2024universal}, each with distinct features. Some architectures support long-range interaction using individually addressed Rydberg lasers for CZ gate execution \cite{Graham22multi,radnaev2024universal}, while others enable dynamic atom shuttling \cite{Lengwenus10coherent,bluvstein2022quantum}, or employ a global Rydberg laser to illuminate the entire plane, allowing parallel execution of CZ gates \cite{bluvstein2022quantum}. These architectural differences have led to distinct compiler designs that leverage each platform's unique capabilities. 
}

\blue{
Tetris \cite{Li23tetris} operates with individually addressed CZ gates. To better exploit long-range interaction, atoms are compactly arranged in a \emph{fixed} 2-dimensional array with distance $d$. For any given interaction radius $R_\text{int}\define r_\text{int}\times d\geq d$, a smaller $d$ implies a larger interaction factor (i.e., $r_\text{int}$) and greater qubit connectivity (cf. Fig.~\ref{fig:AG_diff_rini}). Notably, Tetris adopts a variable restriction area that depends on the inter-qubit distance of the gate, and this area is strictly smaller than the restriction area we define when $R_\text{restr}$ is set equal to $R_\text{int}$. This difference allows Tetris to achieve more parallel CZ gates than DasAtom, even when we set $R_\text{restr}$ to its minimal.
}

\blue{
Enola \cite{tan2024compilation} and Atomique \cite{wang2024atomique} are designed for DPQA architectures, which support atom shuttling during computation and use a global Rydberg laser to apply CZ gates in parallel. To facilitate parallel execution and avoid unwanted  interactions, both compilers assume an atom distance of 2.5 times the interaction radius. Enola uses one SLM (static) array and one AOD (movable) array, applying CZ gates between qubits on the SLM array and allowing atoms to transfer between arrays. This enables qubit routing to be fully accomplished via atom shuttling. Atomique assumes one SLM  and multiple AOD arrays, applying CZ gates between atoms across different arrays and allowing SWAP gates during compilation, without requiring atoms to be transferred between arrays. 
}

\blue{
To leverage both long-range interaction and atom shuttling, DasAtom assumes an NAQC architecture that supports in-computation atom shuttling, similar to DPQA, but with individually addressable Rydberg lasers for executing CZ gates. This setup requires one SLM array and one AOD array, with atoms able to transfer between layers. To facilitate long-range interaction, we assume atoms are initially arranged in a 2-dimensional SLM array with distance $d \leq R_\text{int}$. While no current NAQC devices combine both capabilities, devices demonstrating each individually exist \cite{bluvstein2022quantum,radnaev2024universal}, and there are no fundamental physical limitations preventing the future integration of these two features.
}

\subsection{Overall fidelity comparison}
\label{sec:evaluation}

In our experiments, the common NA hardware parameters we assumed are summarised in Table~\ref{tab:parameter}.
In addition, we assume
\begin{itemize}
    \item Atom distance $d$: 3$\mu$m for Tetris and DasAtom, 15$\mu$m for Enola and Atomique.
    \item $R_{\text{int}} = 6\mu\text{m}$ and $R_\text{restr} = 12\mu\text{m}$.  
\end{itemize}
For DasAtom and Tetris, the default coupling graph is set to grid of $\mathceil{\sqrt{n}} \times \mathceil{\sqrt{n}}$. For Atomique, we use the default configuration of two AOD arrays.

\begin{table*}[htb]
    \caption{Comparison of DasAtom, Tetris, Enola, \blue{and Atomique} on benchmark circuits from \cite{Li23tetris}. `Depth' represents the number of CZ layers in the input circuit, `F' and `RT' denote the fidelity and runtime of the corresponding algorithm, `D' indicates the CZ-depth of the compiled circuit, `M' is the number of move stages, `P' is the number of subcircuits, and `SW' represents the number of SWAPs. \blue{Circuits are ordered by depth.} 
    } 
    \label{tab:tetris}
    \centering
    \begin{adjustbox}{width=\textwidth}
\begin{tabular}{c|c|c|c||c|c|c|c|c||c|c|c|c||c|c|c||c|c|c|c|c}
  \multicolumn{4}{c||}{Circuit Info.} & \multicolumn{5}{c||}{DasAtom} & \multicolumn{4}{c||}{Tetris } & 
  \multicolumn{3}{c||}{Enola (static)} & \multicolumn{5}{c}{\blue{Atomique}} \\

    \hline
        Name & \#Q & \#CZ & Depth & F & M & D & P & RT (s) & F & SW & D & RT (s) &  F & M & RT (s) & F & SW & M & D &  RT(s) \\ \hline
        4mod5-v1\_22 & 5 & 11 & 10 & $\textbf{0.9463}$ & 0 & 11 & 1 & 0.01 & 0.9322 & 1 & 14 & 1.9 & 0.9311 & 20 & 46.38 & 0.9378 & 0 & 2 & 11 & 0.433 \\ \hline
        alu-v0\_27 & 5 & 17 & 15 & \textbf{0.9183} & 0 & 17 & 1 & 0.01 & 0.8911 & 2 & 20 & 4.37 & 0.896 & 32 & 61.92 & 0.9081 & 0 & 5 & 16 & 0.434\\ \hline
        bv\_n16 & 16 & 15 & 15 & 0.9195 & 9 & 15 & 2 & 0.03 & 0.8224 & 8 & 21 & 0.61 & 0.8573 & 30 & 56.54 & $\textbf{0.9244}$ & 0 & 15 & 15 & 0.434 \\ \hline
        qv\_n16\_d5 & 16 & 120 & 15 & \textbf{0.5436} & 8 & 120 & 2 & 0.69 & 0.3762 & 25 & 82 & 13.47 & 0.3435 & 132 & 126.23 & 0.5221 & 2 & 43 & 46 & 0.907 \\ \hline
        mod5mils\_65 & 5 & 16 & 16 & \textbf{0.9229} & 0 & 16 & 1 & 0.01 & 0.9091 & 1 & 19 & 5.2 & 0.9005 & 32 & 56.91 & 0.9127 & 0 & 5 & 16 & 0.63 \\ \hline
        ising\_model\_13 & 13 & 120 & 20 & \textbf{0.5479} & 0 & 120 & 1 & 0.06 & 0.5159 & 4 & 77 & 30.62 & 0.3974 & 80 & 154.6 & 0.5305 & 0 & 41 & 51 & 0.715 \\ \hline
        ising\_model\_10 & 10 & 90 & 20 & \textbf{0.6368} & 0 & 90 & 1 & 0.05 & 0.5907 & 5 & 57 & 24.18 & 0.5342 & 80 & 131.1 & 0.6187 & 0 & 28 & 41 & 0.905 \\ \hline
        ising\_model\_16 & 16 & 150 & 20 & \textbf{0.4713} & 0 & 150 & 1 & 0.1 & 0.4117 & 9 & 88 & 118.1 & 0.2941 & 120 & 140.9 & 0.4473 & 0 & 38 & 52 & 0.792 \\ \hline
        decod24-v2\_43 & 4 & 22 & 22 & \textbf{0.8956} & 0 & 22 & 1 & 0.02 & 0.7705 & 10 & 52 & 7.62 & 0.8717 & 44 & 126.1 & 0.8813 & 0 & 2 & 22 & 0.67 \\ \hline
        4gt13\_92 & 5 & 30 & 26 & \textbf{0.8604} & 0 & 30 & 1 & 0.02 & \textbf{0.8604} & 0 & 30 & 5.22 & 0.8221 & 58 & 108.29 & 0.8424 & 0 & 5 & 26 & 0.641\\ \hline
        qv\_n12\_d10 & 12 & 180 & 30 & \textbf{0.4032} & 8 & 180 & 2 & 0.27 & 0.2622 & 29 & 119 & 15.43 & 0.2383 & 204 & 209 & 0.3334 & 5 & 48 & 89 & 1.35 \\ \hline
        qft\_16 & 16 & 240 & 58 & \textbf{0.2883} & 41 & 240 & 6 & 0.32 & 0.1549 & 44 & 204 & 118.1 & 0.0957 & 380 & 238 & 0.1476 & 17 & 125 & 217 & 1.486 \\ \hline
        rd84\_142 & 15 & 154 & 81 & \textbf{0.4591} & 7 & 154 & 2 & 0.31 & 0.3369 & 21 & 146 & 60.23 & 0.2448 & 238 & 161.2 & 0.338 & 5 & 55 & 118 & 1.168 \\ \hline
        sf\_274 & 6 & 336 & 300 & \textbf{0.1855} & 0 & 336 & 1 & 0.49 & 0.1721 & 5 & 343 & 218 & 0.1012 & 636 & 879.3 & 0.1186 & 3 & 84 & 305 & 1.83 \\ \hline
        sf\_276 & 6 & 336 & 301 & \textbf{0.1855} & 0 & 336 & 1 & 0.42 & 0.1645 & 8 & 341 & 147.4 & 0.102 & 620 & 693.3 & 0.1060 & 5 & 93 & 309 & 1.89 \\ \hline
        con1\_216 & 9 & 415 & 346 & \textbf{0.1237} & 17 & 415 & 3 & 0.48 & 0.0897 & 22 & 415 & 147.6 & 0.0399 & 776 & 965.1 & 0.0462 & 12 & 146 & 378 & 2.295 \\ \hline
        wim\_266 & 11 & 427 & 352 & \textbf{0.1166} & 12 & 427 & 3 & 0.5 & 0.0819 & 24 & 436 & 163.9 & 0.029 & 758 & 549.7 & 0.0359 & 10 & 124 & 386 & 2.294 \\ \hline
        rd53\_130 & 7 & 448 & 383 & \textbf{0.1054} & 9 & 448 & 3 & 0.56 & 0.0807 & 18 & 459 & 206.8 & 0.0404 & 852 & 943.2 & 0.0185 & 20 & 129 & 434 & 2.615 \\ \hline
        f2\_232 & 8 & 525 & 449 & \textbf{0.0715} & 13 & 525 & 3 & 0.54 & 0.0583 & 14 & 523 & 251.6 & 0.0206 & 978 & 900.2 & 0.016 & 13 & 151 & 468 & 2.727 \\ \hline
        cm152a\_212 & 12 & 532 & 461 & \textbf{0.0691} & 7 & 532 & 2 & 1.22 & 0.0571 & 13 & 498 & 295.8 & 0.0101 & 1010 & 598 & 0.0218 & 6 & 184 & 474 & 2.643 \\ \hline
         rd53\_251 & 8 & 564 & 492 & \textbf{0.0587} & 16 & 564 & 4 & 1.07 & 0.0438 & 20 & 577 & 275.2 & 0.0149 & 1089 & 911.3 & 0.0110 & 16 & 178 & 527 & 3.039 \\ \hline
        hwb5\_53 & 6 & 598 & 535 & \textbf{0.0495} & 23 & 598 & 6 & 0.58 & 0.0375 & 19 & 622 & 337.2 & 0.0168 & 1136 & 1354 & 0.004 & 27 & 216 & 609 & 3.356 \\ \hline
        pm1\_249 & 14 & 771 & 634 & \textbf{0.0205} & 23 & 771 & 4 & 1.8 & 0.0122 & 36 & 763 & 385.1 & 8.18E-04 & 1378 & 755 & 0.0013 & 15 & 231 & 686 & 3.944 \\ \hline
        cm42a\_207 & 14 & 771 & 634 & \textbf{0.0205} & 23 & 771 & 4 & 1.32 & 0.012 & 37 & 754 & 0.61 & 8.93E-04 & 1428 & 802.5 & 0.0013 & 15 & 231 & 686 & 4.215 \\ \hline
        dc1\_220 & 11 & 833 & 705 & \textbf{0.0152} & 19 & 833 & 4 & 2.02 & 0.0083 & 41 & 838 & 380.6 & 9.51E-04 & 1548 & 1194 & 0.0010 & 15 & 189 & 750 & 4.25 \\ \hline
        squar5\_261 & 13 & 869 & 720 & \textbf{0.0125} & 31 & 869 & 6 & 1.91 & 5.35E-03 & 58 & 866 & 409.9 & 4.11E-04 & 1630 & 900.2 & 3.95E-04 & 24 & 290 & 790 & 4.664 \\ \hline
        z4\_268 & 11 & 1343 & 1112 & \textbf{0.0012} & 46 & 1343 & 8 & 2.22 & 3.41E-04 & 83 & 1369 & 687.2 & 1.31E-05 & 2500 & 2017 & 7.90E-07 & 44 & 381 & 1227 & 8.308 \\ \hline
        radd\_250 & 13 & 1405 & 1210 & \textbf{8.36E-04} & 54 & 1405 & 9 & 2.19 & 2.06E-04 & 96 & 1430 & 700.6 & 3.08E-06 & 2658 & 1826 & 3.05E-07 & 44 & 427 & 1306 & 8.75 \\ \hline
        adr4\_197 & 13 & 1498 & 1249 & \textbf{5.25E-04} & 52 & 1498 & 8 & 2.01 & 1.06E-04 & 109 & 1519 & 754.3 & 1.39E-06 & 2786 & 1682 & 1.59E-07 & 44 & 446 & 1372 & 9.669 \\ \hline
        sym6\_145 & 7 & 1701 & 1499 & \textbf{0.0002} & 38 & 1701 & 9 & 1.84 & 9.47E-05 & 49 & 1729 & 947.5 & 5.02E-06 & 3286 & 3210 & 4.66E-07 & 37 & 462 & 1582 & 11.47 \\ \hline
        misex1\_241 & 15 & 2100 & 1797 &\textbf{2.54E-05} & 56 & 2100 & 7 & 3.78 & 6.40E-06 & 95 & 2040 & 1017 & 1.50E-09 & 3936 & 2144 & 9.75E-10 & 29 & 625 & 1852 & 15.52 \\ \hline
        rd73\_252 & 10 & 2319 & 1963 & \textbf{8.47E-06} & 87 & 2319 & 14 & 4.024 & 8.79E-07 & 154 & 2317 & 1130 & 6.50E-09 & 4930 & 3011 & 1.04E-12 & 63 & 731 & 2161 & 18.57 \\ \hline
        square\_root\_7 & 15 & 3089 & 2520 & \textbf{1.64E-07} & 147 & 3089 & 17 & 8.855 & 5.99E-09 & 229 & 2940 & 1583 & 1.56E-13 & 5716 & 2940 & 9.73E-15 & 41 & 835 & 2677 & 25.28 \\ \hline \hline
        \blue{\textbf{Geo.Mean}} & - & - & - & \textbf{0.03458} & - & - & - & 0.3824 & 0.01951 & - & - & 84.21 & 0.0033 & - & 472.6 & 0.0018 & - & - & - & 2.345
    \end{tabular}
    \end{adjustbox}
\end{table*}

\vspace*{2mm}

\subsubsection{Results on benchmark circuits used in  \cite{Li23tetris}}
In \cite{Li23tetris}, the authors evaluated, among others, 33 benchmark circuits from RevLib and IBM Qiskit. After proper decomposition, these circuits include qubit counts ranging from 5 to 16 and up to 3,089 CZ gates. We empirically evaluated the performance of the algorithms on these benchmarks, and the results are summarised in Table~\ref{tab:tetris}.

During compilation, we assume that each SWAP gate inserted by Tetris \blue{and Atomique} is decomposed into three CZ gates and several single-qubit gates. In our calculation of \blue{their} overall fidelity, we completely disregard the presence of additional single-qubit gates in the compiled circuit. As a result, the calculated fidelity is slightly higher than the actual value. 

Recall that Enola operates in both static and dynamic modes. For these benchmark circuits, the two modes produce similar overall fidelity, with the dynamic mode performing slightly better. This improvement comes at the cost of a significantly increased runtime. Indeed, for all circuits with CZ depth $\geq 30$, the dynamic method does not complete within 1 hour. \blue{So we only show results for the static method in Table~\ref{tab:tetris}.}

\blue{Atomique allocates qubits to distinct arrays and does not permit atom transfer. Instead, it may introduce SWAP gates during compilation process. Compilation results also indicate the use of ancilla qubits.}

From Table~\ref{tab:tetris}, we observe the following:
\begin{enumerate}
    \item \blue{\textbf{DasAtom consistently outperforms Tetris and Enola in fidelity and surpasses Atomique in all but one case}. On average, DasAtom's fidelity is 1.77, 10.48, and 19.21 times that of Tetris, Enola, and Atomique, respectively. The largest improvement occurs on the circuit `square\_root\_7', with fidelity ratios of 27.38, 1.05E06, and 1.69E07, respectively.}
    \item \blue{For each algorithm and circuit, compilation can be completed within one hour, with DasAtom and Atomique being significantly faster than Tetris and Enola.}    
    
\end{enumerate}

\subsubsection{Results on practical quantum circuits with varying qubit number and topologies} 
\begin{figure*}
    \centering
    \tikzset{
        node_style/.style={circle, draw, minimum size=1cm, font = \huge, ultra thick},
        edge_style/.style={-, ultra thick},
        every node/.style={node_style},
    }
    \subfloat[]{\scalebox{0.35}{\input{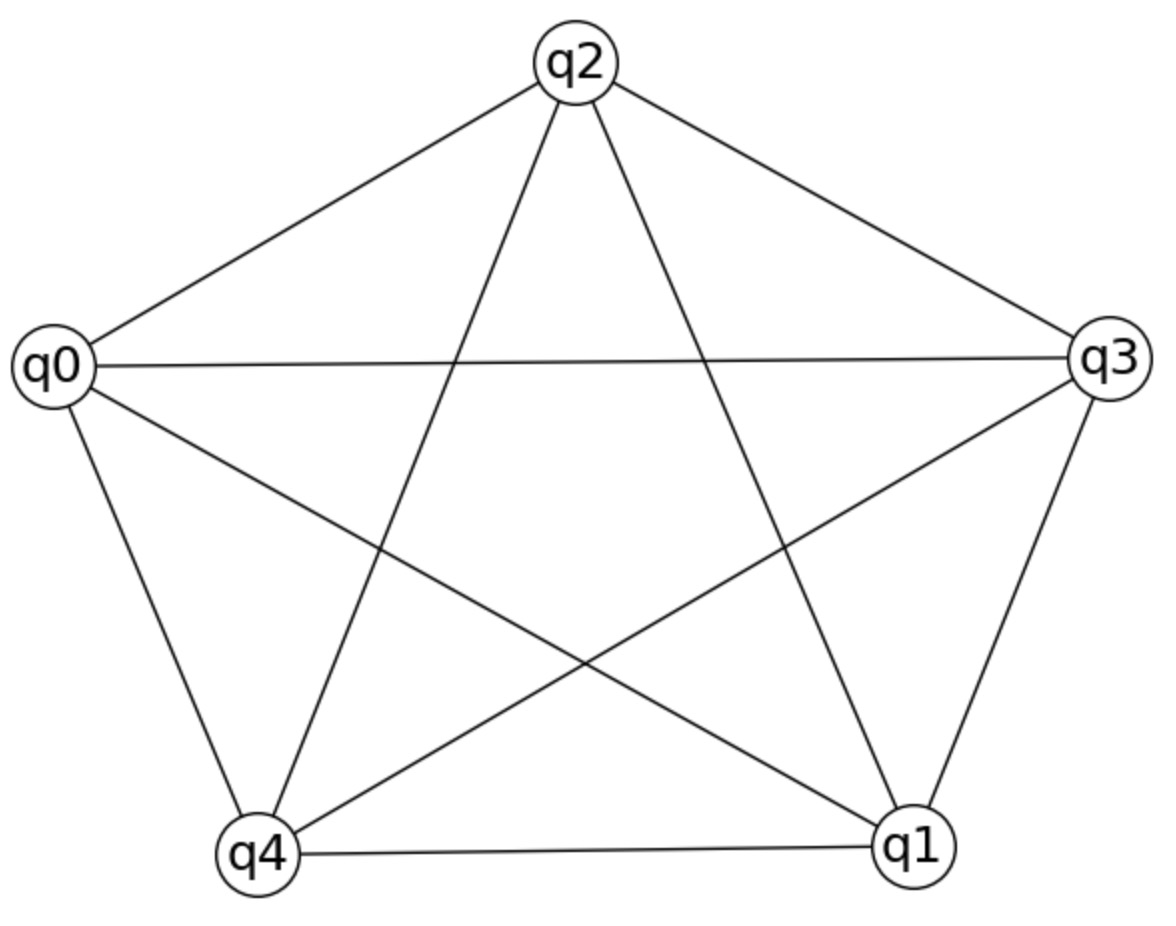}} \quad\quad
    \label{subfig:IG_qft5}}
    \subfloat[]{\scalebox{0.35}{\input{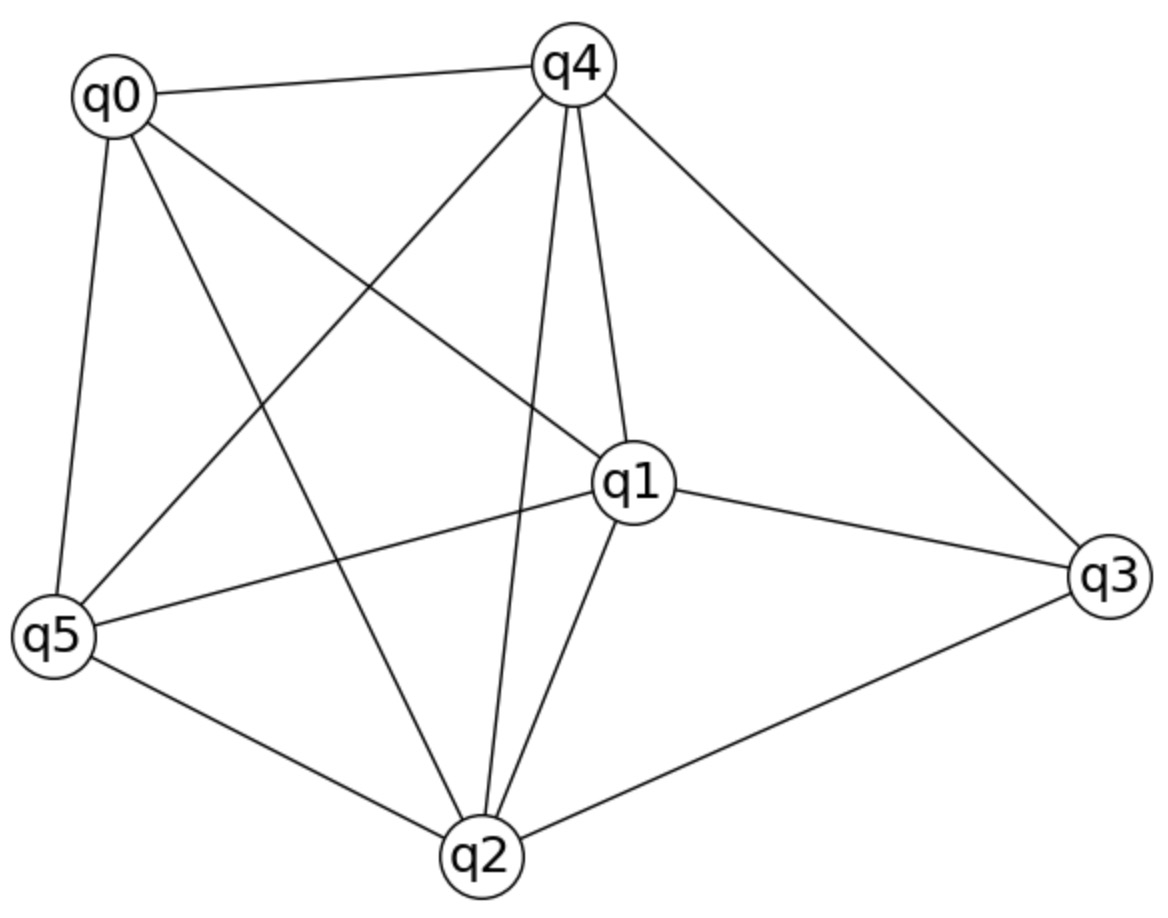}} \quad\quad
    \label{subfig:IG_qv6}}
    \subfloat[]{\scalebox{0.35}{\input{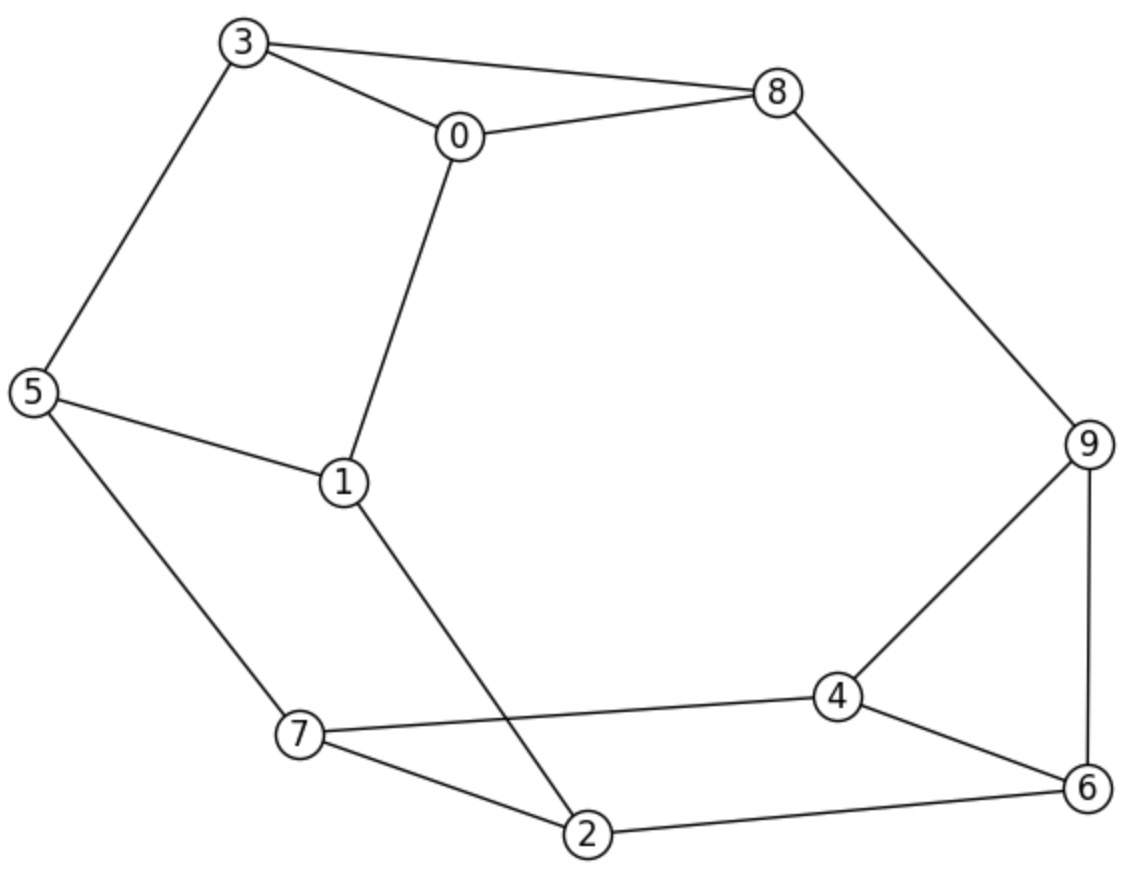}} \quad\quad
    \label{subfig:IG_3reg10}}
    \subfloat[]{\scalebox{0.35}{\input{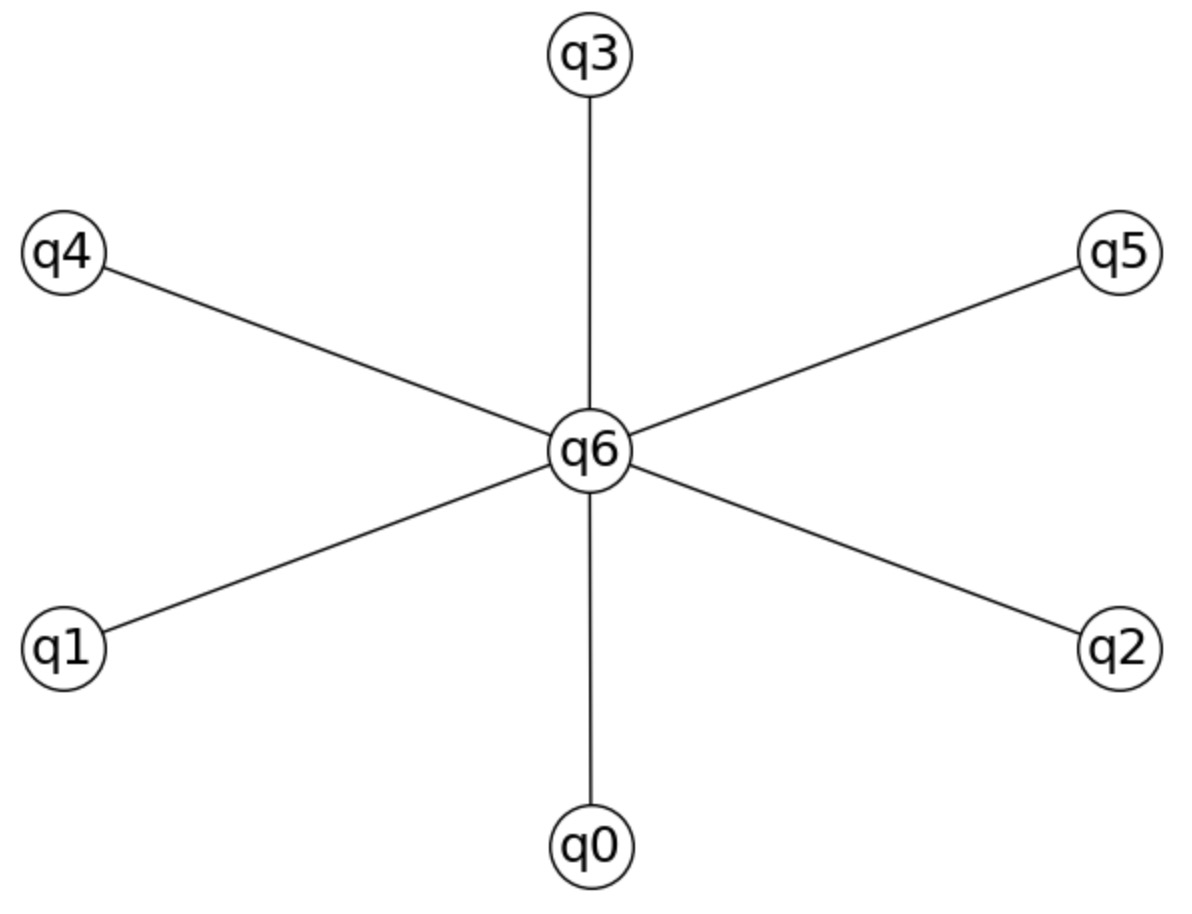}} \quad\quad
    \label{subfig:IG_DJ7}}
    \subfloat[]{\scalebox{0.35}{\input{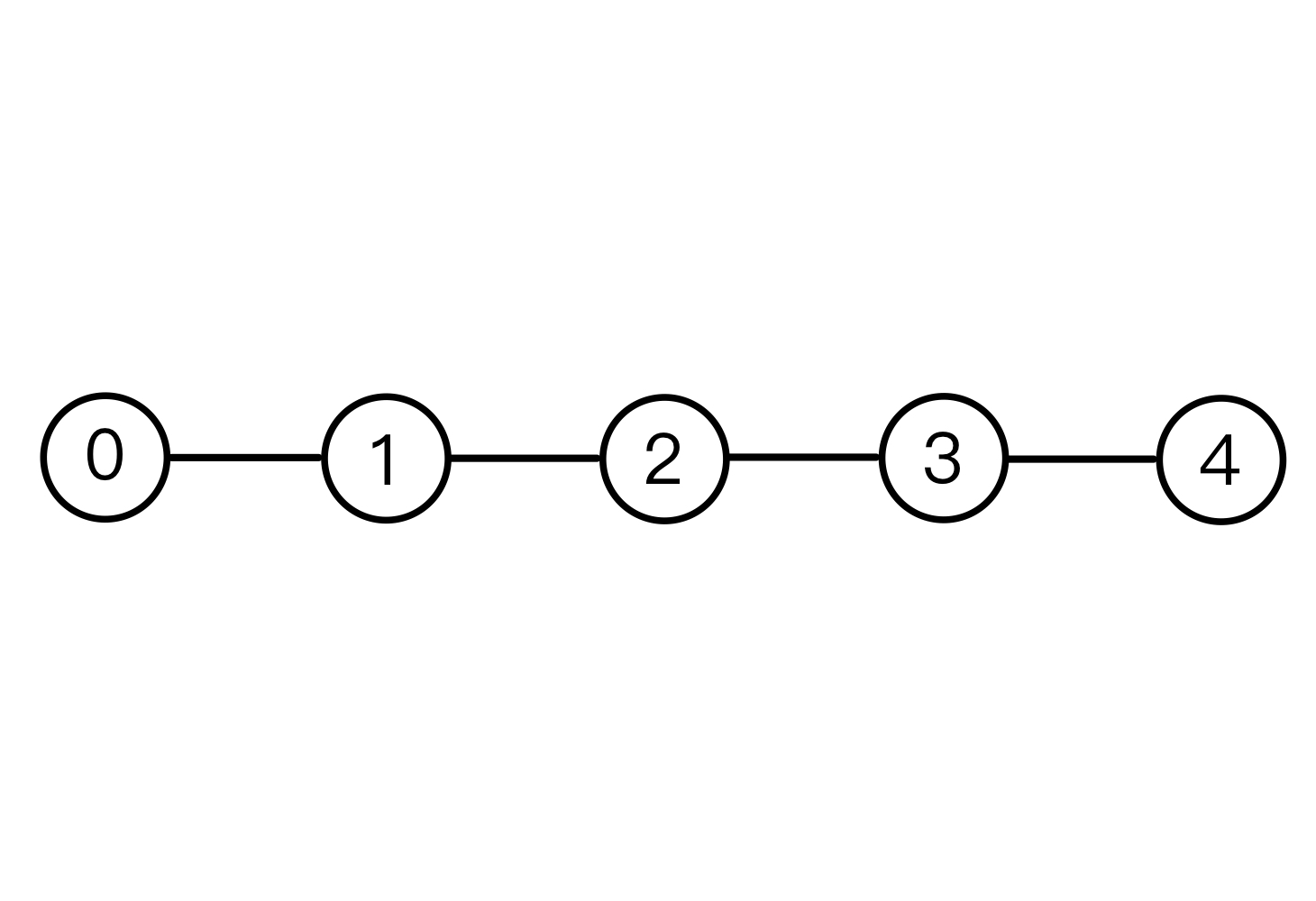}}
    \label{subfig:IG_GHZ5}}
    \caption{\blue{Interaction graphs of (a) QFT-5 and two-local random, (b) QV, (c) 3-regular graph, (d) DJ, and (e) GHZ and W-state}}\label{fig:5intgraphs}
\end{figure*}

To assess the scalability of our algorithm, we also evaluated the performance of the \blue{four} algorithms on larger circuits, extracted from MQTBench\footnote{https://www.cda.cit.tum.de/mqtbench/} and \cite{tan2024compilation}. These circuits include DJ, GHZ, QFT, W-state, QV, two-local random circuits, and 3-regular MaxCut QAOA circuits, which have very different interaction graphs (cf.~Fig.~\ref{fig:5intgraphs}).
Since the dynamic mode of Enola performs exponentially better than the static mode as the number of qubits increases, we focus solely on the dynamic mode in these experiments.  

\begin{table*}[!ht]
    \centering
    \caption{Comparison among DasAtom, Tetris, Enola, \blue{and Atomique} on 20-qubit quantum benchmark circuits with varying topologies, where the columns have the same interpretation as Table~\ref{tab:tetris} \blue{and `a.c.' denotes `almost complete'}.}\label{tab:20q}
    \begin{adjustbox}{width=\textwidth}
    \begin{tabular}{c|c|c|c||c|c|c|c||c|c|c||c|c||c|c|c|c}
    \multicolumn{4}{c||}{Circuit Info.} & \multicolumn{4}{c||}{DasAtom} & \multicolumn{3}{c||}{Tetris} & 
  \multicolumn{2}{c||}{Enola (dynamic)} & \multicolumn{4}{c}{\blue{Atomique}} \\
    \hline
        Name  & \#CZ & Depth & topology & F & M & D & P & F & SW & D & F & M & F & SW & M & D  \\ \hline
        QV  & 600 & 60 & a.c. & \textbf{0.0456} & 61 & 579 & 6 & 0.0055 & 146 & 356 & 0.0039 & 508 & 0.027 & 22 & 185 & 230  \\ \hline
        2-local random  & 570 & 77 & a.c. & \textbf{0.0467} & 166 & 540 & 18 & 0.0013 & 250 & 549 & 0.0037 & 603 & 0.0208 & 37 & 285 & 341  \\ \hline
        QFT   & 410 & 77 & complete & \textbf{0.1176} & 68 & 398 & 8 & 0.0237 & 112 & 324 & 0.0178 & 424 & 0.0606 & 18 & 157 & 252  \\ \hline
        W-state  & 38 & 21 & linear & \textbf{0.8265} & 0 & 38 & 1 & 0.6696 & 14 & 42 & 0.6845 & 55& 0.7873 & 0 & 28 & 38 \\ \hline
        3-regular graph  & 30 & 8 & 3-regular & \textbf{0.8603} & 0 & 30 & 1 & 0.5997 & 24 & 35 & 0.7522 & 32 & 0.8502 & 0 & 9 & 11  \\ \hline
        DJ  & 19 & 19 & star & \textbf{0.9025} & 6 & 19 & 2 & 0.7255 & 15 & 31 & 0.8044 & 37 & 0.9024 & 0 & 19 & 19 \\ \hline
        GHZ  & 19 & 19 & linear & \textbf{0.9091} & 0 & 19 & 1 & 0.7365 & 14 & 28 & 0.8119 & 37 & 0.8868 & 0 & 14 & 19 \\ \hline \hline
        \textbf{\blue{Geo.Mean}} & - & - & - & \textbf{0.2832} & - & - & - & 0.0865 & - & - & 0.0979 & - & 0.2104 & - & - & - 
    \end{tabular}
    \end{adjustbox}
\end{table*}

Table~\ref{tab:20q} summarises the circuit information and performance details of the algorithms on seven 20-qubit circuits. As shown in the table, DasAtom consistently outperforms the other \blue{three} in overall fidelity. In addition, the fidelity of DasAtom is strongly correlated  with the number of CZ gates in the circuit. \blue{Table~\ref{tab:geo_mean_F} also presents the geometric mean of fidelity for each type of circuit with varying qubit counts. The table indicates that (i) DasAtom significantly outperforms both Tetris and Enola, and (ii) Atomique performs comparably to, or even slightly better than, DasAtom on DJ, W-state, and 3-regular circuits. However,  on QFT circuits, DasAtom's fidelity is, on average,  7.9x of that of Atomique. Notably, Atomique also outperforms Enola on these generic circuits; as shown in \cite{tan2024compilation}, Enola excels when leveraging gate commutativity.}

\begin{table}[!ht]
\caption{Geometric mean fidelity of various circuits}
\label{tab:geo_mean_F}
    \centering
    \blue{
    \begin{tabular}{c|c||c|c|c|c}
        Circuit & \#Q  & DasAtom & Tetris & Enola & Atomique \\ \hline
        QFT & 5-50 & \textbf{0.0066} & 3.65E-04 & 2.63E-06 & 8.31E-04 \\ \hline
        two local & 5-30 & \textbf{0.0613} & 0.0044 & 0.0044 & 0.0265 \\ \hline
        QV & 5-25 & \textbf{0.1411} & 0.0393 & 0.0398 & 0.1002 \\ \hline
        DJ & 5-50 & 0.8412 & 0.6791 & 0.6413 & \textbf{0.8615} \\ \hline
        W-state & 5-50 & \textbf{0.7665} & 0.5424 & 0.4763 & 0.6945 \\ \hline
        3-regular & 10-60 & \textbf{0.8548} & 0.2547 & 0.2785 & 0.7890
    \end{tabular}
}
\end{table}

\blue{We now provide a detailed comparison of these algorithms' performance across the different circuit types.}

\paragraph{QFT circuits} These circuits also have complete interaction graphs, see Fig.~\ref{subfig:IG_qft5}. In Fig.~\ref{subfig:qft_F}, the y-axis denotes the ratio of DasAtom's fidelity compared to that of Tetris, Enola, \blue{and Atomique}, plotted on a logarithmic scale. The results clearly demonstrate that the ratio increases exponentially with the number of qubits, ranging from 5 to 50. In terms of runtime, Fig.~\ref{subfig:qft_rt} shows that Enola and Tetris are three and two orders of magnitude slower than DasAtom, respectively, \blue{while Atomique's runtime is closer to that of DasAtom.}

\begin{figure*}
\centering
    \pgfplotsset{
        Qubits_axis/.style={
            xlabel={qubit number},
            ylabel={fidelity ratio (log scale)},
            xtick={5,10,15,20,25,30,35,40,45,50},
            xticklabel style={
                /pgf/number format/fixed,
                /pgf/number format/precision=3,
                rotate=45,
                anchor=east
            },
            width = 8cm,
            height = 6cm,
            grid=both,
            mark repeat=1,
            xlabel style = {font =\Large},
            ylabel style = {font =\Large},                  
        },
        rt_axis/.style={
            xlabel={qubit number},
            ylabel={runtime (log scale)},
            xtick={5,10,15,20,25,30,35,40,45,50},
            xticklabel style={
                /pgf/number format/fixed,
                /pgf/number format/precision=3,
                rotate=45,
                anchor=east
            },
            width = 8cm,
            height = 6cm,
            grid=both,
            mark repeat=1,
            xlabel style = {font =\Large},
            ylabel style = {font =\Large},
        },
        swap_axis/.style={
            xlabel={qubit number},
            ylabel={number of gates},
            xtick={5,10,15,20,25,30,35,40,45,50},
            xticklabel style={
                /pgf/number format/fixed,
                /pgf/number format/precision=3,
                rotate=45,
                anchor=east
            },
            width = 8cm,
            height = 6cm,
            grid=both,
            mark repeat=1,
            xlabel style = {font =\Large},
            ylabel style = {font =\Large},  
        },
        move_dis_axis/.style={
         xlabel={qubit number},
            ylabel={distance (log scale)},
            xtick={5,10,15,20,25,30,35,40,45,50},
            xticklabel style={
                /pgf/number format/fixed,
                /pgf/number format/precision=3,
                rotate=45,
                anchor=east
            },
            width = 8cm,
            height = 6cm,
            grid=both,
            mark repeat=1,
            xlabel style = {font =\Large},
            ylabel style = {font =\Large}, 
        },
        move_stage_axis/.style={
        xlabel={qubit number},
            ylabel={\#move stage (log scale)},
            xtick={5,10,15,20,25,30,35,40,45,50},
            xticklabel style={
                /pgf/number format/fixed,
                /pgf/number format/precision=3,
                rotate=45,
                anchor=east
            },
            width = 8cm,
            height = 6cm,
            grid=major,
            mark repeat=1,
            xlabel style = {font =\Large},
            ylabel style = {font =\Large}, 
        },
        atom_trans_axis/.style={
            xlabel={qubit number},
            ylabel={number of atom transfer},
            xtick={5,10,15,20,25,30,35,40,45,50},
            xticklabel style={
                /pgf/number format/fixed,
                /pgf/number format/precision=3,
                rotate=45,
                anchor=east
            },
            width = 8cm,
            height = 6cm,
            grid=major,
            mark repeat=1,
            xlabel style = {font =\Large},
            ylabel style = {font =\Large}, 
        },
        t_axis/.style={
        xlabel={qubit number},
            ylabel={time (log scale)},
            xtick={5,10,15,20,25,30,35,40,45,50},
            xticklabel style={
                /pgf/number format/fixed,
                /pgf/number format/precision=3,
                rotate=45,
                anchor=east
            },
            width = 8cm,
            height = 6cm,
            grid=major,
            mark repeat=1,
            xlabel style = {font =\Large},
            ylabel style = {font =\Large}, 
        }
    }
    \subfloat[Fidelity ratio]{
    \scalebox{0.5}{
    \input{tex/tex_methods_comp/tikz_QFT_F_comp}
    }
    \label{subfig:qft_F}
    }
    \hfill
    \subfloat[Runtime (s)]{
    \scalebox{0.5}{
    \input{tex/tex_methods_comp/tikz_QFT_rt_comp}
    }
    \label{subfig:qft_rt}
    }
    \hfill
    \subfloat[SWAP count]{
    \scalebox{0.5}{
    \input{tex/tex_diff_variables/tikz_SWAP}
    }
    }
    \hfill
    \subfloat[Move stage count]{
    \scalebox{0.5}{
    \input{tex/tex_diff_variables/tikz_move_stage}
    }
    }
    \hfill
    \subfloat[Total movement distance ($\mu$m)]{
    \scalebox{0.5}{
    \input{tex/tex_diff_variables/tikz_move_dis}
    }
    }
    \hfill
    \subfloat[Number of atom transfer]{
    \scalebox{0.5}{
    \input{tex/tex_diff_variables/tikz_atom_trans}
    }
    }
    \hfill
    \subfloat[$T_\text{idle}$ ($\mu$s)]{
    \scalebox{0.5}{
    \input{tex/tex_diff_variables/tikz_t_idle}
    }
    }
    \hfill
    \subfloat[$T$  ($\mu$s)]{
    \scalebox{0.5}{
    \input{tex/tex_diff_variables/tikz_t_total}
    }
    }
\caption{\blue{Change of key performance indicators of QFT circuits under different algorithms}} \label{fig:QFTchange}
\end{figure*}

\paragraph{Two-local random circuits} These circuits have complete interaction graphs. Fig.~\ref{subfig:two_local_F} shows that DasAtom's fidelity outperforms of Tetris, Enola, \blue{and Atomique}, with performance ratios increasing exponentially as the qubit count ranges from 5 to 30. \blue{For Atomique, the increase in the performance ratio is  more gradual.}  Note that only results for circuits with up to 30 qubits are shown here, as Enola's runtime is very slow for larger circuits.

\begin{figure*}
    \centering
    \pgfplotsset{
        Qubits_axis/.style={
            xlabel={qubit number},
            ylabel={fidelity},
            xtick={5,10,15,20,25,30,35,40,45,50},
            xticklabel style={
                /pgf/number format/fixed,
                /pgf/number format/precision=3,
                rotate=45,
                anchor=east
            },
            width = 8cm,
            height = 6cm,
            grid=both,
            mark repeat=1,
            xlabel style = {font =\Large},
            ylabel style = {font =\Large},
        },
        two_local_axis/.style={
            xlabel={qubit number},
            ylabel={fidelity ratio (log scale)},
            xtick = {5,10,15,20,25,30},
            xticklabel style={
                /pgf/number format/fixed,
                /pgf/number format/precision=3,
                rotate=45,
                anchor=east
            },
            ytick = {1,10,100,1000,10000},
            width = 8cm,
            height = 6cm,
            grid=major,
            mark repeat=1,
            xlabel style = {font =\Large},
            ylabel style = {font =\Large}, 
        },
        qv_axis/.style={
            xlabel={qubit number},
            ylabel={fidelity ratio},
            xtick = {5,10,15,20,25},
            xticklabel style={
                /pgf/number format/fixed,
                /pgf/number format/precision=3,
                rotate=45,
                anchor=east
            },
            width = 8cm,
            height = 6cm,
            grid=both,
            mark repeat=1,
            xlabel style = {font =\Large},
            ylabel style = {font =\Large},
        },
        3_regular_axis/.style={
            xlabel={qubit number},
            ylabel={fidelity},
            xtick = {10,20,30,40,50,60},
            xticklabel style={
                /pgf/number format/fixed,
                /pgf/number format/precision=3,
                rotate=45,
                anchor=east
            },
            width = 8cm,
            height = 6cm,
            grid=both,
            mark repeat=1,
            xlabel style = {font =\Large},
            ylabel style = {font =\Large},
        }
    }

    \subfloat[two local random]{
    \scalebox{0.6}{
        \input{tex/tex_methods_comp/tikz_two_local_F_comp}
    }
    \label{subfig:two_local_F}
    }
    \hspace{0.3cm}
    \subfloat[quantum volume]{
    \scalebox{0.6}{
    \input{tex/tex_methods_comp/tikz_qv_F_comp}
    }
    \label{subfig:qv_F}
    }
    \hspace{0.3cm}
    \subfloat[3-regular]{
    \scalebox{0.6}{
    \input{tex/tex_methods_comp/tikz_3_regular_F_comp}
    }
    \label{subfig:3_reg_F}
    }

    \vspace{0.8cm}
    \subfloat[DJ]{
    \scalebox{0.6}{
    \input{tex/tex_methods_comp/tikz_DJ_F_comp}
    }
    \label{subfig:dj_F}
    }
    \hspace{0.3cm}
    \subfloat[GHZ]{
    \scalebox{0.6}{
    \input{tex/tex_methods_comp/tikz_GHZ_F_comp}
    }
    \label{subfig:ghz_F}
    }
    \hspace{0.3cm}
    \subfloat[W-state]{
    \scalebox{0.6}{
    \input{tex/tex_methods_comp/tikz_Wstate_F_comp}
    }
    \label{subfig:W_F}
    }
    \caption{\blue{Fidelity comparison on two-local random, QV, 3-regular, DJ, GHZ, and W-state circuits}}
    \label{fig:all_methods}
\end{figure*}

\paragraph{Quantum Volume circuits}  These circuits have (nearly) complete interaction graphs. Fig.~\ref{subfig:qv_F} shows that DasAtom's fidelity outperforms that of Tetris and Enola, with performance ratios increasing exponentially with qubit count. \blue{Meanwhile, DasAtom achieves an average fidelity improvement of 1.4 times over Atomique.}

\paragraph{3-Regular MaxCut QAOA circuits} These circuits feature 3-regular interaction graphs. Fig.~\ref{subfig:3_reg_F} shows that the fidelity of all four algorithms decrease as the qubit count increases. DasAtom significantly outperforms both Tetris and Enola, \blue{while its fidelity remains close to that of Atomique.}

\paragraph{Deutsch-Jozsa circuits} These circuits have star-like interaction graphs.  Fig.~\ref{subfig:dj_F} shows how fidelity decreases as the qubit count increases for the \blue{four} algorithms. DasAtom outperforms both Tetris and Enola by a large margin, \blue{but is slightly outperformed by Atomique. This may be  because the star-like graph is a bipartite graph, allowing the central qubit to be placed on one AOD array, with all other qubits on the SLM array, thus avoiding intra-array two-qubit interactions.}

\paragraph{GHZ circuits and W-state circuits} These circuits have linear interaction graphs.  Fig.s~\ref{subfig:ghz_F} and~\ref{subfig:W_F} show how the fidelity decreases as the qubit count increases for \blue{all} algorithms. \blue{DasAtom clearly outperforms the other three.}

\subsection{\blue{Analysis}}

\blue{The experiments demonstrate that DasAtom outperforms the other three algorithms, achieving exponentially better results in QFT and QV circuits. In this subsection, we address two key questions: Which hardware parameters most significantly impact DasAtom's performance? What factors enable DasAtom to outperform the other algorithms?
}
\vspace*{2mm}
\subsubsection{\blue{Ablation studies on hardware parameters}}
\blue{
\paragraph{Interaction and restriction radii}
We start by evaluating DasAtom's performance with different interaction radii $r_\text{int}\in \set{1,\sqrt{2},2}$ on DJ circuits (5 to 45 qubits) and QFT circuits (5 to 20 qubits). The results indicate that setting $r_\text{int}=2$ yields slightly better performance than other values, justifying our choice of $R_\text{int}=2d$ in the evaluations presented in Sec.~\ref{sec:evaluation}. 
}

\blue{
To analyse the effects of different interaction and restriction radii on DasAtom, we examined the QFT-30 circuit, varying $r_\text{int}$ and $r_\text{restr}$ from $1$ to $8$, with the condition that $r_\text{restr}\geq r_\text{int}$. Notably, when $r_\text{int}\geq 9$, the architecture graph becomes complete on $G(6,6)$, rendering further increases in $r_\text{int}$ unnecessary. As shown in Fig.~\ref{fig:rint}, increasing $r_\text{int}$ from 1 to 8 has a more pronounced effect on Tetris than on DasAtom: the fidelity of Tetris rises from 2.3E-06 to 0.01, while DasAtom's fidelity increases from 0.007 to 0.01. This difference is partly related to the fact that Tetris uses a smaller  restriction area than DasAtom, even when $r_{\text{restr}}=r_{\text{int}}$.
}

\blue{
For DasAtom, we also experimented with different values of $r_{\text{restr}}$ for each $r_{\text{int}}$. The findings show that as $r_{\text{restr}}$ increases, fidelity gradually decreases. Once $r_{\text{restr}}$ reaches a certain threshold, the fidelity stabilises because all gates must be executed serially. However, the reduction in fidelity due to increasing $r_{\text{restr}}$ is nearly negligible.
}

\begin{figure}
    \centering
    \pgfplotsset{
    rint_axis/.style={
            xlabel={$r_{\text{int}}$},
            ylabel={fidelity},
            xtick={1,2,3,4,5,6,7,8},
            xticklabel style={
                /pgf/number format/fixed,
                /pgf/number format/precision=3,
                rotate=45,
                anchor=east
            },
            width = 8cm,
            height = 6cm,
            grid=major,
            legend entries={DasAtom, Tetris},
            mark repeat=1,
            legend pos=south east,
            xlabel style = {font =\Large},
            ylabel style = {font =\Large},
        }  
    }
    \scalebox{0.6}{\input{tex/tex_data/tikz_rint_range}}
    \caption{\blue{Fidelity comparison on QFT-30 with various interaction radii, where values at $r_\text{int}=1,5,8$ are explicitly shown.}}
    \label{fig:rint}
\end{figure}

\begin{figure*}
    \centering
    \pgfplotsset{
        F_CZ_axis/.style={
            ylabel={fidelity},
            xtick={0.95,0.96,0.98,0.99,0.999},
            xticklabel style={
                /pgf/number format/fixed,
                /pgf/number format/precision=3,
                rotate=45,
                anchor=east
            },
            width = 8cm,
            height = 6cm,
            grid=both,
            mark repeat=1                   
            ,
            xlabel style = {font =\Large},
            ylabel style = {font =\Large},
        },
        F_atom_trans_axis/.style={
            xtick={0.991,0.993,0.995,0.997,0.999,1},
            xticklabel style={
                /pgf/number format/fixed,
                /pgf/number format/precision=3,
                rotate=45,
                anchor=east
            },
            width = 8cm,
            height = 6cm,
            grid=both,
            mark repeat=1,
            xlabel style = {font =\Large},
            ylabel style = {font =\Large},
            scaled y ticks=false
        },
        T2_axis/.style={
            xtick={0.15, 1.5, 5, 10, 15},
            xticklabel style={
                /pgf/number format/fixed,
                /pgf/number format/precision=3,
                rotate=45,
                anchor=east
            },
            width = 8cm,
            height = 6cm,
            grid=both,
            mark repeat=1                    
            ,
            xlabel style = {font =\Large},
            ylabel style = {font =\Large},
            ytick={0, 0.05,0.1},
            yticklabel style={
            /pgf/number format/fixed,
            /pgf/number format/precision=2
            },
            scaled y ticks=false
        },
        dis_axis/.style={
            xtick={1, 3, 5, 7, 9, 11, 13, 15, 17, 20},
            xticklabel style={
                /pgf/number format/fixed,
                /pgf/number format/precision=3,
                rotate=45,
                anchor=east
            },
            width = 8cm,
            height = 6cm,
            grid=both,
            mark repeat=1,
            xlabel style = {font =\Large},
            ylabel style = {font =\Large}, 
            ytick={0, 0.05,0.1},
            yticklabel style={
            /pgf/number format/fixed,
            /pgf/number format/precision=2
            },
            scaled y ticks=false
        }
    }
    \subfloat[$f_{\text{trans}}$]{
        \scalebox{0.55}{\input{tex/tex_data/tikz_F_trans_QFT_benchmark}}
    }
    \hfill
    \subfloat[$f_{\text{cz}}$]{
        \scalebox{0.55}{\input{tex/tex_data/tikz_F_cz_QFT_benchmark}}
    }
    \hfill
    \subfloat[$T_2$ (s)]{
        \scalebox{0.55}{\input{tex/tex_data/tikz_T2_QFT_benchmark}}
    }
    \hfill
    \subfloat[Atom distance ($\mu$m)]{
        \scalebox{0.55}{\input{tex/tex_data/tikz_dis_QFT_benchmark}}
    }
\caption{Ablation studies on QFT-20 for hardware parameters $f_\text{cz},f_\text{trans},T_2$, and atom distance $d$.}\label{fig:ablation}
\end{figure*}

\vspace*{2mm}
\blue{
\paragraph{The most important factors}
From Eq.~\ref{eq:appr.success_probability}, the approximate success probability of the compiled circuit,  $P(C) = \exp\big(\!-\frac{T_{\text{idle}}}{T_2}\big) \times f_\text{cz}^m \times f_\text{trans}^{s}$ exponentially depends on $T_\text{idle}$ and polynomially on $f_\text{cz}$ and $f_\text{trans}$, raised to the powers of the CZ count $m$ and atom transfer count $s$, respectively. To identify the parameters that most affect performance, we conducted ablation studies on the QFT-20 circuit, focusing on hardware parameters $f_\text{trans},f_\text{cz},T_2$, and atom distance $d$. The results, shown in Fig.~\ref{fig:ablation}, reveal the following:
\begin{itemize}
    \item $f_\text{cz}$ has the most significant impact on all  algorithms.
    \item $f_\text{trans}$ also affects DasAtom and Enola significantly, but has no impact on the other two algorithms.
    \item Coherence time $T_2$ has a strong influence on all algorithms except Tetris, when $T_2$ varies from 0.15s to 15s.
    \item Increasing the atom distance from 1 to 20 reduces DasAtom's fidelity by only 3.4\%. This parameter clearly affects Atomique and Enola, but has no impact on Tetris' fidelity.
\end{itemize}
}

\blue{Like atom distance, other parameters, including atom movement speed $v$, and CZ and atom transfer durations ($t_\text{cz}$ and $t_\text{trans}$), have relatively limited impact on DasAtom's fidelity. Specifically, reducing $v$ from 0.55 to 0.25, increasing $t_\text{cz}$ from 0.2 to 0.5, or increasing $t_\text{trans}$ from 20 $\mu$s to 40 $\mu$s decreases fidelity by approximately $1.4\%$, $0.15\%$, and $11.8\%$, respectively.}

\blue{
Similar findings are observed for QV-20. However, due to the simpler topologies of the 20-qubit GHZ and 3-regular graph circuits, the impact of $f_\text{cz}$ and $f_\text{trans}$ is somewhat less pronounced, though the overall trend remains similar to that for QFT-20 and QV-20. 
}

\vspace*{2mm}
\blue{
\subsubsection{Why DasAtom performs better?}
We analyse here why DasAtom outperforms the other algorithms.}

\vspace*{2mm}
\paragraph{SWAP gates have low fidelity}\label{sec:swap}
Based on our hardware parameter settings (see Table.~\ref{tab:parameter}), the fidelity cost of inserting a SWAP gate can be approximated as $f_\text{cz}^3=0.995^3=0.985$. In contrast, the fidelity cost of moving an atom at a speed of $0.55$ $\mu\text{m}/\mu\text{s}$ over a distance $x\mu\text{m}$ can be estimated as $\exp\left(-\frac{2\times T_{\text{trans}}+x/0.55}{T_2}\right)$, where $2 \cdot T_{\text{trans}}$ represents the time for the atom to transition between SLM to AOD traps at both the initial and target positions. A simple calculation shows that the fidelity drops to 0.985 when $x\approx 12,384$ $\mu\text{m}$. Additionally, for swapping the positions of two atoms, the distance can be up to 6,202 $\mu\text{m}$ for the fidelity to remain at 0.985. This illustrates that SWAP gates incur significant fidelity costs. \blue{As shown in Fig.~\ref{fig:QFTchange}c, Tetris has a much rapid increase in SWAP count compared to Atomique. 
}

\blue{\paragraph{Atom transfer and atom shuttling are slow} 
In contrast, atom transfer and shuttling can be up to 100x slower than SWAP gates. This implies that the overall circuit execution time $T$, and consequently the qubit idle time $T_\text{idle}$, may be significantly prolonged for algorithms dependent on atom shuttling. As shown in Fig.~\ref{fig:QFTchange}e-h, Enola exhibits the fastest increases in total movement distance, atom transfer count, $T_\text{idle}$, and $T$ among the three algorithms that use atom shuttling.
}

\vspace*{2mm}
\blue{
\begin{remark}
Fig.~\ref{fig:QFTchange}a demonstrates an exponential increase in  DasAtom's fidelity ratio compared to Tetris, Enola, and Atomique for QFT circuits. Since this evaluation assumes  $f_\text{trans}=1$, the approximate success probability  (cf. Eq.~\ref{eq:appr.success_probability}) is essentially
$ P(C) = \exp\big(\!-\frac{T_{\text{idle}}}{T_2}\big) \times f_\text{cz}^m $. Here, $m$ (the CZ gate count) generally increases with the qubit count $n$. DasAtom's exponential fidelity improvement over Tetris (Atomique) is primarily (partly) due to the additional SWAP gates inserted, which increase at least linearly with $n$ (cf. Fig.~\ref{fig:QFTchange}c). For Enola (Atomique), the increase is primarily (partly) driven by higher $T_\text{idle}$ values as $n$ grows (cf. Eq.~\ref{eq:T_idle} and Fig.~\ref{fig:QFTchange}g), caused by increased atom transfers and total movement distance (cf. Eq.~\ref{eq:T} and Fig.~\ref{fig:QFTchange}e,f). Note that if $f_\text{trans}<1$, DasAtom's fidelity ratio compared to Enola will be even higher, as Enola's increasing atom transfer count $s$ (appearing as an exponent of $f_\text{trans}$ in Eq.~\ref{eq:appr.success_probability}) grows faster than that of DasAtom (cf.~\ref{fig:QFTchange}f).
\end{remark}
}

\section{Further discussions}
The evaluations in the previous section demonstrate that, despite its simplicity,  the DAC scheme is highly effective. This effectiveness can be attributed to two primary factors: first, atom shuttling incurs significantly lower fidelity costs than SWAP gate insertion (cf. Sec.~\ref{sec:swap}); second, long-range interactions enable dense qubit connectivity, \blue{minimising the need for mapping transitions and ultimately lowering atom movement costs.}

\blue{
To implement DasAtom, we assume the NAQC architecture has the following capabilities: high-fidelity atom shuttling during computation, long-range interaction, and individually applied CZ gates. Notably, a significant portion of DasAtom's improvement over Tetris stems from the hardware capability of in-computation atom shuttling. A rough estimate indicates that, when simulating SWAP gates with atom shuttling (cf.~\cite{Schmid24qst}), DasAtom outperforms Tetris by 3.14x on QFT-50. By contrast, Fig.~\ref{fig:QFTchange}a shows an improvement of 8,108.5x under the default settings.
}

\blue{
DasAtom assumes a compact atom arrangement with atom distance smaller than the interaction radius, while Atomique and Enola assume an atom distance that is 2.5x the interaction radius. This configuration results in shorter atom shuttling distances for DasAtom. Recall that the total movement distance $D$ can contribute significantly to the circuit execution time $T$ and qubit idle time $T_\text{idle}$ (cf. Eq.~\ref{eq:T_idle} and Eq.~\ref{eq:T}). One might wonder if this compact atom spacing is the primary reason for DasAtom's exponential performance advantage over Enola as the qubit count increases on QFT circuits. However, as shown in Fig.~\ref{fig:ablation}d, increasing the atom distance from 1 to 20 $\mu$m only reduces DasAtom's fidelity on QFT-20 by about 7\%. The primary reason for DasAtom's success over Enola lies in its exponentially shorter total movement distance and lower atom transfer count (cf. Fig.~\ref{fig:QFTchange}e,f), which together result in reduced circuit execution time $T$ and shorter qubit idle time $T_\text{idle}$. 
}

\subsection{\blue{Implications}}

\blue{
 Although a denser atom configuration is generally advantageous, atom distance does not significantly impact DasAtom's improvement over other algorithms. In fact, a smaller atom distance translates to a larger interaction factor $r_\text{int}$. As shown in Fig.~\ref{fig:rint} and Fig.~\ref{fig:ablation}d, changes in fidelity remain modest as $r_\text{int}$ or atom distance varies for DasAtom. 
}

\blue{
Our chosen architecture parameters are optimistic, and there are scenarios where Tetris may outperform DasAtom. For instance, Fig.~\ref{fig:tetris_larger} shows that for the QFT-50 circuit, Tetris achieves higher fidelity than DasAtom under certain conditions, such as when $f_\text{trans}\leq 0.99$ and $f_\text{cz}\geq 0.98$. In general, Tetris performs comparatively better when $f_\text{cz}$ is close to 1, $f_\text{trans}$ is low, $T_2$ is small, and a larger $r_\text{int}$ is available. 
}

\blue{
Enola and Atomique are suitable when using a DPQA architecture, where individually applied CZ gates are unavailable. This is particularly advantageous for certain circuits in QAOA and quantum simulation, where gate commutativity can be exploited. For generic circuits, Enola's fidelity improves faster with higher values of $f_\text{trans}$ and $T_2$ (cf. Fig.~\ref{fig:ablation}a,c). However, due to the exponential increase in total movement distance and atom transfer count with the qubit count (cf. Fig.~\ref{fig:QFTchange}e,f), Enola is unlikely to outperform DasAtom when CZ gates are individually addressable. 
}

\begin{figure}
    \centering
    \scalebox{0.5}{
    \input{tex/T2fixed_tetris}
    }
    \caption{\blue{Shaded areas show where Tetris achieves higher fidelity than DasAtom for QFT-20 (teal cross-hatch) and QFT-50 (orange horizontal lines).}}
    \label{fig:tetris_larger}
\end{figure}

\blue{
When atom transfer fidelity is relatively low but CZ gate fidelity is high, Atomique is a more suitable choice, especially for circuits with moderate depth (e.g., $\leq 100$, cf. Table~\ref{tab:tetris}). Additionally, Atomique is scalable, but it is also worth noting that it requires multiple AOD arrays and ancilla qubits.
}

\subsection{\blue{Limitation and potential improvement}}
It is important to note that our implementation is not yet fully optimised. Specifically, the mappings for each subcircuit could be designed to enable smoother transitions, making the process more efficient. For example, the mapping of the second part of QFT-5 shown in Fig.~\ref{fig:qft5_p2}c is not optimal. A better mapping would be to place $q_4$ at $(0,2)$ and $q_1$ at $(1,0)$, requiring only an exchange between the positions of $q_0$ and $q_4$. 

Although our algorithm relies on subgraph isomorphism checks, which are theoretically NP-hard, this has not proven to be a significant obstacle in most cases. By leveraging the Rustworkx implementation of VF2, we successfully compiled the 500-qubit QFT in about one hour. However, this implementation does not perform as well on certain circuits, like QV, especially when the number of qubits exceeds 25. To handle circuits with thousands of qubits, more efficient (though potentially approximate) subgraph isomorphism algorithms \blue{(like PathLAD+ \cite{Wang24pathlad+})} will be necessary. Given that our target architecture is grid-based, this should not pose a major challenge. However, as shown in the fidelity ratio curves in Fig.~\ref{fig:QFTchange}a, the primary concern for compilation on NAQC hardware should be low fidelity rather than scalability.

\blue{Our divide-and-shuttle method could be integrated with the multi-AOD array approach introduced in Atomique \cite{wang2024atomique}. In Atomique, program qubits are distributed across different arrays. Interactions between qubits in different arrays are performed via atom shuttling, while SWAP gates are used for interactions between qubits within the same array. However, as shown in Table~\ref{tab:tetris}, Atomique's performance degrades significantly even compared to Enola when the circuit depth exceeds 300. A hybrid strategy might involve partitioning the circuit into subcircuits, employing Atomique within each subcircuit, and utilising atom shuttling to connect consecutive subcircuits. Although this would introduce atom transfers, it has the potential to enhance overall fidelity by reducing the number of inserted SWAP gates.
}

\section{Conclusion}
We have introduced DasAtom, a novel algorithm designed to optimise the execution of quantum circuits on neutral atom platforms by exploiting long-range interactions and atom shuttling. Unlike previous methods such as SWAP-based Tetris and atom move-based Enola, DasAtom achieves significant fidelity improvements through partitioning circuits into subcircuits and dynamically adjusting qubit mappings. Our results demonstrate that for a 30-qubit QFT,  DasAtom outperforms Enola by 415.8x and Tetris by 10.6x, illustrating the substantial gains in fidelity that can be achieved with this method.

DasAtom's ability to scale its performance with the increasing number of qubits highlights its potential as a crucial tool in the future of NAQC. As quantum circuits grow in complexity and size, the exponential improvement in fidelity offered by DasAtom will become increasingly valuable. This work not only provides a pathway to more efficient quantum circuit execution on NAQC platforms but \blue{could also inspire hardware experts to explore and experiment with new architectures}. Future work will focus on expanding DasAtom's capabilities to handle even larger and more complex quantum circuits.

\section*{Acknowledgement}
We gratefully acknowledge the anonymous reviewers for their constructive feedback, which has significantly enhanced the presentation of this work.

\bibliographystyle{IEEEtran} 
\bibliography{na,naplus,qct_lsj}
\end{document}

%% file: tex/arch_ex.tex
\begin{tikzpicture}
    \begin{scope}[line width=0.5mm]
    \tikzset{
        useNode/.style={circle, draw=black, ,minimum size=8pt, inner sep=1pt, line width=0.2 mm},
    }

    \foreach \x in {0,1,2} {
        \foreach \y in {0,1,2} {
            \node[useNode] (q\x\y) at (3*\x,3*\y) {(\x,\y)};
        }
    }

    \foreach \x in {0,1,2} {
        \foreach \y in {0,1,2} {
            \pgfmathtruncatemacro{\xplusone}{\x+1}
            \pgfmathtruncatemacro{\yplusone}{\y+1}
            
            \ifnum\x<2 \draw[black] (q\x\y) -- (q\xplusone\y); \fi
            \ifnum\y<2 \draw[black] (q\x\y) -- (q\x\yplusone); \fi
        }
    }
    
    \foreach \x in {0,1,2} {
        \foreach \y in {0,1,2} {
            \pgfmathtruncatemacro{\xplusone}{\x+1}
            \pgfmathtruncatemacro{\xminusone}{\x-1}
            \pgfmathtruncatemacro{\yplusone}{\y+1}
            
            \ifnum\x<2 \ifnum\y<2 \draw[color={rgb,255:red,217;green,80;blue,61}, dotted] (q\x\y) -- (q\xplusone\yplusone); \fi \fi
            \ifnum\x>0 \ifnum\y<2 \draw[color={rgb,255:red,217;green,80;blue,61}, dotted] (q\x\y) -- (q\xminusone\yplusone); \fi \fi
        }
    }

    \draw[color={rgb,255:red,38;green,115;blue,239}] (q00) to [bend left] (q02);
    \draw[color={rgb,255:red,38;green,115;blue,239}] (q02) to [bend left] (q22);
    \draw[color={rgb,255:red,38;green,115;blue,239}] (q22) to [bend left] (q20);
    \draw[color={rgb,255:red,38;green,115;blue,239}] (q20) to [bend left] (q00);
    \draw[color={rgb,255:red,38;green,115;blue,239}] (q12) to [bend left] (q10);
    \draw[color={rgb,255:red,38;green,115;blue,239}] (q21) to [bend left] (q01);
    \end{scope}
\end{tikzpicture}

%% file: tex/mov_fig.tex

\begin{tikzpicture}
    \tikzstyle{unoccupied}=[fill=gray!30, circle, minimum size=6pt, inner sep=0pt]
    \tikzstyle{occupied}=[fill=black, circle, minimum size=6pt, inner sep=0pt]
    \tikzstyle{aod_circle}=[fill={rgb,255:red,128; blue,128}, opacity=0.3]
    \tikzstyle{row_aod_line}=[color={rgb,255:red,217;green,80;blue,61},thick, dashed]
    \tikzstyle{col_aod_line}=[color={rgb,255:red,217;green,80;blue,61},thick, dashed]

    \begin{scope}
        \draw[->, thick] (-0.7, -0.7) -- (3.8, -0.7) node[right] {x};
        \draw[->, thick] (-0.7, -0.7) -- (-0.7, 3.8) node[above] {y};

        \foreach \x in {0, 1, 2, 3} {
            \node at (\x, -0.9) {\x};
        }
        \foreach \y in {0, 1, 2, 3} {
            \node at (-0.9, \y) {\y};
        }
    
        \foreach \x in {0, 1, 2, 3} {
            \foreach \y in {0, 1, 2, 3} {
                \node[unoccupied] at (\x, \y) {};
            }
        }
        
        \foreach [count=\i from 0] \x/\y in {1/3, 0/2, 1/0, 3/1, 0/0} {
            \node[occupied, label=below left:{$q_{\i}$}] at (\x, \y) {};
        }
        
        
        
        \node[below] at ($(1.5,-1.4)$) {\small (a)};
    \end{scope}

    \begin{scope}[shift={(6,0)}]
        \draw[->, thick] (-0.7, -0.7) -- (3.8, -0.7) node[right] {x};
        \draw[->, thick] (-0.7, -0.7) -- (-0.7, 3.8) node[above] {y};

        \foreach \x in {0, 1, 2, 3} {
            \node at (\x, -0.9) {\x};
        }
        \foreach \y in {0, 1, 2, 3} {
            \node at (-0.9, \y) {\y};
        }
    
        \foreach \x in {0, 1, 2, 3} {
            \foreach \y in {0, 1, 2, 3} {
                \node[unoccupied] at (\x, \y) {};
            }
        }
        
        \foreach [count=\i from 0] \x/\y in {1/3, 0/2, 1/0, 3/1, 0/0} {
            \node[occupied, label=below left:{$q_{\i}$}] at (\x, \y) {};
        }
        
        \foreach \x/\y in {0/2, 1/0} {
            \fill[aod_circle] (\x, \y) circle (0.2);
        }
        
        \draw[row_aod_line] (-0.8, 0) -- (3.4, 0);
        \draw[row_aod_line] (-0.8, 2) -- (3.4, 2);
        \draw[col_aod_line] (0, -0.8) -- (0, 3.4);
        \draw[col_aod_line] (1, -0.8) -- (1, 3.4);
        \draw[col_aod_line,->,thick,solid] (0,2.3) -- (0.8,2.3);
        \draw[row_aod_line,->,thick,solid] (2.3,0) -- (2.3,0.8);
        
        \node[below] at ($(1.5,-1.5)$) {\small (b)};
    \end{scope}

    \begin{scope}[shift={(12,0)}]
        \draw[->, thick] (-0.7, -0.7) -- (3.8, -0.7) node[right] {x};
        \draw[->, thick] (-0.7, -0.7) -- (-0.7, 3.8) node[above] {y};

        \foreach \x in {0, 1, 2, 3} {
            \node at (\x, -0.9) {\x};
        }
        \foreach \y in {0, 1, 2, 3} {
            \node at (-0.9, \y) {\y};
        }
    
        \foreach \x in {0, 1, 2, 3} {
            \foreach \y in {0, 1, 2, 3} {
                \node[unoccupied] at (\x, \y) {};
            }
        }
        
        \foreach [count=\i from 0] \x/\y in {1/3, 2/3, 3/2, 3/1, 0/0} {
            \node[occupied, label=below left:{$q_{\i}$}] at (\x, \y) {};
        }
        
        \foreach \x/\y in {2/3, 3/2} {
            \fill[aod_circle] (\x, \y) circle (0.2);
        }
        
        \draw[row_aod_line] (-0.8, 2) -- (3.4, 2);
        \draw[row_aod_line] (-0.8, 3) -- (3.4, 3);
        \draw[col_aod_line] (2, -0.8) -- (2, 3.4);
        \draw[col_aod_line] (3, -0.8) -- (3, 3.4);
        
        \node[below] at ($(1.5,-1.5)$) {\small (c)};
    \end{scope}
    \begin{scope}[shift={(4,4.5)}]
        \node[unoccupied, label=right:{Optical trap}] at (0, 0) {};
        \node[occupied, label=right:{SLM atom}] at (3, 0) {};
        \fill[aod_circle] (6, 0) circle (0.2);
        \node at (6, 0) [occupied,label=right:{AOD atom}] {};
    \end{scope}
\end{tikzpicture}


%% file: tex/qft5_embedding1.tex
\begin{tikzpicture}
    \definecolor{lightgray}{rgb}{0.8,0.8,0.8}
    \definecolor{darkgreen}{rgb}{0  ,0.5,0  }
    \tikzset{
        unuseNode/.style={
            circle,
            draw={rgb,255:red,59;green,91;blue,168},
            dashed, 
            minimum size=8pt,
            inner sep=0pt
        },
        useNode/.style={circle, draw=none, fill={rgb,255:red,59;green,91;blue,168}, minimum size=8pt, inner sep=0pt},
        fontNode1/.style={above right, font=\Large },
        fontNode2/.style={right, midway,font=\small}
    }
    
    

    
    
    \draw[lightgray, dashed, xstep =3, ystep =3] (0,0) grid (6,6);
    
    \draw[->, thick] (-0.2,-0.2) -- (6.4,-0.2) node[right] {\Large x};
    \draw[->, thick] (-0.2,-0.2) -- (-0.2,6.4) node[above] {\Large y};

    \foreach \x in {0,3} {
        \foreach \y in {0,3} {
            \draw[lightgray, dashed] (\x,\y) -- (\x+3,\y+3); 
            \draw[lightgray, dashed] (\x+3,\y) -- (\x,\y+3); 
        }
    }
    
    \foreach \x in {0,1,2} {
        \node[below] at ({3*\x}, -0.2) {\Large \x};
    }

    \foreach \y in {0,1,2} {
        \node[left] at (-0.2, {3*\y}) {\Large \y};
    }

    \node[useNode] (q0) at (0,6) {};
    \node[fontNode1] at (q0) {$q0$};
    \node[useNode] (q1) at (3,0) {};
    \node[fontNode1] at (q1) {$q1$};
    \node[useNode] (q2) at (0,0) {};
    \node[fontNode1] at (q2) {$q2$};
    \node[useNode] (q3) at (3,3) {};
    \node[fontNode1] at (q3) {$q3$};
    \node[useNode] (q4) at (0,3) {};
    \node[fontNode1] at (q4) {$q4$};
\end{tikzpicture}

%% file: tex/qft5_interaction1.tex
\tikzset{
     node_style/.style={circle, draw, minimum size=1cm, font = \huge, ultra thick},
        edge_style/.style={-, ultra thick},
        every node/.style={node_style},
}

\begin{tikzpicture}
    \node (q0) at (0,6) {$q0$};
    \node (q1) at (3,0) {$q1$};
    \node (q2) at (0,0) {$q2$};
    \node (q3) at (3,3) {$q3$};
    \node (q4) at (0,3) {$q4$};

    \foreach \i/\j in {q0/q4, q0/q3, q4/q3, q4/q2, q4/q1, q3/q2, q1/q3, q1/q2} {
        \draw[edge_style] (\i) -- (\j);
    }
\end{tikzpicture}

%% file: tex/qft5_embedding2.tex
\begin{tikzpicture}
    \definecolor{lightgray}{rgb}{0.8,0.8,0.8}
    \definecolor{darkgreen}{rgb}{0  ,0.5,0  }
    \tikzset{
        unuseNode/.style={
            circle,
            draw={rgb,255:red,59;green,91;blue,168},
            dashed, 
            minimum size=8pt,
            inner sep=0pt
        },
        useNode/.style={circle, draw=none, fill={rgb,255:red,59;green,91;blue,168}, minimum size=8pt, inner sep=0pt},
        fontNode1/.style={above right, font=\Large },
        fontNode2/.style={right, midway,font=\small}
    }
    \draw[lightgray, dashed, xstep =3, ystep =3] (0,0) grid (6,6);
    
    \draw[->, thick] (-0.2,-0.2) -- (6.4,-0.2) node[right] {\Large x};
    \draw[->, thick] (-0.2,-0.2) -- (-0.2,6.4) node[above] {\Large y};

    \foreach \x in {0,3} {
        \foreach \y in {0,3} {
            \draw[lightgray, dashed] (\x,\y) -- (\x+3,\y+3); 
            \draw[lightgray, dashed] (\x+3,\y) -- (\x,\y+3); 
        }
    }
    
    \foreach \x in {0,1,2} {
        \node[below] at ({3*\x}, -0.2) {\Large \x};
    }

    \foreach \y in {0,1,2} {
        \node[left] at (-0.2, {3*\y}) {\Large \y};
    }

    \node[useNode] (q0) at (0,3) {};
    \node[fontNode1] at (q0) {$q0$};
    \node[useNode] (q1) at (0,6) {};
    \node[fontNode1] at (q1) {$q1$};
    \node[useNode] (q2) at (0,0) {};
    \node[fontNode1] at (q2) {$q2$};
    \node[useNode] (q3) at (3,3) {};
    \node[fontNode1] at (q3) {$q3$};
    \node[useNode] (q4) at (3,0) {};
    \node[fontNode1] at (q4) {$q4$};
\end{tikzpicture}

%% file: tex/qft5_interaction2.tex
\tikzset{
     node_style/.style={circle, draw, minimum size=1cm, font = \huge, ultra thick},
        edge_style/.style={-, ultra thick},
        every node/.style={node_style},
}

\begin{tikzpicture}
    \node(q0) at (0,3) {$q0$};
    \node(q1) at (0,6) {$q1$};
    \node(q2) at (0,0) {$q2$};
    \node(q3) at (3,3) {$q3$};
    \node(q4) at (3,0) {$q4$};

    \foreach \i/\j in {q0/q1, q0/q2} {
        \draw[edge_style] (\i) -- (\j);
    }
\end{tikzpicture}

%% file: tex/qft5_move.tex
\begin{tikzpicture}
    \definecolor{lightgray}{rgb}{0.8,0.8,0.8}
    \definecolor{darkgreen}{rgb}{0  ,0.5,0  }
    \tikzset{
        unuseNode/.style={
            circle,
            draw={rgb,255:red,59;green,91;blue,168},
            dashed, 
            minimum size=8pt,
            inner sep=0pt
        },
        useNode/.style={circle, draw=none, fill={rgb,255:red,59;green,91;blue,168}, minimum size=8pt, inner sep=0pt},
        fontNode1/.style={above right, font=\Large },
        fontNode2/.style={right, midway,font=\small}
    }
    \draw[lightgray, dashed, xstep =3, ystep =3] (0,0) grid (6,6);
    
    \draw[->, thick] (-0.2,-0.2) -- (6.4,-0.2) node[right] {\Large x};
    \draw[->, thick] (-0.2,-0.2) -- (-0.2,6.4) node[above] {\Large y};

    \foreach \x in {0,3} {
        \foreach \y in {0,3} {
            \draw[lightgray, dashed] (\x,\y) -- (\x+3,\y+3); 
            \draw[lightgray, dashed] (\x+3,\y) -- (\x,\y+3); 
        }
    }
    
    \foreach \x in {0,1,2} {
        \node[below] at ({3*\x}, -0.2) {\Large \x};
    }

    \foreach \y in {0,1,2} {
        \node[left] at (-0.2, {3*\y}) {\Large \y};
    }

    \node[useNode] (q0) at (0,6) {};
    \node[fontNode1] at (q0) {$q0$};
    \node[unuseNode] (q1') at (0.5,6) {};
    \node[useNode] (q1) at (3,0) {};
    \node[fontNode1] at (q1) {$q1$};
    \node[unuseNode] (q4') at (2.5,0) {};
    \node[useNode] (q2) at (0,0) {};
    \node[fontNode1] at (q2) {$q2$};
    \node[useNode] (q3) at (3,3) {};
    \node[fontNode1] at (q3) {$q3$};
    \node[useNode] (q4) at (0,3) {};
    \node[fontNode1] at (q4) {$q4$};
    \node[unuseNode] (q0') at (0,3.5) {};

    \draw[->,line width = 0.3mm,color = orange,dashed] (q0) -- (q0') node[fontNode2] {step 2};
    \draw[->,line width = 0.3mm,color = darkgreen,dashed] (q1) -- (q1') node[fontNode2] {step 1};
    \draw[->,line width = 0.3mm,color = purple,dashed] (q4) -- (q4') node[fontNode2] {step 3};
\end{tikzpicture}

%% file: tex/mov_conflict_graph.tex
\begin{tikzpicture}

    \begin{scope}[shift={(10, 1)}]
        \node[draw, circle, minimum size=10pt, font = \Large, very thick] (M0) at (0, 0) {$m_0$};
        \node[draw, circle, minimum size=10pt, font = \Large, very thick] (M1) at (2, 0) {$m_1$};
        \node[draw, circle, minimum size=10pt, font = \Large, very thick] (M4) at (1, 1.5) {$m_4$};

        \draw[very thick,-] (M0) -- (M1);
        \draw[very thick,-] (M1) -- (M4);
        \draw[very thick,-] (M0) -- (M4);

    \end{scope}
\end{tikzpicture}

%% file: Figures/IG_qft_5.tex

\begin{tikzpicture}
    \node (q0) at (180:3cm) {q0};
    \node (q1) at (240:3cm) {q1};
    \node (q2) at (300:3cm) {q2};
    \node (q3) at (0:3cm) {q3};
    \node (q4) at (90:2cm) {q4};

    \foreach \i in {0, 1, 2, 3, 4} {
        \foreach \j in {0, 1, 2, 3, 4} {
            \ifnum\i<\j
                \draw[edge_style] (q\i) -- (q\j);
            \fi
        }
    }

\end{tikzpicture}


%% file: Figures/IG_qv6.tex

\begin{tikzpicture}

    \node (q0) at (300:3cm) {q0};
    \node (q1) at (0:3cm) {q1};
    \node (q2) at (61:2.4cm) {q2};
    \node (q4) at (180:3cm) {q4};
    \node (q5) at (240:3cm) {q5};
    \node (q3) at (-2, 2) {q3};

    \foreach \i/\j in {q0/q1, q0/q2, q0/q4, q0/q5, q1/q2, q1/q3, q1/q4, q1/q5, q2/q3, q2/q4, q2/q5, q3/q4,  q4/q5} {
        \draw[edge_style] (\i) -- (\j);
    }

\end{tikzpicture}


%% file: Figures/IG_3_reg_10.tex

\begin{tikzpicture}

    \node (q0) at (0, 2) {q0};
    \node (q1) at (-1, 0) {q1};
    \node (q2) at (0, -3) {q2};
    \node (q3) at (-2, 3) {q3};
    \node (q4) at (1, -1) {q4};
    \node (q5) at (-3, 1) {q5};
    \node (q6) at (3, -2) {q6};
    \node (q7) at (-2, -2) {q7};
    \node (q8) at (2, 3) {q8};
    \node (q9) at (3, 1) {q9};

    \foreach \i/\j in {q0/q3, q0/q8, q0/q1, q1/q5, q1/q2, q2/q7, q2/q6, q3/q5, q4/q9, q4/q6, q4/q7, q5/q7, q8/q3, q8/q9, q9/q6} {
        \draw[edge_style] (\i) -- (\j);
    }

\end{tikzpicture}


%% file: Figures/IG_dj7.tex

\begin{tikzpicture}

    \node (q6) at (0, 0) {q6};

    \foreach \i in {0,..., 5} {
        \node (q\i) at ({60 * \i + 180}:3cm) {q\i};
    }

    \foreach \i in {0, ..., 5} {
        \draw[edge_style] (q6) -- (q\i);
    }

\end{tikzpicture}


%% file: Figures/IG_ghz5.tex

\begin{tikzpicture}

    \node (q0) at (0, 0) {q0};
    \foreach \i in {1, 2, 3, 4} {
        \node (q\i) [below=0.3cm of q\the\numexpr\i-1\relax] {q\i};
        \draw[edge_style] (q\the\numexpr\i-1\relax) -- (q\i);
    }

\end{tikzpicture}


%% file: tex/tex_methods_comp/tikz_QFT_F_comp.tex
\begin{tikzpicture}

\begin{axis}[Qubits_axis,ymode=log,
legend entries={DasAtom, Tetris, Enola, Atomique},
legend pos=north west,
ytick = {1,1000,10000000,100000000000},
]

\addplot [DasAtomStyle]
table{
5	1
6	1
7	1
8	1
9	1
10	1
11	1
12	1
13	1
14	1
15	1
16	1
17	1
18	1
19	1
20	1
21	1
22	1
23	1
24	1
25	1
26	1
27	1
28	1
29	1
30	1
31	1
32	1
33	1
34	1
35	1
36	1
37	1
38	1
39	1
40	1
41	1
42	1
43	1
44	1
45	1
46	1
47	1
48	1
49	1
50	1
};

\addplot [TetrisStyle]
table {%
5	1.015168585
6	1.060142397
7	1.092173276
8	1.159523236
9	1.171734706
10	1.399654668
11	1.65091988
12	1.743721324
13	2.077567173
14	1.855794862
15	2.277509862
16	2.283456522
17	3.370774466
18	3.703762127
19	4.613826456
20	4.959166851
21	4.972015096
22	7.127651927
23	10.1117646
24	9.193106243
25	13.62812752
26	10.90946012
27	13.20894824
28	11.40555955
29	12.58802397
30	10.64204164
31	12.52170821
32	11.48419328
33	25.61534862
34	19.90240538
35	29.34321135
36	23.89878549
37	91.71105556
38	166.0730276
39	87.93389064
40	90.93052378
41	257.2820459
42	166.7172887
43	383.7996871
44	280.0514354
45	601.0068441
46	392.9308282
47	955.1020583
48	962.0707773
49	394.5774031
50	8108.52472
};
    
\addplot [EnolaStyle]
table {%
5	1.034111141
6	1.058241037
7	1.087737505
8	1.151445598
9	1.209120593
10	1.29712467
11	1.381335133
12	1.520725406
13	1.71358355
14	2.00296324
15	2.252773016
16	3.985499469
17	3.38870618
18	4.048609604
19	4.977555473
20	6.521403735
21	8.930910128
22	11.4109372
23	16.45147335
24	24.16078352
25	35.63961296
26	57.17200994
27	91.9598126
28	144.0084363
29	243.7222275
30	415.8208538
31	790.6739614
32	1413.647197
33	2599.323834
34	6179.007349
35	16567.90259
36	37657.88221
37	101204.8882
38	194418.6653
39	509398.9308
40	1505342.775
41	4074095.027
42	10749459.35
43	32076120.49
44	170163866.7
45	375677939.1
46	1490592652
47	7523049748
48	18389292384
49	1.85365e+11
50	6.19371e+11
};

\addplot [AtomiqueStyle]
table {%
5	1.055701387
6	1.085154798
7	1.120281238
8	1.16393692
9	1.186800571
10	1.253144121
11	1.274344564
12	1.249728773
13	1.426993383
14	1.332542399
15	1.460058831
16	1.770952018
17	1.632414952
18	1.649863336
19	1.77487859
20	1.942112882
21	2.035021385
22	2.258523885
23	2.719094345
24	3.976337328
25	2.971240024
26	2.116097645
27	3.110960741
28	2.915828429
29	3.022966598
30	16.1435914
31	4.533594129
32	7.532105692
33	31.47773694
34	10.07447412
35	19.63023319
36	20.87583936
37	6.863816475
38	13.37367376
39	22.19036257
40	32.15456883
41	12.74521903
42	28.37554911
43	319.5998827
44	37.50232387
45	227.9442282
46	65.70805348
47	81.66407138
48	45.42915384
49	2060.318122
50	80591.99434
};

\end{axis}
\end{tikzpicture}

%% file: tex/tex_data/legand.tex
\begin{tikzpicture}
\draw[gray!40,fill=gray!10] (-1.4em, 0.2em) rectangle (7em, 5.2em);

\node at (0, 4em) {
    \tikz \draw[DasAtomStyle] (0, 0.5em) -- (2em, 0.5em);
};
\node[anchor=west] at (2.5em, 4em) {DasAtom};

\node at (0, 3em) {
    \tikz \draw[TetrisStyle] (0, 0.5em) -- (2em, 0.5em);
};
\node[anchor=west] at (2.5em, 3em) {Tetris};

\node at (0, 2em) {
    \tikz \draw[EnolaStyle] (0, 0.5em) -- (2em, 0.5em);
};
\node[anchor=west] at (2.5em, 2em) {Enola};

\node at (0, 1em) {
    \tikz \draw[AtomiqueStyle] (0, 0.5em) -- (2em, 0.5em);
};
\node[anchor=west] at (2.5em, 1em) {Atomique};

\end{tikzpicture}

%% file: tex/tex_methods_comp/tikz_QFT_rt_comp.tex
\begin{tikzpicture}
\begin{axis}[rt_axis, ymode=log,ytick={1e-2,1e0,1e2,1e4,1e5}]
\addplot [DasAtomStyle]
table {%
5	0.081439495
6	0.033262491
7	0.033680439
8	0.041443825
9	0.05156827
10	0.068287611
11	0.109350681
12	0.115610123
13	0.179630041
14	0.199267864
15	0.19664526
16	0.21411705
17	0.544114113
18	0.588293076
19	0.526881933
20	0.652358294
21	0.603479385
22	0.651375771
23	0.687244892
24	0.764049768
25	1.110395908
26	1.589700699
27	1.673766136
28	1.757907629
29	2.301149368
30	1.895470142
31	2.406918764
32	2.104747534
33	2.598798752
34	2.248964787
35	3.019013882
36	2.691742659
37	6.31317687
38	6.231184959
39	6.619459867
40	6.470872402
41	6.656517506
42	6.725443363
43	6.952340126
44	6.978361845
45	7.274305344
46	7.220304251
47	7.551631212
48	7.525385857
49	8.126078129
50	11.12960362
};

\addplot [TetrisStyle]
table {%
5	4.55142
6	8.52185
7	12.3402
8	16.9503
9	17.47
10	33.8433
11	47.8553
12	59.7952
13	80.1599
14	92.7757
15	99.6806
16	131.237
17	171.461
18	188.34
19	212.127
20	217.218
21	211.061
22	242.742
23	274.549
24	315.331
25	345.33
26	527.354
27	585.641
28	623.912
29	659.546
30	728.298
31	739.384
32	714.312
33	941.568
34	903.731
35	1019.77
36	1253.23
37	1364.31
38	1260.58
39	807.752
40	948.527
41	1049.49
42	1006.59
43	1149.21
44	1004.68
45	1237.19
46	1298.16
47	1451.42
48	1247.67
49	1356.03
50	1538.26
};

\addplot [EnolaStyle]
table {%
5	562.6864305
6	809.8276086
7	1039.094843
8	1378.740061
9	1662.467413
10	2127.537686
11	2483.864078
12	3023.531153
13	3475.338817
14	4044.159832
15	4586.664693
16	5114.395787
17	5804.250089
18	6701.82888
19	7315.394698
20	8267.891322
21	9146.799642
22	10040.49667
23	10868.07459
24	12120.7502
25	12693.53947
26	13581.41137
27	14863.87197
28	16388.65771
29	17612.11246
30	18671.24272
31	19933.60586
32	21241.1573
33	22481.4379
34	24123.75358
35	25918.28625
36	32029.23848
37	33627.51432
38	36643.79622
39	30988.60408
40	32210.43685
41	34301.98096
42	35270.0015
43	37246.56132
44	38997.62587
45	42063.84257
46	42964.09508
47	165554.8258
48	52882.2927
49	45822.39639
50	47235.86046
};

\addplot [AtomiqueStyle]
table {%
5	0.856870413
6	0.623314619
7	1.052027464
8	0.712203264
9	1.086867332
10	0.867548943
11	1.336059809
12	1.117481232
13	1.45745945
14	1.245637655
15	1.604957819
16	1.478187561
17	1.97254777
18	1.758034945
19	2.329418182
20	2.238646507
21	2.768344164
22	2.732204437
23	2.854877234
24	3.057801008
25	3.272812128
26	3.325596094
27	3.598107815
28	3.517098904
29	4.08165741
30	4.913783073
31	5.087067127
32	5.647842884
33	5.699200392
34	5.385308981
35	5.612904072
36	7.201058388
37	6.780175209
38	6.933322191
39	7.32270813
40	7.806434393
41	7.37445116
42	8.047029495
43	8.207630873
44	9.543311596
45	9.767385721
46	8.902109861
47	9.71711278
48	10.61761236
49	11.76319504
50	11.74981165
};

\end{axis}
\end{tikzpicture}

%% file: tex/tex_diff_variables/tikz_SWAP.tex
\begin{tikzpicture}
\begin{axis}[swap_axis, legend entries={Tetris, Atomique},legend pos=north west]

\addplot [TetrisStyle]
table {%
5	1
6	4
7	6
8	10
9	11
10	23
11	34
12	38
13	50
14	43
15	57
16	58
17	84
18	91
19	106
20	112
21	113
22	139
23	162
24	158
25	185
26	170
27	183
28	176
29	184
30	174
31	187
32	183
33	238
34	223
35	252
36	240
37	330
38	373
39	333
40	338
41	411
42	383
43	441
44	423
45	477
46	453
47	514
48	519
49	464
50	666
};

\addplot [AtomiqueStyle]
table {%
5	3
6	5
7	5
8	7
9	8
10	8
11	10
12	10
13	12
14	21
15	17
16	17
17	20
18	21
19	25
20	20
21	23
22	22
23	24
24	24
25	28
26	30
27	26
28	27
29	38
30	30
31	30
32	31
33	75
34	48
35	37
36	37
37	45
38	40
39	38
40	50
41	46
42	74
43	58
44	49
45	70
46	76
47	80
48	104
49	72
50	100
};

\end{axis}
\end{tikzpicture}

%% file: tex/tex_diff_variables/tikz_move_stage.tex
\begin{tikzpicture}
\begin{axis}[move_stage_axis, legend entries={DasAtom, Enola, Atomique}, legend pos=north west,ymode=log]
\addplot [DasAtomStyle]
table {%
5	0
6	5
7	5
8	5
9	13
10	16
11	15
12	21
13	26
14	33
15	37
16	47
17	46
18	53
19	55
20	67
21	73
22	92
23	85
24	104
25	107
26	103
27	101
28	121
29	126
30	133
31	144
32	154
33	159
34	168
35	183
36	189
37	186
38	204
39	210
40	223
41	237
42	239
43	245
44	257
45	266
46	284
47	286
48	302
49	313
50	310
};

\addplot [EnolaStyle]
table {%
5	42
6	57
7	71
8	106
9	117
10	144
11	164
12	192
13	224
14	236
15	273
16	311
17	346
18	368
19	399
20	465
21	466
22	512
23	532
24	575
25	623
26	636
27	712
28	743
29	800
30	837
31	887
32	932
33	959
34	1013
35	1052
36	1155
37	1174
38	1212
39	1262
40	1314
41	1409
42	1453
43	1500
44	1566
45	1587
46	1670
47	1716
48	1761
49	1892
50	1931
};

\addplot [AtomiqueStyle]
table {%
5	14
6	24
7	28
8	39
9	52
10	56
11	73
12	74
13	107
14	134
15	123
16	137
17	152
18	204
19	199
20	193
21	267
22	293
23	312
24	280
25	314
26	311
27	331
28	442
29	400
30	519
31	557
32	508
33	648
34	589
35	568
36	589
37	618
38	699
39	783
40	784
41	793
42	929
43	901
44	1017
45	1068
46	925
47	1046
48	1165
49	1111
50	1083
};

\end{axis}
\end{tikzpicture}

%% file: tex/tex_diff_variables/tikz_move_dis.tex
\begin{tikzpicture}
\begin{axis}[move_dis_axis,  legend entries={DasAtom, Enola, Atomique}, legend pos=north west, ymode=log,
ytick={1e1,1e4,1e7,1e9},ymax=1e10]
\addplot [DasAtomStyle]
table {%
5	0
6	23.65904855
7	23.65904855
8	24.90168924
9	63.04601917
10	96.7542231
11	91.0672889
12	139.1852848
13	170.9543348
14	216.000354
15	243.7364436
16	324.5256318
17	311.1969425
18	380.0388586
19	399.0606081
20	473.6943614
21	531.3341119
22	678.6531212
23	639.651264
24	821.2053188
25	875.2982267
26	757.0670635
27	774.2339073
28	926.3286629
29	1103.455604
30	1054.220271
31	1217.995438
32	1210.663477
33	1368.218372
34	1369.11537
35	1587.321977
36	1617.899971
37	1704.004152
38	1794.977479
39	1980.862913
40	1976.129426
41	2242.048114
42	2089.219384
43	2265.126348
44	2275.380434
45	2461.703941
46	2501.937173
47	2595.346257
48	2747.261428
49	3002.13726
50	3086.993937
};

\addplot [EnolaStyle]
table {%
5	172525.0046
6	316114.5282
7	468791.2741
8	1070644.128
9	1671540.562
10	2494749.969
11	3177845.179
12	4643145.063
13	6471761.32
14	7720173.562
15	11068728.3
16	14185099.82
17	17450632
18	20670483.99
19	26002267.55
20	35043360.09
21	38911827.24
22	46199037.36
23	51424671.14
24	63212159.13
25	75891596.19
26	83693399.87
27	100336538.1
28	114478853.1
29	132894264
30	147589285.6
31	172175101.7
32	195434727.9
33	209292920.5
34	244660276.5
35	258252163.9
36	310239470.2
37	338002922.4
38	369000993.6
39	410283272.2
40	448716248.7
41	508309768.3
42	553780747.6
43	607364738.3
44	652209785.5
45	692736845.6
46	778000895.2
47	821433154.5
48	875249959.8
49	984874267.9
50	1048053383
};

\addplot [AtomiqueStyle]
table {%
5	89.09545
6	114.5513
7	199.4041
8	296.9848
9	500.6316
10	487.9037
11	725.4916
12	789.1312
13	1383.101
14	1663.115
15	1480.682
16	1748.816
17	2271.307
18	2774.687
19	1643.015
20	3127.962
21	3747.639
22	3663.563
23	4324.121
24	3977.427
25	4364.129
26	3593.661
27	4478.517
28	6466.729
29	4855.844
30	5507.524
31	6274.03
32	5363.011
33	9183.098
34	6580.787
35	7019.629
36	7463.748
37	5620.42
38	8400.1
39	7763.697
40	8559.865
41	10864.37
42	12035.31
43	12491.72
44	11922.21
45	13621.09
46	10493.35
47	12955.84
48	14420.94
49	11478.07
50	12938.23
};

\end{axis}
\end{tikzpicture}

%% file: tex/tex_diff_variables/tikz_atom_trans.tex
\begin{tikzpicture}
\begin{axis}[atom_trans_axis,legend entries={DasAtom, Enola}, legend pos=north west,scaled y ticks=false, 
ytick = {0,10,100,1000,10000},ymode=log,grid=major
]
\addplot [DasAtomStyle]
table {%
5	0
6	24
7	24
8	32
9	60
10	76
11	80
12	120
13	132
14	172
15	204
16	260
17	276
18	344
19	340
20	472
21	496
22	612
23	580
24	748
25	788
26	856
27	928
28	1032
29	1100
30	1212
31	1292
32	1400
33	1472
34	1564
35	1644
36	1744
37	1612
38	1828
39	1916
40	2092
41	2120
42	2348
43	2372
44	2640
45	2692
46	2872
47	2916
48	3204
49	3300
50	3160
};

\addplot [EnolaStyle]
table {%
5	100
6	152
7	200
8	268
9	332
10	416
11	496
12	596
13	692
14	808
15	920
16	1052
17	1180
18	1328
19	1472
20	1636
21	1796
22	1976
23	2152
24	2348
25	2540
26	2752
27	2960
28	3188
29	3412
30	3656
31	3896
32	4156
33	4412
34	4688
35	4960
36	5252
37	5540
38	5848
39	6152
40	6476
41	6796
42	7128
43	7448
44	7780
45	8100
46	8432
47	8752
48	9084
49	9404
50	9736
};

\end{axis}
\end{tikzpicture}

%% file: tex/tex_diff_variables/tikz_t_idle.tex
\begin{tikzpicture}
\begin{axis}[t_axis
,legend entries={DasAtom, Tetris, Enola, Atomique}, legend pos=south east,
legend style={
        font=\small, 
        nodes={scale=0.8, transform shape} 
    }
,ymode=log
]
\addplot [DasAtomStyle]
table {%
5	20.8
6	2697.098711
7	3162.315163
8	3657.406389
9	10526.06213
10	14748.16769
11	15271.34578
12	23526.76985
13	31498.33882
14	42985.99083
15	51694.15755
16	70392.74565
17	73112.41459
18	89861.03537
19	98683.7301
20	125935.2495
21	144676.9025
22	191074.7248
23	185428.6529
24	238104.0139
25	256689.083
26	253250.0248
27	259713.2464
28	322188.5501
29	354843.8046
30	381511.9239
31	431000.6519
32	470416.2023
33	508047.5023
34	548345.2229
35	620667.1985
36	658048.9072
37	672486.0066
38	752076.8258
39	804081.5884
40	866402.5037
41	954150.2958
42	972958.9893
43	1030871.096
44	1098442.835
45	1171413.941
46	1267594.727
47	1311041.335
48	1414202.597
49	1509916.938
50	1535708.812
};

\addplot [TetrisStyle]
table {%
5	23.2
6	47.4
7	57.6
8	92.4
9	133.2
10	167.2
11	227.4
12	297.6
13	372
14	454.4
15	558.6
16	654.8
17	703
18	865.2
19	945.6
20	1146.8
21	1266
22	1476.4
23	1671.8
24	1784.4
25	1951.8
26	2173
27	2355.6
28	2590.8
29	2827.6
30	3210.6
31	3524.4
32	3560.6
33	4005.6
34	4248.8
35	4885.6
36	5245.2
37	5037.8
38	5442
39	5997.6
40	6001.2
41	7195.2
42	7024
43	7524.2
44	8491.4
45	9037.6
46	9325.2
47	9330.6
48	10802.2
49	10452.6
50	10843.4
};

\addplot [EnolaStyle]
table {%
5	53935.94549
6	87651.64904
7	123648.9928
8	213436.9901
9	301528.8122
10	409089.5894
11	506703.4632
12	670572.5304
13	857256.9841
14	1006546.773
15	1293230.183
16	1558301.376
17	1831469.867
18	2119181.029
19	2501419.212
20	3053211.489
21	3386646.029
22	3867823.6
23	4259250.84
24	4934160.856
25	5642544.793
26	6173763.123
27	6987946.571
28	7749598.108
29	8639413.631
30	9407005.288
31	10481728.38
32	11564383.09
33	12339934.35
34	13755648.27
35	14531009.31
36	16349168.85
37	17534712.63
38	18850856.01
39	20376064.49
40	21890800.08
41	23837327.69
42	25467824.13
43	27327647.18
44	28918916.45
45	30580708.07
46	32975722.3
47	34713689.33
48	36537080.24
49	39465478.67
50	41574759.64
};

\addplot [AtomiqueStyle]
table {%
5	29801
6	46298
7	62972
8	100428
9	176736
10	165059
11	192491
12	244310
13	231629
14	509864
15	496458
16	647272
17	985960
18	641995
19	1692417
20	1015388
21	1115070
22	1300209
23	1465791
24	1680141
25	1593901
26	1853940
27	2562251
28	2246905
29	3201040
30	2745636
31	3375634
32	4344014
33	5916853
34	6592345
35	6176500
36	6910880
37	7043993
38	6936009
39	6677012
40	8635886
41	8528257
42	8890964
43	10025480
44	9342913
45	10308330
46	16468170
47	14826360
48	13724100
49	14159610
50	14665340
};

\end{axis}
\end{tikzpicture}

%% file: tex/tex_diff_variables/tikz_t_total.tex
\begin{tikzpicture}
\begin{axis}[t_axis,  
,legend entries={DasAtom, Tetris, Enola, Atomique}, legend pos=south east,
legend style={
        font=\tiny, 
        nodes={scale=0.8, transform shape} 
    },
ytick={1e1,1e2,1e4,1e6},
ymode=log,
]
\addplot [DasAtomStyle]
table {%
5	5.2
6	450.8164519
7	453.2164519
8	458.8757986
9	1171.429126
10	1476.916769
11	1390.576889
12	1963.064154
13	2425.626063
14	3073.327916
15	3449.35717
16	4402.846603
17	4304.212623
18	4995.979743
19	5197.764742
20	6300.862475
21	6893.662022
22	8689.714766
23	8066.802298
24	9925.90058
25	10272.65132
26	9745.68557
27	9624.498013
28	11512.43393
29	12241.88292
30	12723.16413
31	13909.53716
32	14707.00632
33	15402.06977
34	16134.70067
35	17740.43996
36	18286.43631
37	18182.78937
38	19799.19542
39	20625.36893
40	21668.16259
41	23280.25112
42	23174.1807
43	23982.41154
44	24973.45533
45	26040.42535
46	27565.57668
47	27903.81138
48	29472.02078
49	30824.23138
50	30723.91625
};

\addplot [TetrisStyle]
table {%
5	5.8
6	9.6
7	10.2
8	14
9	17.4
10	20.2
11	24.8
12	29.2
13	33.6
14	37.2
15	42.6
16	46.4
17	47.8
18	54.8
19	57
20	64.8
21	67.8
22	75.4
23	81.6
24	83.2
25	87.6
26	92.8
27	96.8
28	102
29	107.2
30	116.6
31	123.6
32	121.2
33	132.4
34	135.8
35	151
36	157
37	149
38	156.8
39	166.8
40	163.2
41	189.8
42	181.2
43	189.8
44	207.6
45	216.2
46	217.8
47	214.4
48	241
49	228.6
50	234.6
};

\addplot [EnolaStyle]
table {%
5	10788.2291
6	14609.90817
7	17665.59898
8	26681.32376
9	33505.06802
10	40911.05894
11	46066.22393
12	55883.5442
13	65945.52185
14	71899.09806
15	86218.42553
16	97397.13599
17	107737.004
18	117735.9794
19	131657.527
20	152664.6744
21	161273.1442
22	175814.6636
23	185189.5061
24	205594.9357
25	225706.8797
26	237457.7278
27	258818.3248
28	276777.061
29	297916.7045
30	313572.9429
31	338126.5606
32	361393.4715
33	373944.0954
34	404584.7903
35	415178.7859
36	454150.8792
37	473918.6441
38	496082.8581
39	522471.0843
40	547278.1021
41	581406.5289
42	606385.2555
43	635535.3436
44	657256.9465
45	679580.2948
46	716872.6978
47	738598.4497
48	761198.6384
49	805427.5321
50	831504.9327
};

\addplot [AtomiqueStyle]
table {%
5	4967.991735
6	7718.075088
7	7873.152931
8	12555.77245
9	14729.83927
10	15007.6976
11	16043.47556
12	20362.18394
13	15444.72884
14	31869.84573
15	26132.34836
16	30825.46465
17	41084.6485
18	32103.68547
19	48357.30035
20	44151.40396
21	42891.28827
22	43344.02376
23	52354.03906
24	64625.88631
25	56930.17945
26	63934.32998
27	82658.55786
28	72486.48832
29	91463.80677
30	72258.87974
31	93773.52668
32	114322.129
33	113790.7413
34	140268.2676
35	147065.9609
36	153581.4508
37	156539.9459
38	161309.9084
39	145159.6119
40	172724.79
41	189524.9988
42	161661.1775
43	192805.6221
44	169878.7474
45	184085.2268
46	257322.4229
47	247114.0731
48	196065.8986
49	244140.0244
50	257296.2584
};

\end{axis}
\end{tikzpicture}

%% file: tex/tex_methods_comp/tikz_two_local_F_comp.tex
\begin{tikzpicture}
\begin{axis}[two_local_axis, ymode=log,
            legend entries={DasAtom, Tetris, Enola, Atomique},
            legend pos=north west,
]

\addplot [DasAtomStyle]
table{
5	1
6	1
7	1
8	1
9	1
10	1
11	1
12	1
13	1
14	1
15	1
16	1
17	1
18	1
19	1
20	1
21	1
22	1
23	1
24	1
25	1
26	1
27	1
28	1
29	1
30  1
};

\addplot [TetrisStyle]
table {%
5	1.062026953
6	1.138969623
7	1.224278887
8	1.329878109
9	1.448595457
10	1.743227753
11	1.949760401
12	2.638164954
13	3.17394329
14	4.684328142
15	5.159859367
16	6.102898543
17	12.14004195
18	17.7995482
19	23.16156828
20	34.94685708
21	49.53343818
22	103.4190684
23	90.21736168
24	178.9833596
25	356.2835078
26	47.3143599
27	45.58914675
28	108.926747
29	161.521212
30	184.6967296
};

\addplot [EnolaStyle]
table {%
5	1.039501872
6	1.066498621
7	1.114509329
8	1.174253606
9	1.262412209
10	1.379963262
11	1.546577005
12	1.723473771
13	2.078455569
14	2.420527137
15	2.936840978
16	3.631588562
17	4.888374911
18	6.283346169
19	9.147446315
20	12.62585406
21	18.59659304
22	27.61227023
23	46.03863006
24	78.94205799
25	131.4022219
26	311.2811315
27	510.0403188
28	1026.213457
29	1816.260756
30	4514.106911
};

\addplot [AtomiqueStyle]
table {
5	1.121929048
6	1.196044245
7	1.26687195
8	1.386532353
9	1.36110899
10	1.435045565
11	1.501798069
12	1.584374296
13	1.729391207
14	1.737507431
15	1.993256256
16	1.874256811
17	2.437516974
18	2.497198333
19	2.636138589
20	2.257158797
21	2.630063779
22	3.038027471
23	3.174560367
24	2.921482715
25	3.15864786
26	4.914515599
27	3.821501367
28	3.315751593
29	5.044721525
30	9.302740295
};

\end{axis}
\end{tikzpicture}

%% file: tex/tex_methods_comp/tikz_qv_F_comp.tex
\begin{tikzpicture}
\begin{axis}[qv_axis]

\addplot [DasAtomStyle]
table{
5	1
6	1
7	1
8	1
9	1
10	1
11	1
12	1
13	1
14	1
15	1
16	1
17	1
18	1
19	1
20	1
21	1
22	1
23	1
24	1
25	1
};

\addplot [TetrisStyle]
table {%
5	1.015170751
6	1.04616728
7	1.030539912
8	1.142199997
9	1.107373917
10	1.517158633
11	1.68235158
12	1.776513032
13	1.592392538
14	2.104883577
15	2.056324274
16	1.959903811
17	4.116046106
18	5.160105808
19	6.674250907
20	8.304711278
21	9.167549113
22	14.29236431
23	15.39231399
24	22.28344782
25	99.78124858
};

\addplot [EnolaStyle]
table {%
5	1.036750228
6	1.07615759
7	1.103027812
8	1.188555749
9	1.239093852
10	1.395336746
11	1.470363795
12	1.714066178
13	1.876429584
14	2.352769677
15	2.618866969
16	3.515628551
17	4.212041353
18	6.063120194
19	7.869917731
20	11.8400372
21	16.28352027
22	27.35792297
23	39.47176455
24	71.90629012
25	112.3761314
};

\addplot [AtomiqueStyle]
table {
5	1.009266084
6	1.064544743
7	1.065527397
8	1.081376491
9	1.079244242
10	1.122711234
11	1.121559577
12	1.268417833
13	1.21956407
14	1.291405157
15	1.270301603
16	1.488138729
17	1.473209227
18	1.721693294
19	1.74626874
20	1.691107651
21	1.739166578
22	2.105130575
23	1.578366475
24	2.603129441
25	1.847703172
};

\end{axis}
\end{tikzpicture}

%% file: tex/tex_methods_comp/tikz_3_regular_F_comp.tex
\begin{tikzpicture}
\begin{axis}[3_regular_axis]
\addplot [DasAtomStyle]
table {%
10	0.927552273
12	0.913700764
14	0.900054665
16	0.886610952
18	0.873366644
20	0.860318805
30	0.797946
40	0.7028174
50	0.6363548
60	0.5735872
};

\addplot [TetrisStyle]
table {%
10	0.860366984
12	0.834870459
14	0.810127133
16	0.786119229
18	0.697006452
20	0.599681659
30	0.500619788
40	0.328531457
50	0.177320888
60	0.121729298
};

\addplot [EnolaStyle]
table {%
10	0.891126583
12	0.870820206
14	0.839560314
16	0.805943573
18	0.779989638
20	0.752170226
30	0.597580875
40	0.425384854
50	0.268869999
60	0.18906919
};

\addplot [AtomiqueStyle]
table {
10	0.894543074
12	0.910570783
14	0.874506109
16	0.861591485
18	0.851098797
20	0.850177643
30	0.791906243
40	0.709652654
50	0.642153495
60	0.583356059
};

\end{axis}
\end{tikzpicture}

%% file: tex/tex_methods_comp/tikz_DJ_F_comp.tex
\begin{tikzpicture}
\begin{axis}[Qubits_axis]
\addplot [DasAtomStyle]
table {%
5	0.98014741
6	0.975245502
7	0.970367852
8	0.965514339
9	0.960684846
10	0.955879255
11	0.951097449
12	0.943614445
13	0.937223808
14	0.929850306
15	0.925489706
16	0.919478505
17	0.920986407
18	0.91425893
19	0.907406187
20	0.902471426
21	0.897741137
22	0.890694303
23	0.883306923
24	0.880778544
25	0.877372087
26	0.861759849
27	0.857962077
28	0.848543508
29	0.840520791
30	0.838288731
31	0.831138037
32	0.822954261
33	0.815851063
34	0.808690631
35	0.803520973
36	0.80260605
37	0.791706252
38	0.771926724
39	0.764490746
40	0.756786526
41	0.753269853
42	0.743520448
43	0.735847505
44	0.728515974
45	0.720784154
46	0.717815836
47	0.710494186
48	0.702098294
49	0.703723916
50	0.688366493
};

\addplot [TetrisStyle]
table {%
5	0.980145319
6	0.946346883
7	0.955882696
8	0.936905074
9	0.946343476
10	0.900064986
11	0.882189307
12	0.864679714
13	0.873385975
14	0.843276577
15	0.864668647
16	0.822394451
17	0.806038666
18	0.766635118
19	0.707545845
20	0.725497628
21	0.711096517
22	0.751399896
23	0.693477961
24	0.669572378
25	0.646469319
26	0.72546116
27	0.721844345
28	0.751374349
29	0.72546116
30	0.679680086
31	0.636791136
32	0.652961492
33	0.649674942
34	0.666169266
35	0.643210043
36	0.679676823
37	0.656182247
38	0.56172629
39	0.550579633
40	0.531589554
41	0.545062819
42	0.518422234
43	0.515823451
44	0.536927903
45	0.473684054
46	0.500544567
47	0.468962245
48	0.503036794
49	0.478422148
50	0.374185432
};

\addplot [EnolaStyle]
table {%
5	0.974874739
6	0.967151084
7	0.958893115
8	0.950080921
9	0.940756878
10	0.930777488
11	0.920590813
12	0.909410208
13	0.897443009
14	0.885662078
15	0.872950116
16	0.860255759
17	0.84807589
18	0.833877799
19	0.819204605
20	0.804404709
21	0.78937492
22	0.774533335
23	0.759164697
24	0.743460088
25	0.727476026
26	0.711878498
27	0.694469044
28	0.680721964
29	0.662703496
30	0.647523726
31	0.627858709
32	0.612131469
33	0.59457677
34	0.576689566
35	0.560316562
36	0.544590007
37	0.52799871
38	0.510116674
39	0.496158644
40	0.478487202
41	0.460755619
42	0.446576664
43	0.429333037
44	0.413902153
45	0.399603679
46	0.385339923
47	0.368555177
48	0.355093048
49	0.340685406
50	0.326166027
};

\addplot [AtomiqueStyle]
table {
5	0.979971008
6	0.974944631
7	0.969844021
8	0.964759916
9	0.959662724
10	0.954160307
11	0.948894115
12	0.943013342
13	0.938130344
14	0.934484423
15	0.929259213
16	0.924361723
17	0.919425381
18	0.913783413
19	0.909570336
20	0.902352714
21	0.896679605
22	0.891893582
23	0.885740722
24	0.881068045
25	0.877181397
26	0.871263305
27	0.866316354
28	0.861366399
29	0.856518249
30	0.850693114
31	0.845031312
32	0.841067814
33	0.834883759
34	0.829281041
35	0.826750005
36	0.821650459
37	0.814239422
38	0.810466485
39	0.807774757
40	0.797855059
41	0.794170568
42	0.792063972
43	0.784265374
44	0.782159159
45	0.776156567
46	0.769588735
47	0.764208074
48	0.758963582
49	0.751109216
50	0.745908386
};

\end{axis}
\end{tikzpicture}

%% file: tex/tex_methods_comp/tikz_GHZ_F_comp.tex
\begin{tikzpicture}
\begin{axis}[Qubits_axis]
\addplot [DasAtomStyle]
table {%
5	0.98014741
6	0.975245502
7	0.970367852
8	0.965514339
9	0.960684846
10	0.955879255
11	0.951097449
12	0.946339312
13	0.941604728
14	0.936893581
15	0.932205757
16	0.927541142
17	0.922899622
18	0.918281083
19	0.913685414
20	0.909112502
21	0.904562236
22	0.900034504
23	0.895529197
24	0.891046205
25	0.886585418
26	0.882146728
27	0.877730025
28	0.873335204
29	0.868962155
30	0.864610773
31	0.860280952
32	0.855972585
33	0.851685568
34	0.847419796
35	0.843175165
36	0.83895157
37	0.83474891
38	0.830567081
39	0.826405982
40	0.82226551
41	0.818145565
42	0.814046045
43	0.809966851
44	0.805907883
45	0.801869042
46	0.797850229
47	0.793851346
48	0.789872295
49	0.785912978
50	0.7819733
};

\addplot [TetrisStyle]
table {%
5	0.980145319
6	0.975243552
7	0.955880402
8	0.951099985
9	0.946343476
10	0.886627975
11	0.895559049
12	0.839062096
13	0.873393428
14	0.869022869
15	0.851765517
16	0.847503186
17	0.758985605
18	0.755180709
19	0.718256192
20	0.736496901
21	0.700471918
22	0.729143247
23	0.714665006
24	0.679710174
25	0.599652076
26	0.693438295
27	0.689997691
28	0.573161047
29	0.693454937
30	0.630436286
31	0.608705663
32	0.605653494
33	0.602605059
34	0.636788079
35	0.578935779
36	0.62101229
37	0.508138204
38	0.513247756
39	0.48815898
40	0.471324935
41	0.505574458
42	0.419994974
43	0.381817081
44	0.441580482
45	0.41371585
46	0.417884134
47	0.391504558
48	0.44600911
49	0.393466724
50	0.341909197
};

\addplot [EnolaStyle]
table {%
5	0.974896979
6	0.967202018
7	0.958923741
8	0.950280398
9	0.94097909
10	0.931224511
11	0.921306777
12	0.910602514
13	0.899610759
14	0.888107489
15	0.876432151
16	0.863929752
17	0.851356689
18	0.838784125
19	0.825380973
20	0.811881824
21	0.798554332
22	0.784402801
23	0.770864401
24	0.755818096
25	0.740645032
26	0.726836187
27	0.711587149
28	0.696806216
29	0.681077484
30	0.666885589
31	0.651361819
32	0.635701129
33	0.620894716
34	0.605122619
35	0.590071544
36	0.574970156
37	0.559921975
38	0.546555924
39	0.529783144
40	0.514896828
41	0.500220118
42	0.484188928
43	0.47188357
44	0.456247448
45	0.441477326
46	0.428439806
47	0.414228971
48	0.400974885
49	0.386370614
50	0.373961741
};

\addplot [AtomiqueStyle]
table {
5	0.979971
6	0.973866
7	0.969949
8	0.964998
9	0.959885
10	0.949575
11	0.945958
12	0.942266
13	0.932479
14	0.925192
15	0.916275
16	0.910268
17	0.916476
18	0.910503
19	0.895779
20	0.886814
21	0.89162
22	0.883707
23	0.882338
24	0.862582
25	0.851023
26	0.860781
27	0.855476
28	0.845825
29	0.840284
30	0.829321
31	0.823337
32	0.81796
33	0.804475
34	0.80198
35	0.782728
36	0.775471
37	0.788595
38	0.766137
39	0.774298
40	0.733671
41	0.755167
42	0.757903
43	0.754095
44	0.720625
45	0.736188
46	0.729889
47	0.716167
48	0.700213
49	0.704064
50	0.696459
};

\end{axis}
\end{tikzpicture}

%% file: tex/tex_methods_comp/tikz_Wstate_F_comp.tex
\begin{tikzpicture}
\begin{axis}[Qubits_axis]
\addplot [DasAtomStyle]
table {%
5	0.960688945
6	0.95110379
7	0.941613767
8	0.932217938
9	0.922915373
10	0.91370515
11	0.904586358
12	0.895558094
13	0.886619464
14	0.877769583
15	0.869007574
16	0.86033257
17	0.851743712
18	0.843240148
19	0.834821036
20	0.826485541
21	0.818232838
22	0.810062109
23	0.801972543
24	0.79396334
25	0.786033704
26	0.778182849
27	0.770409998
28	0.762714378
29	0.755095227
30	0.74755179
31	0.740083316
32	0.732689067
33	0.725368307
34	0.718120311
35	0.710944358
36	0.703839738
37	0.696805743
38	0.689841677
39	0.682946847
40	0.676120569
41	0.669362165
42	0.662670964
43	0.6560463
44	0.649487516
45	0.642993961
46	0.636564988
47	0.630199959
48	0.623898242
49	0.617659209
50	0.611482241
};

\addplot [TetrisStyle]
table {%
5	0.960684846
6	0.951101127
7	0.927556354
8	0.91830202
9	0.909141109
10	0.847510418
11	0.839053146
12	0.806068221
13	0.822392697
14	0.814189547
15	0.770506562
16	0.774371743
17	0.690010663
18	0.683130266
19	0.608731229
20	0.669561218
21	0.633643255
22	0.656263881
23	0.602655278
24	0.624173816
25	0.539719216
26	0.611740912
27	0.605653494
28	0.556142794
29	0.584751851
30	0.545105989
31	0.539668701
32	0.510728502
33	0.461973889
34	0.508157853
35	0.534267697
36	0.468973187
37	0.381807103
38	0.401414222
39	0.368638168
40	0.370497834
41	0.372335256
42	0.312403319
43	0.304683094
44	0.361301825
45	0.368614085
46	0.300122055
47	0.288323594
48	0.307742794
49	0.291219008
50	0.279730598
};

\addplot [EnolaStyle]
table {%
5	0.951126993
6	0.935404091
7	0.923217324
8	0.907431264
9	0.891502645
10	0.867382625
11	0.855543721
12	0.840528684
13	0.819563576
14	0.802638863
15	0.77713216
16	0.754674023
17	0.739608071
18	0.727975206
19	0.690034056
20	0.684462087
21	0.652099843
22	0.63786159
23	0.611067638
24	0.5924958
25	0.572724605
26	0.55195522
27	0.532566517
28	0.505875121
29	0.498745974
30	0.481738803
31	0.454384396
32	0.436084675
33	0.405799542
34	0.408495898
35	0.37556884
36	0.365143725
37	0.347710508
38	0.325399092
39	0.316305581
40	0.298103166
41	0.286485327
42	0.269358589
43	0.250975778
44	0.245884807
45	0.22267979
46	0.215936647
47	0.196621217
48	0.190801333
49	0.178137827
50	0.167392184
};

\addplot [AtomiqueStyle]
table {
5	0.959432652
6	0.948443359
7	0.939585871
8	0.928398938
9	0.919977343
10	0.905213522
11	0.896909034
12	0.884160063
13	0.869749965
14	0.858442981
15	0.842147286
16	0.831312821
17	0.832204798
18	0.815434245
19	0.802655678
20	0.787328615
21	0.7852988
22	0.772925941
23	0.766269427
24	0.74478347
25	0.729268765
26	0.726049134
27	0.712865237
28	0.703874174
29	0.694885317
30	0.680980469
31	0.663977948
32	0.659022747
33	0.638600724
34	0.633390461
35	0.620248586
36	0.609139732
37	0.60638428
38	0.58765006
39	0.592779007
40	0.557862732
41	0.549198901
42	0.558029995
43	0.549138798
44	0.529046638
45	0.53413132
46	0.508487621
47	0.503983427
48	0.485867498
49	0.467086613
50	0.468081418
};

\end{axis}
\end{tikzpicture}

%% file: tex/tex_data/tikz_rint_range.tex
\begin{tikzpicture}
\begin{axis}[rint_axis,
ytick={0,0.005,0.011},
yticklabel style={
    /pgf/number format/fixed,
    /pgf/number format/precision=3
    },
scaled y ticks=false
]
\addplot [DasAtomStyle]
table {
1	0.00694589
2	0.007900367
3	0.008728971
4	0.009499564
5	0.009876617
6	0.010152719
7	0.010152719
8	0.010152719
};
\addlegendentry{DasAtom}

\addplot [TetrisStyle]
table {
1	2.34747E-06
2	7.34644E-05
3	0.00052667
4	0.001919292
5	0.006685212
6	0.00916674
7	0.010031532
8	0.010182688
};
\addlegendentry{Tetris}

\addplot[only marks, mark=o,line width=1pt, color={rgb,255:red,0;green,0;blue,0}] 
    coordinates {
        (1,0.00694589) (5,0.009876617) 
    };
\addplot[only marks, mark=|,  mark size=5pt,line width=0.75pt, color={rgb,255:red,92;green,167;blue,199},ultra thick] 
    coordinates {
    (1,2.34747E-06) (5,0.006685212)
    };

\node at (axis cs:1,0.00694589) [anchor=north west, font=\small] {0.006946};
\node at (axis cs:1,2.34747E-06) [anchor=north west, font=\small] {2.347e-6};
\node at (axis cs:5,0.009876617) [anchor=south east, font=\small] {0.009877};
\node at (axis cs:5,0.006685212) [anchor=north east, font=\small] {0.006685};

\end{axis}

    \begin{axis}[
        at={(5cm, 2.4cm)}, 
        anchor=north west, 
        width=3cm, 
        height=2.5cm, 
        xtick={7,8},
        xticklabel style={
                font=\footnotesize,
                /pgf/number format/fixed,
                /pgf/number format/precision=4,
                rotate=45,
                anchor=east
            },
        yticklabel style={
        /pgf/number format/fixed,
        /pgf/number format/precision=5
    },
    ytick={0.01005,0.01015},
    grid = both,
        scaled y ticks=false,
        axis background/.style={fill=white}, 
    ]
\addplot [DasAtomStyle]
table {
7	0.010152719
8	0.010152719
};

\addplot [TetrisStyle]
table {
7	0.010031532
8	0.010182688
};
\end{axis}
\end{tikzpicture}

%% file: tex/tex_data/tikz_F_trans_QFT_benchmark.tex
\begin{tikzpicture}
\begin{axis}[F_atom_trans_axis,ymode=log,
legend entries={DasAtom,Tetris, Enola, Atomique},
legend pos=south east,
]
\addplot [DasAtomStyle]
table {%
0.991	0.001651072
0.992	0.002657727
0.993	0.004276086
0.994	0.006876612
0.995	0.011053383
0.996	0.017758605
0.997	0.028517788
0.998	0.045773762
0.999	0.073436431
1	      0.1177609
};

\addplot [TetrisStyle]
table {%
0.991	0.023751235
0.992	0.023751235
0.993	0.023751235
0.994	0.023751235
0.995	0.023751235
0.996	0.023751235
0.997	0.023751235
0.998	0.023751235
0.999	0.023751235
1	0.023751235
};

\addplot [EnolaStyle]
table {%
0.991	6.71372e-09
0.992	3.49591e-08
0.993	1.81734e-07
0.994	9.43169e-07
0.995	4.8868e-06
0.996	2.5278e-05
0.997	0.00013054
0.998	0.000673024
0.999	0.003464204
1	0.017801835
};

\addplot [AtomiqueStyle]
table {%
0.991	0.048179571
0.992	0.048179571
0.993	0.048179571
0.994	0.048179571
0.995	0.048179571
0.996	0.048179571
0.997	0.048179571
0.998	0.048179571
0.999	0.048179571
1	0.048179571
};

\end{axis}
\end{tikzpicture}

%% file: tex/tex_data/tikz_F_cz_QFT_benchmark.tex
\begin{tikzpicture}
\begin{axis}[F_CZ_axis,ymode=log]
\addplot [DasAtomStyle]
table {%
0.95 6.764195935658048e-10
0.96 4.951577227488511e-08
0.97 3.4669618269011044e-06
0.98 0.00023239621186436929
0.99 0.014926837183020377
0.991 0.02258064534281901
0.993 0.051609528785428925
0.995 0.11776090008372139
0.997 0.2682581458421339
0.998 0.4046317325127107
0.999 0.6100819602511983
0.9999 0.8825331888319055
};

\addplot [TetrisStyle]
table {%
0.95 2.407007884506807e-17
0.96 5.943028526994131e-14
0.97 1.353261511225944e-10
0.98 2.846560600769755e-07
0.99 0.0005540170585716745
0.991 0.0011765603630107997
0.993 0.005294285348299142
0.995 0.02375123538372682
0.997 0.10623217367012125
0.998 0.22441494097715448
0.999 0.47372050543212835
0.9999 0.9274018950907676
};

\addplot [EnolaStyle]
table {%
0.95 1.0225388803239163e-10
0.96 7.485265480472717e-09
0.97 5.240982517843539e-07
0.98 3.5131176644159965e-05
0.99 0.0022564797834198434
0.991 0.0034135007361508897
0.993 0.007801777222345272
0.995 0.017801834846735642
0.997 0.040552400713463374
0.998 0.061167902680929975
0.999 0.0922256733061934
0.9999 0.13341193947379063
};

\addplot [AtomiqueStyle]
table {%
0.95 1.7223026292342555e-11
0.96 2.363167511628847e-09
0.97 3.081272329875911e-07
0.98 3.821822864242171e-05
0.99 0.004513946834858784
0.991 0.0072549193160830455
0.993 0.018713790861195394
0.995 0.04817957107394073
0.997 0.12380534231393599
0.998 0.19832131064066677
0.999 0.3175370787103337
0.9999 0.48484410267478034
};

\end{axis}

    \begin{axis}[
        at={(4cm, 2.3cm)}, 
        anchor=north west, 
        width=4cm, 
        height=3cm, 
        xtick={0.998,0.999,0.9999},
        xticklabel style={
                font=\footnotesize,
                /pgf/number format/fixed,
                /pgf/number format/precision=4,
                rotate=45,
                anchor=east
            },
        ytick = {0, 0.4, 0.7, 1},
        axis background/.style={fill=white}, 
    ]
    \addplot [DasAtomStyle]
table {%
0.998 0.4046317325127107
0.999 0.6100819602511983
0.9999 0.8825331888319055
};

\addplot [TetrisStyle]
table {%
0.998 0.22441494097715448
0.999 0.47372050543212835
0.9999 0.9274018950907676
};

\addplot [EnolaStyle]
table {%
0.998 0.061167902680929975
0.999 0.0922256733061934
0.9999 0.13341193947379063
};

\addplot [AtomiqueStyle]
table {%
0.998 0.19832131064066677
0.999 0.3175370787103337
0.9999 0.48484410267478034
};
    \end{axis}
\end{tikzpicture}

%% file: tex/tex_data/tikz_T2_QFT_benchmark.tex
\begin{tikzpicture}
\begin{axis}[T2_axis]
\addplot [DasAtomStyle]
table {%
0.15	0.055315038
0.5	0.099558223
1	0.112919806
1.5	0.1177609
5	0.124889092
10	0.126471831
15	0.127003856
};

\addplot [TetrisStyle]
table {%
0.15	0.023588369
0.5	0.023714946
1	0.023742158
1.5	0.023751235
5	0.02376395
10	0.023766675
15	0.023767584
};

\addplot [EnolaStyle]
table {%
0.15	3.44731e-10
0.5	0.000343928
1	0.006636905
1.5	0.017801835
5	0.070853875
10	0.095260609
15	0.105138862
};

\addplot [AtomiqueStyle]
table {%
0.15	0.000108892
0.5	0.012442077
1	0.034345546
1.5	0.048179571
5	0.077384226
10	0.085654503
15	0.088603208
};

\end{axis}
\end{tikzpicture}

%% file: tex/tex_data/tikz_dis_QFT_benchmark.tex
\begin{tikzpicture}
\begin{axis}[dis_axis]
\addplot [DasAtomStyle]
table {%
1	0.11238292
2	0.111953559
3	0.111525838
4	0.111099752
5	0.110675294
6	0.110252457
7	0.109831236
8	0.109411624
9	0.108993615
10	0.108577203
11	0.108162382
12	0.107749146
13	0.107337488
14	0.106927404
15	0.106518886
16	0.106111929
17	0.105706526
18	0.105302673
19	0.104900362
20	0.104499588
};

\addplot [TetrisStyle]
table {%
1	0.023751235
2	0.023751235
3	0.023751235
4	0.023751235
5	0.023751235
6	0.023751235
7	0.023751235
8	0.023751235
9	0.023751235
10	0.023751235
11	0.023751235
12	0.023751235
13	0.023751235
14	0.023751235
15	0.023751235
16	0.023751235
17	0.023751235
18	0.023751235
19	0.023751235
20	0.023751235
};

\addplot [EnolaStyle]
table {%
1	0.052363273
2	0.044819981
3	0.039777254
4	0.035969491
5	0.032917995
6	0.030382667
7	0.028223905
8	0.026352635
9	0.024708239
10	0.023247461
11	0.021938285
12	0.020756318
13	0.01968253
14	0.018701784
15	0.017801835
16	0.016972636
17	0.016205846
18	0.015494459
19	0.014832542
20	0.01421502
};

\addplot [AtomiqueStyle]
table {%
1	0.051063551
2	0.049600605
3	0.048179571
4	0.046799249
5	0.045458473
6	0.04415611
7	0.042891058
8	0.04166225
9	0.040468646
10	0.039309239
11	0.038183048
12	0.037089122
13	0.036026536
14	0.034994393
15	0.03399182
16	0.033017971
17	0.032072022
18	0.031153174
19	0.03026065
20	0.029393697
};

\end{axis}
\end{tikzpicture}

%% file: tex/T2fixed_tetris.tex

\begin{tikzpicture}
    \begin{axis}[
        xlabel={$f_{\text{cz}}$},
        ylabel={$f_{\text{trans}}$},
        yticklabel style={
            /pgf/number format/fixed,
            /pgf/number format/precision=3,
        },
        xticklabel style={
            /pgf/number format/fixed,
            /pgf/number format/precision=3,
        },
        xticklabels={0.95,,0.96,,0.97,,0.98,,0.99,,1},
        yticklabels={0.96,,0.97,,0.98,,0.99,,1},
        scaled x ticks=false,
        scaled y ticks=false,
        xmin=0.95, xmax=1,
        ymin=0.96, ymax=1,
        xtick={0.950,0.955,0.960,0.965,0.970,0.975,0.980,0.985,0.990,0.995,1},
        ytick={0.960,0.965,0.970,0.975,0.980,0.985,0.990,0.995,1},
        grid=major,
        width=10cm,
        height=8cm,
        legend style={at={(0.5,1.05)}, anchor=south, legend columns=2, nodes={scale=1.2, transform shape}},
        xlabel style = {font =\huge},
        ylabel style = {font =\huge},
    ]
    
    \def\TtwoShort{150000} 
    \def\TtwoLong{1500000} 
    
    \path[fill opacity=0.4,pattern=crosshatch, pattern color=teal, thick, draw=teal]
    (0.95, 0.96) --
    plot[domain=0.95:1, samples=200, smooth]
    ({\x}, {exp((ln(\x)*336 + (125850.85/\TtwoShort))/472)}) --
    (1, 0.96) -- cycle;
    \addlegendimage{area legend,fill=teal, pattern=crosshatch, pattern color=teal, thick, draw=teal,opacity=0.8}
    \addlegendentry{QFT20}
    

    \path [fill opacity=1,pattern=horizontal lines, pattern color={rgb,255:red,222;green,164;blue,65}, thick, draw={rgb,255:red,222;green,164;blue,65}]
    (0.95, 0.96) --
    plot[domain=0.95:1, samples=200, smooth]
    ({\x}, {exp((ln(\x)*1998 + (1524863.41/\TtwoShort))/3160)}) --
    (1, 0.96) -- cycle;
    \addlegendimage{area legend, fill=orange, pattern=horizontal lines, pattern color={rgb,255:red,222;green,164;blue,65}, thick, draw={rgb,255:red,222;green,164;blue,65},opacity=1}
    \addlegendentry{QFT50}


    


    \end{axis}
\end{tikzpicture}


%% file: main.bbl
\begin{thebibliography}{10}
\providecommand{\url}[1]{#1}
\csname url@samestyle\endcsname
\providecommand{\newblock}{\relax}
\providecommand{\bibinfo}[2]{#2}
\providecommand{\BIBentrySTDinterwordspacing}{\spaceskip=0pt\relax}
\providecommand{\BIBentryALTinterwordstretchfactor}{4}
\providecommand{\BIBentryALTinterwordspacing}{\spaceskip=\fontdimen2\font plus
\BIBentryALTinterwordstretchfactor\fontdimen3\font minus
  \fontdimen4\font\relax}
\providecommand{\BIBforeignlanguage}[2]{{%
\expandafter\ifx\csname l@#1\endcsname\relax
\typeout{** WARNING: IEEEtran.bst: No hyphenation pattern has been}%
\typeout{** loaded for the language `#1'. Using the pattern for}%
\typeout{** the default language instead.}%
\else
\language=\csname l@#1\endcsname
\fi
#2}}
\providecommand{\BIBdecl}{\relax}
\BIBdecl

\bibitem{Shor94}
P.~Shor, ``Algorithms for quantum computation: discrete logarithms and
  factoring,'' in \emph{Proceedings 35th Annual Symposium on Foundations of
  Computer Science}, 1994, pp. 124--134.

\bibitem{cao2019quantum}
Y.~Cao, J.~Romero, J.~P. Olson, M.~Degroote, P.~D. Johnson, M.~Kieferov{\'a},
  I.~D. Kivlichan, T.~Menke, B.~Peropadre, N.~P. Sawaya \emph{et~al.},
  ``Quantum chemistry in the age of quantum computing,'' \emph{Chemical
  {R}eviews}, vol. 119, no.~19, pp. 10\,856--10\,915, 2019.

\bibitem{schuld2015introduction}
M.~Schuld, I.~Sinayskiy, and F.~Petruccione, ``An introduction to quantum
  machine learning,'' \emph{Contemporary Physics}, vol.~56, no.~2, pp.
  172--185, 2015.

\bibitem{henriet2020quantum}
L.~Henriet, L.~Beguin, A.~Signoles, T.~Lahaye, A.~Browaeys, G.-O. Reymond, and
  C.~Jurczak, ``Quantum computing with neutral atoms,'' \emph{Quantum}, vol.~4,
  p. 327, 2020.

\bibitem{bluvstein2022quantum}
D.~Bluvstein, H.~Levine, G.~Semeghini, T.~T. Wang, S.~Ebadi, M.~Kalinowski,
  A.~Keesling, N.~Maskara, H.~Pichler, M.~Greiner \emph{et~al.}, ``A quantum
  processor based on coherent transport of entangled atom arrays,''
  \emph{Nature}, vol. 604, no. 7906, pp. 451--456, 2022.

\bibitem{radnaev2024universal}
\BIBentryALTinterwordspacing
A.~G. Radnaev, W.~C. Chung, D.~C. Cole, D.~Mason, T.~G. Ballance, M.~J.
  Bedalov, D.~A. Belknap, M.~R. Berman, M.~Blakely, I.~L. Bloomfield, P.~D.
  Buttler, C.~Campbell, A.~Chopinaud, E.~Copenhaver, M.~K. Dawes, S.~Y.
  Eubanks, A.~J. Friss, D.~M. Garcia, J.~Gilbert, M.~Gillette, P.~Goiporia,
  P.~Gokhale, J.~Goldwin, D.~Goodwin, T.~M. Graham, C.~Guttormsson, G.~T.
  Hickman, L.~Hurtley, M.~Iliev, E.~B. Jones, R.~A. Jones, K.~W. Kuper, T.~B.
  Lewis, M.~T. Lichtman, F.~Majdeteimouri, J.~J. Mason, J.~K. McMaster, J.~A.
  Miles, P.~T. Mitchell, J.~D. Murphree, N.~A. Neff-Mallon, T.~Oh, V.~Omole,
  C.~P. Simon, N.~Pederson, M.~A. Perlin, A.~Reiter, R.~Rines, P.~Romlow, A.~M.
  Scott, D.~Stiefvater, J.~R. Tanner, A.~K. Tucker, I.~V. Vinogradov, M.~L.
  Warter, M.~Yeo, M.~Saffman, and T.~W. Noel, ``A universal neutral-atom
  quantum computer with individual optical addressing and non-destructive
  readout,'' 2024. [Online]. Available: \url{https://arxiv.org/abs/2408.08288}
\BIBentrySTDinterwordspacing

\bibitem{Lengwenus10coherent}
\BIBentryALTinterwordspacing
A.~Lengwenus, J.~Kruse, M.~Schlosser, S.~Tichelmann, and G.~Birkl, ``Coherent
  transport of atomic quantum states in a scalable shift register,''
  \emph{Phys. Rev. Lett.}, vol. 105, p. 170502, Oct. 2010. [Online]. Available:
  \url{https://link.aps.org/doi/10.1103/PhysRevLett.105.170502}
\BIBentrySTDinterwordspacing

\bibitem{Baker21isca-neutral_atom}
J.~M. Baker, A.~Litteken, C.~Duckering, H.~Hoffmann, H.~Bernien, and F.~T.
  Chong, ``Exploiting long-distance interactions and tolerating atom loss in
  neutral atom quantum architectures,'' in \emph{2021 ACM/IEEE 48th Annual
  International Symposium on Computer Architecture (ISCA)}, 2021, pp. 818--831.

\bibitem{Patel22isca-geyser}
\BIBentryALTinterwordspacing
T.~Patel, D.~Silver, and D.~Tiwari, ``Geyser: a compilation framework for
  quantum computing with neutral atoms,'' in \emph{Proceedings of the 49th
  Annual International Symposium on Computer Architecture}, ser. ISCA'22.\hskip
  1em plus 0.5em minus 0.4em\relax New York, NY, USA: Association for Computing
  Machinery, 2022, p. 383–395. [Online]. Available:
  \url{https://doi.org/10.1145/3470496.3527428}
\BIBentrySTDinterwordspacing

\bibitem{Li23tetris}
\BIBentryALTinterwordspacing
Y.~Li, Y.~Zhang, M.~Chen, X.~Li, and P.~Xu, ``Timing-aware qubit mapping and
  gate scheduling adapted to neutral atom quantum computing,'' \emph{{IEEE}
  Trans. Comput. Aided Des. Integr. Circuits Syst.}, vol.~42, no.~11, pp.
  3768--3780, 2023. [Online]. Available:
  \url{https://doi.org/10.1109/TCAD.2023.3261244}
\BIBentrySTDinterwordspacing

\bibitem{Tan22reconfigurable}
\BIBentryALTinterwordspacing
B.~Tan, D.~Bluvstein, M.~D. Lukin, and J.~Cong, ``Qubit mapping for
  reconfigurable atom arrays,'' in \emph{Proceedings of the 41st IEEE/ACM
  International Conference on Computer-Aided Design}, ser. ICCAD '22.\hskip 1em
  plus 0.5em minus 0.4em\relax New York, NY, USA: Association for Computing
  Machinery, 2022. [Online]. Available:
  \url{https://doi.org/10.1145/3508352.3549331}
\BIBentrySTDinterwordspacing

\bibitem{Tan2024compilingquantum}
\BIBentryALTinterwordspacing
D.~B. Tan, D.~Bluvstein, M.~D. Lukin, and J.~Cong, ``Compiling {Q}uantum
  {C}ircuits for {D}ynamically {F}ield-{P}rogrammable {N}eutral {A}toms {A}rray
  {P}rocessors,'' \emph{{Quantum}}, vol.~8, p. 1281, Mar. 2024. [Online].
  Available: \url{https://doi.org/10.22331/q-2024-03-14-1281}
\BIBentrySTDinterwordspacing

\bibitem{wang2024atomique}
\BIBentryALTinterwordspacing
H.~Wang, P.~Liu, D.~Tan, Y.~Liu, J.~Gu, D.~Z. Pan, J.~Cong, U.~A. Acar, and
  S.~Han, ``Atomique: A quantum compiler for reconfigurable neutral atom
  arrays,'' in \emph{2024 ACM/IEEE 51st Annual International Symposium on
  Computer Architecture (ISCA)}.\hskip 1em plus 0.5em minus 0.4em\relax Los
  Alamitos, CA, USA: IEEE Computer Society, Jul. 2024, pp. 293--309. [Online].
  Available:
  \url{https://doi.ieeecomputersociety.org/10.1109/ISCA59077.2024.00030}
\BIBentrySTDinterwordspacing

\bibitem{Wang23Q-Pilot}
H.~Wang, D.~Tan, P.~Liu, Y.~Liu, J.~Gu, J.~Cong, and S.~Han, ``Q-pilot: Field
  programmable qubit array compilation with flying ancillas,'' in
  \emph{Proceedings of the 61st ACM/IEEE Design Automation Conference
  (DAC)}.\hskip 1em plus 0.5em minus 0.4em\relax ACM/IEEE, 2024.

\bibitem{tan2024compilation}
D.~B. Tan, W.-H. Lin, and J.~Cong, ``Compilation for dynamically
  field-programmable qubit arrays with efficient and provably near-optimal
  scheduling,'' \emph{arXiv:2405.15095}, 2024.

\bibitem{Zulehner+18_Astar}
A.~Zulehner, A.~Paler, and R.~Wille, ``An efficient methodology for mapping
  quantum circuits to the {IBM QX} architectures,'' \emph{IEEE Transactions on
  Computer-Aided Design of Integrated Circuits and Systems}, vol.~38, no.~7,
  pp. 1226--1236, 2018.

\bibitem{Li+19-sabre}
\BIBentryALTinterwordspacing
G.~Li, Y.~Ding, and Y.~Xie, ``Tackling the qubit mapping problem for {NISQ}-era
  quantum devices,'' in \emph{Proceedings of the Twenty-Fourth International
  Conference on Architectural Support for Programming Languages and Operating
  Systems, {ASPLOS} 2019, Providence, RI, USA, April 13-17, 2019}, I.~Bahar,
  M.~Herlihy, E.~Witchel, and A.~R. Lebeck, Eds.\hskip 1em plus 0.5em minus
  0.4em\relax {ACM}, 2019, pp. 1001--1014. [Online]. Available:
  \url{https://doi.org/10.1145/3297858.3304023}
\BIBentrySTDinterwordspacing

\bibitem{Sivarajah20qst_tket}
S.~Sivarajah, S.~Dilkes, A.~Cowtan, W.~Simmons, A.~Edgington, and R.~Duncan,
  ``t$|\text{ket}\rangle$: {A} retargetable compiler for {NISQ} devices,''
  \emph{Quantum Science and Technology}, vol.~6, no.~1, p. 014003, 2020.

\bibitem{LiZF21_fidls}
\BIBentryALTinterwordspacing
S.~Li, X.~Zhou, and Y.~Feng, ``Qubit mapping based on subgraph isomorphism and
  filtered depth-limited search,'' \emph{{IEEE} Trans. Computers}, vol.~70,
  no.~11, pp. 1777--1788, 2021. [Online]. Available:
  \url{https://doi.org/10.1109/TC.2020.3023247}
\BIBentrySTDinterwordspacing

\bibitem{ZhouFL20_MCTS_iccad}
X.~Zhou, Y.~Feng, and S.~Li, ``A monte carlo tree search framework for quantum
  circuit transformation,'' in \emph{2020 IEEE/ACM International Conference on
  Computer-Aided Design (ICCAD)}.\hskip 1em plus 0.5em minus 0.4em\relax IEEE,
  2020, pp. 1--7.

\bibitem{farhi2014quantum}
E.~Farhi, J.~Goldstone, and S.~Gutmann, ``A quantum approximate optimization
  algorithm,'' \emph{arXiv:1411.4028}, 2014.

\bibitem{grimm2000optical}
R.~Grimm, M.~Weidem{\"u}ller, and Y.~B. Ovchinnikov, ``Optical dipole traps for
  neutral atoms,'' in \emph{Advances in atomic, molecular, and optical
  physics}.\hskip 1em plus 0.5em minus 0.4em\relax Elsevier, 2000, vol.~42, pp.
  95--170.

\bibitem{Schmid24qst}
\BIBentryALTinterwordspacing
L.~Schmid, D.~F. Locher, M.~Rispler, S.~Blatt, J.~Zeiher, M.~Müller, and
  R.~Wille, ``Computational capabilities and compiler development for neutral
  atom quantum processors—connecting tool developers and hardware experts,''
  \emph{Quantum Science and Technology}, vol.~9, no.~3, p. 033001, Apr. 2024.
  [Online]. Available: \url{https://dx.doi.org/10.1088/2058-9565/ad33ac}
\BIBentrySTDinterwordspacing

\bibitem{endres2016atom}
M.~Endres, H.~Bernien, A.~Keesling, H.~Levine, E.~R. Anschuetz, A.~Krajenbrink,
  C.~Senko, V.~Vuletic, M.~Greiner, and M.~D. Lukin, ``Atom-by-atom assembly of
  defect-free one-dimensional cold atom arrays,'' \emph{Science}, vol. 354, no.
  6315, pp. 1024--1027, 2016.

\bibitem{barredo2016atom}
D.~Barredo, S.~De~L{\'e}s{\'e}leuc, V.~Lienhard, T.~Lahaye, and A.~Browaeys,
  ``An atom-by-atom assembler of defect-free arbitrary two-dimensional atomic
  arrays,'' \emph{Science}, vol. 354, no. 6315, pp. 1021--1023, 2016.

\bibitem{barredo2018synthetic}
D.~Barredo, V.~Lienhard, S.~De~Leseleuc, T.~Lahaye, and A.~Browaeys,
  ``Synthetic three-dimensional atomic structures assembled atom by atom,''
  \emph{Nature}, vol. 561, no. 7721, pp. 79--82, 2018.

\bibitem{Schlosser23scalable}
\BIBentryALTinterwordspacing
M.~Schlosser, S.~Tichelmann, D.~Sch\"affner, D.~O. de~Mello, M.~Hambach,
  J.~Sch\"utz, and G.~Birkl, ``Scalable multilayer architecture of assembled
  single-atom qubit arrays in a three-dimensional talbot tweezer lattice,''
  \emph{Phys. Rev. Lett.}, vol. 130, p. 180601, May 2023. [Online]. Available:
  \url{https://link.aps.org/doi/10.1103/PhysRevLett.130.180601}
\BIBentrySTDinterwordspacing

\bibitem{jaksch2000fast}
D.~Jaksch, J.~I. Cirac, P.~Zoller, S.~L. Rolston, R.~C{\^o}t{\'e}, and M.~D.
  Lukin, ``Fast quantum gates for neutral atoms,'' \emph{Physical Review
  Letters}, vol.~85, no.~10, p. 2208, 2000.

\bibitem{Zhou+20_SAHS}
X.~Zhou, S.~Li, and Y.~Feng, ``Quantum circuit transformation based on
  simulated annealing and heuristic search,'' \emph{IEEE Transactions on
  Computer-Aided Design of Integrated Circuits and Systems}, vol.~39, no.~12,
  pp. 4683--4694, 2020.

\bibitem{lao22tcad-timing}
L.~Lao, H.~van Someren, I.~Ashraf, and C.~G. Almudever, ``Timing and
  resource-aware mapping of quantum circuits to superconducting processors,''
  \emph{IEEE Transactions on Computer-Aided Design of Integrated Circuits and
  Systems}, vol.~41, no.~2, pp. 359--371, 2022.

\bibitem{zhu22iterated_local_search}
P.~Zhu, S.~Feng, and Z.~Guan, ``An iterated local search methodology for the
  qubit mapping problem,'' \emph{IEEE Transactions on Computer-Aided Design of
  Integrated Circuits and Systems}, vol.~41, no.~8, pp. 2587--2597, 2022.

\bibitem{Brandhofer_iccad21}
\BIBentryALTinterwordspacing
S.~Brandhofer, I.~Polian, and H.~P. B{\"{u}}chler, ``Optimal mapping for
  near-term quantum architectures based on rydberg atoms,'' in \emph{{IEEE/ACM}
  International Conference On Computer Aided Design, {ICCAD} 2021, Munich,
  Germany, November 1-4, 2021}.\hskip 1em plus 0.5em minus 0.4em\relax {IEEE},
  2021, pp. 1--7. [Online]. Available:
  \url{https://doi.org/10.1109/ICCAD51958.2021.9643490}
\BIBentrySTDinterwordspacing

\bibitem{schmid2023hybrid}
L.~Schmid, S.~Park, S.~Kang, and R.~Wille, ``Hybrid circuit mapping: Leveraging
  the full spectrum of computational capabilities of neutral atom quantum
  computers,'' in \emph{Proceedings of the 61st ACM/IEEE Design Automation
  Conference (DAC)}.\hskip 1em plus 0.5em minus 0.4em\relax ACM/IEEE, 2024.

\bibitem{siraichi+19_bmt}
M.~Y. Siraichi, V.~F.~d. Santos, C.~Collange, and F.~M.~Q. Pereira, ``Qubit
  allocation as a combination of subgraph isomorphism and token swapping,''
  \emph{Proceedings of the ACM on Programming Languages}, vol.~3, no. OOPSLA,
  pp. 1--29, 2019.

\bibitem{iccad22/Wu_robust}
\BIBentryALTinterwordspacing
T.-A. Wu, Y.-J. Jiang, and S.-Y. Fang, ``A robust quantum layout synthesis
  algorithm with a qubit mapping checker,'' ser. ICCAD '22.\hskip 1em plus
  0.5em minus 0.4em\relax New York, NY, USA: Association for Computing
  Machinery, 2022. [Online]. Available:
  \url{https://doi.org/10.1145/3508352.3549394}
\BIBentrySTDinterwordspacing

\bibitem{huang2024adac}
\BIBentryALTinterwordspacing
Y.~Huang, X.~Zhou, F.~Meng, and S.~Li, ``Qubit mapping: The adaptive
  divide-and-conquer approach,'' \emph{arXiv}, vol. {2409.04752}, 2024.
  [Online]. Available: \url{https://arxiv.org/abs/2409.04752}
\BIBentrySTDinterwordspacing

\bibitem{nottingham24circuit}
\BIBentryALTinterwordspacing
N.~Nottingham, M.~A. Perlin, D.~Shah, R.~White, H.~Bernien, F.~T. Chong, and
  J.~M. Baker, ``Circuit decompositions and scheduling for neutral atom devices
  with limited local addressability,'' in \emph{IEEE International Conference
  on Quantum Computing and Engineering (QCE) 2024}, 2024. [Online]. Available:
  \url{https://arxiv.org/abs/2307.14996}
\BIBentrySTDinterwordspacing

\bibitem{Cordella+04-vf2}
\BIBentryALTinterwordspacing
L.~P. Cordella, P.~Foggia, C.~Sansone, and M.~Vento, ``A (sub)graph isomorphism
  algorithm for matching large graphs,'' \emph{{IEEE} Trans. Pattern Anal.
  Mach. Intell.}, vol.~26, no.~10, pp. 1367--1372, 2004. [Online]. Available:
  \url{https://doi.org/10.1109/TPAMI.2004.75}
\BIBentrySTDinterwordspacing

\bibitem{Evered23high}
\BIBentryALTinterwordspacing
S.~J. Evered, D.~Bluvstein, M.~Kalinowski, S.~Ebadi, T.~Manovitz, H.~Zhou,
  S.~H. Li, A.~A. Geim, T.~T. Wang, N.~Maskara, H.~Levine, G.~Semeghini,
  M.~Greiner, V.~Vuletić, and M.~D. Lukin, ``High-fidelity parallel entangling
  gates on a neutral-atom quantum computer,'' \emph{Nature}, vol. 622, no.
  7982, p. 268–272, Oct. 2023. [Online]. Available:
  \url{http://dx.doi.org/10.1038/s41586-023-06481-y}
\BIBentrySTDinterwordspacing

\bibitem{Graham22multi}
\BIBentryALTinterwordspacing
T.~M. Graham, Y.~Song, J.~Scott, C.~Poole, L.~Phuttitarn, K.~Jooya, P.~Eichler,
  X.~Jiang, A.~Marra, B.~Grinkemeyer, M.~Kwon, M.~Ebert, J.~Cherek, M.~T.
  Lichtman, M.~Gillette, J.~Gilbert, D.~Bowman, T.~Ballance, C.~Campbell, E.~D.
  Dahl, O.~Crawford, N.~S. Blunt, B.~Rogers, T.~Noel, and M.~Saffman,
  ``Multi-qubit entanglement and algorithms on a neutral-atom quantum
  computer,'' \emph{Nature}, vol. 604, no. 7906, 4 2022. [Online]. Available:
  \url{https://www.osti.gov/biblio/1870309}
\BIBentrySTDinterwordspacing

\bibitem{Wang24pathlad+}
\BIBentryALTinterwordspacing
Y.~Wang, C.~Jin, and S.~Cai, ``Pathlad+: Towards effective exact methods for
  subgraph isomorphism problem,'' \emph{Artificial Intelligence}, vol. 337, p.
  104219, 2024. [Online]. Available:
  \url{https://www.sciencedirect.com/science/article/pii/S0004370224001553}
\BIBentrySTDinterwordspacing

\end{thebibliography}
